\documentclass[12pt]{article}
\usepackage[utf8]{inputenc}
\usepackage[english]{babel}
\usepackage[centertags]{amsmath}
\usepackage{amssymb}
\usepackage{bm}
\usepackage{amsbsy}
\usepackage{graphicx}
\usepackage{appendix}
\usepackage{mathrsfs}
\usepackage{epsfig}
\usepackage{color}
\usepackage{euscript}
\usepackage{ulem}
\usepackage{multirow}
\usepackage{cite}

\definecolor{dgreen}{rgb}{0,0.6,0}

\definecolor{darkblue}{rgb}{0., 0, 1}

\definecolor{purple}{rgb}{0.65,0.,0.78}

\definecolor{orange}{rgb}{0.89,0.3,0.12}

\usepackage{jheppubm}

\newcommand{\nn}{\nonumber}
\newcommand{\be}{\begin{equation}}
\newcommand{\ee}{\end{equation}}
\newcommand{\bea}{\begin{eqnarray}}
\newcommand{\eea}{\end{eqnarray}}

\newcommand{\fb}{\mathfrak{b}}
\newcommand{\fc}{\mathfrak{c}}
\newcommand{\ff}{\mathfrak{f}}

\newcommand{\fg}{\mathfrak{g}}

\newcommand{\fz}{\mathfrak{z}}

\newcommand{\fG}{\mathfrak{G}}

\newcommand{\cV}{{\cal V}}

\numberwithin{equation}{section}

\title{
 Running Coupling for Holographic QCD with Heavy and Light Quarks: Isotropic case
}

\author{Irina Ya. Aref'eva$^a$, Ali Hajilou$^a$, Pavel Slepov$^a$ and  Marina Usova$^a$}

\affiliation{$^a$Steklov Mathematical Institute, Russian Academy of  Sciences, \\ Gubkina str. 8, 119991, Moscow, Russia}

\emailAdd{arefeva@mi-ras.ru}
\emailAdd{hajilou@mi-ras.ru}
\emailAdd{slepov@mi-ras.ru}
\emailAdd{usovamk@mi-ras.ru}

 \abstract{We consider the running coupling constant in  holographic models supported by Einstein-dilaton-Maxwell action for heavy and light quarks.  To obtain the dependence of the running coupling constant $\alpha$ on temperature and chemical potential we impose boundary conditions on the dilaton field that depend on the position of the horizon.  We use two types of boundary conditions: a simple boundary condition with the dilaton field vanishing at the horizon and a boundary condition that ensures an agreement with lattice calculations of string tension between quarks at zero chemical potential. The location of the 1st  order phase transitions in $(\mu,T)$-plane does not depend on the dilaton boundary conditions for light and heavy quarks. At these  phase transitions,  the function $\alpha$  undergoes jumps depending on temperature and chemical potential. We also show that for the second boundary conditions the running coupling decreases with a temperature increase, and the dependence on temperature and chemical potential both for light and heavy quarks is actually specified in QGP phase by functions of one variable, demonstrating in this sense auto-model behavior.\\

}


\begin{document}

\maketitle

\newpage

\section{Introduction}

The running coupling constant is related to the $\beta$-function via the renormalization group (RG) flow \cite{BogSchirkov,Wilson:1973jj,ModernTextBook}.
The dependence of coupling constants of a physical system on the energy scale can be described by $\beta$-function of the theory \cite{GellMannLow,{Callan:1970yg,Symanzik:1970rt,Wilson:1973jj}}. Holographic duality describes a correspondence between a class of strongly coupled field theories and weakly-coupled gravitational theories \cite{Maldacena:1997re,Casalderrey-Solana:2011dxg,Arefeva:2014kyw}. Then, one can investigate the strongly coupled regime of gauge theories via holography. 
The studies of running coupling in holographic models were extensively performed in \cite{Gursoy:2007cb, Gursoy:2007er,Kiritsis:2014kua,Kiritsis:2016kog,Gursoy:2018umf,Ghosh:2017big} in context of the QCD applications.\\

Investigation of the QCD phase transitions diagram in (temperature, chemical potential)-plane is a challenging and very important task in high energy physics. This is not only an issue of the fundamental interest but has also important implications on the description of the early evolution of the Universe \cite{Rubakov-book} and on the study of the interior of compact stellar objects \cite{NS,Lovato:2022vgq}. One of the goals of the experiments on LHC, RHIC, NICA and FAIR is to study the phase structure of the strong-interaction matter  under extreme conditions. Standard theoretical methods for performing QCD calculations, such as perturbation theory, no longer work in the strongly coupled regime. The lattice theory has a sign problem for non-zero chemical
potential calculations and cannot provide suitable information at its current level. Therefore, to describe physics of the strongly
coupled  QGP produced in heavy ion collisions
(HIC) at RHIC, LHC and NICA (e.g. see \cite{Du:2024wjm}) and future experiments, we need a
non-perturbative approach \cite{Casalderrey-Solana:2011dxg,
  Arefeva:2014kyw}. 
  \\

\begin{figure}[t!]
  \centering
\includegraphics[scale=0.46]{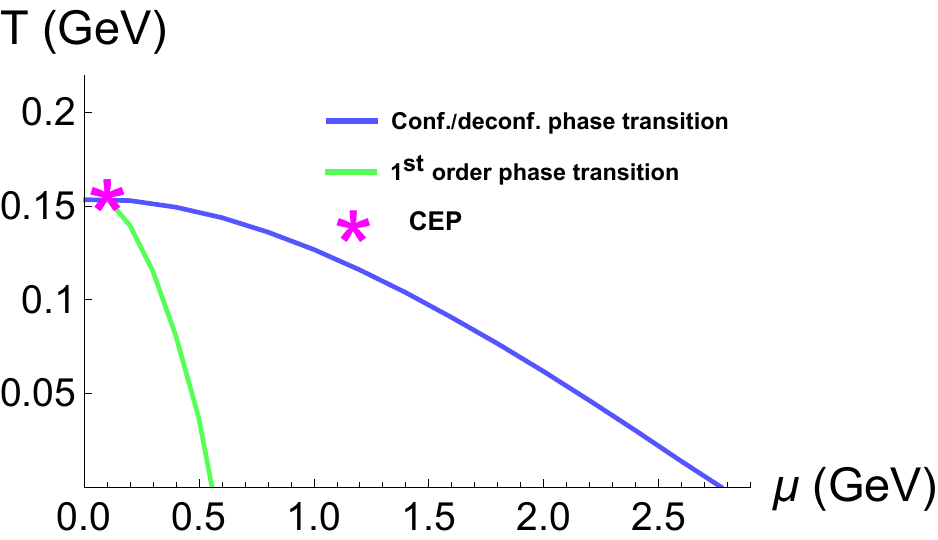} 
\quad\includegraphics[scale=0.51]{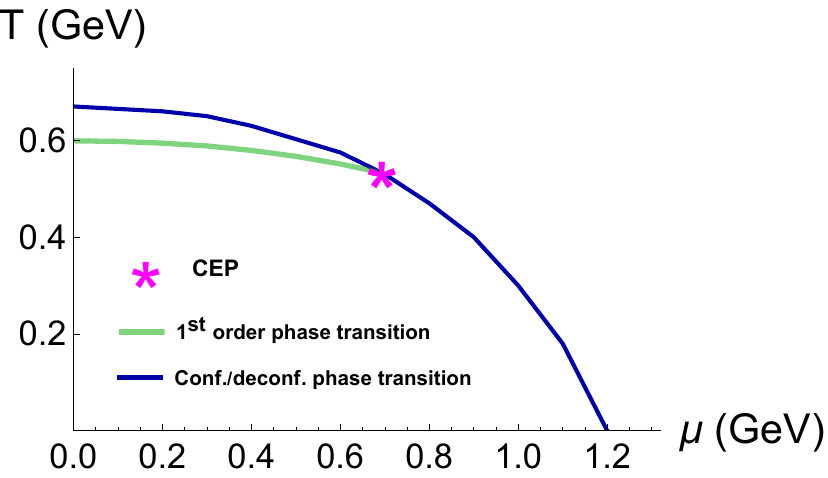}
\\
  A\hspace{200pt}B\\
\caption{Phase structure (A) for light  and (B) for heavy quarks model. The magenta star shows the critical end point (CEP). 
  }
 \label{Fig:PD-LH-HQ-paint}
\end{figure}

  It is expected that the QCD phase diagram structure essentially depends on the quark masses, in particular from lattice calculations \cite{Brown:1990ev,
  Philipsen:2016hkv,
  Guenther:2020jwe,
  Aarts:2023vsf}, and some effective phenomenological approach, e.g. see  \cite{Fu:2019hdw,Dumm:2021vop,
  Du:2024wjm}, it is expected that the QCD phase diagrams have the form presented in Fig.\,\ref{Fig:PD-LH-HQ-paint}. Here regions of a hadronic phase are filled by brown color, the regions where a quark-gluon phase is realized is filled by blue color. The intermediate regions are filled by green color. The hadronic phase is bounded from above mostly by the 1st  order phase transition for light quarks model (the light green  line in Fig.\,\ref{Fig:PD-LH-HQ-paint}A) and both by the 1st  order phase transition line (the light green  line in Fig.\,\ref{Fig:PD-LH-HQ-paint}B) and by the line dividing confinement and deconfinement phases  (the blue  line in Fig.\,\ref{Fig:PD-LH-HQ-paint}) for heavy quarks model. These plots represent an  essential difference between light and heavy quarks.
\\
 
The holographic QCD
models constructed  in \cite{Yang:2015aia,Arefeva:2018hyo,Li:2017tdz,Arefeva:2020byn,Arefeva:2022bhx, Hajilou:2021wmz} 
reproduce phase diagram features for heavy and light quarks at small chemical potential and
predict new phenomena for finite chemical potential, in particular, the locations of the critical end points (CEP). Here CEP corresponds to the end point of the 1st  order phase transition line. The goal of this paper is to study a behavior of the running coupling $\alpha$ at different phases: quark confinement phases (i.e. hadronic and quarkyonic) and  QGP phase, -- as well as near critical lines. 
\\

\begin{figure}[t!]
  \centering
\includegraphics[scale=0.44]{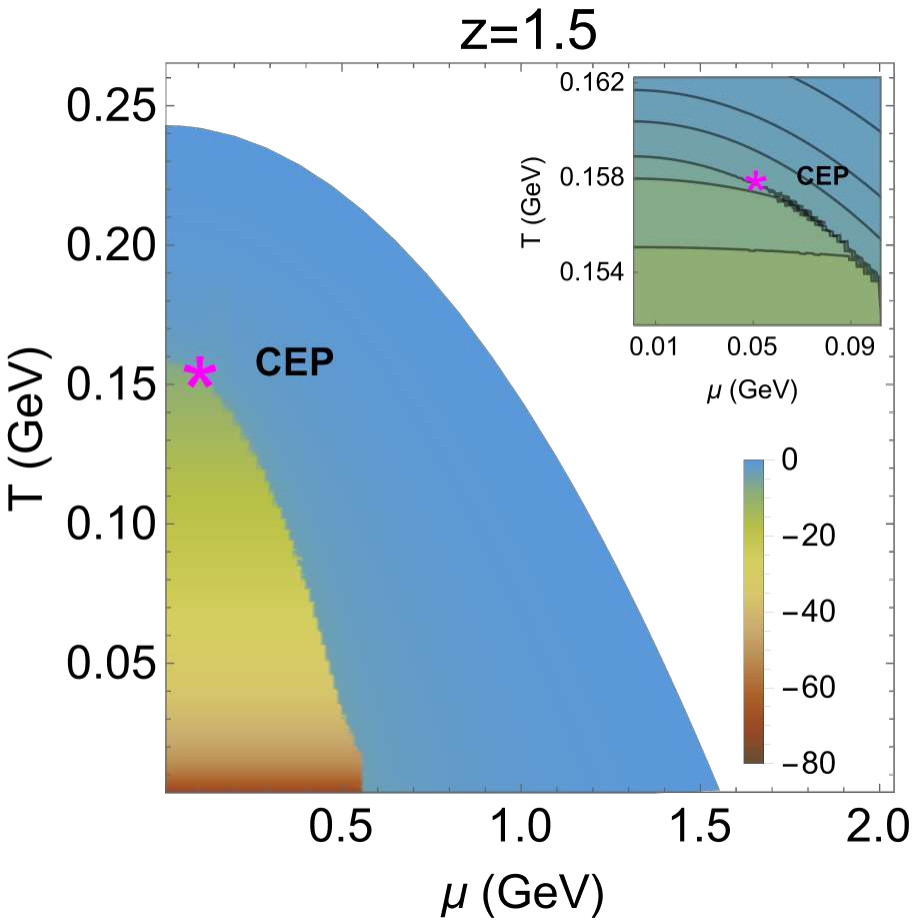}
 \qquad  \quad 
  \includegraphics[scale=0.54]{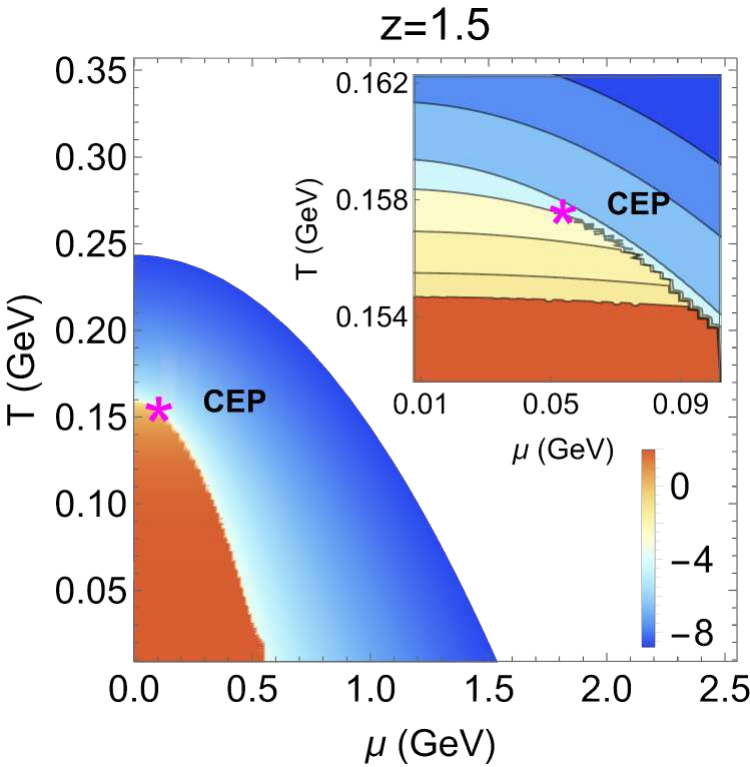}\\
  A\hspace{200pt}B\\\,\\
\includegraphics[scale=0.44]{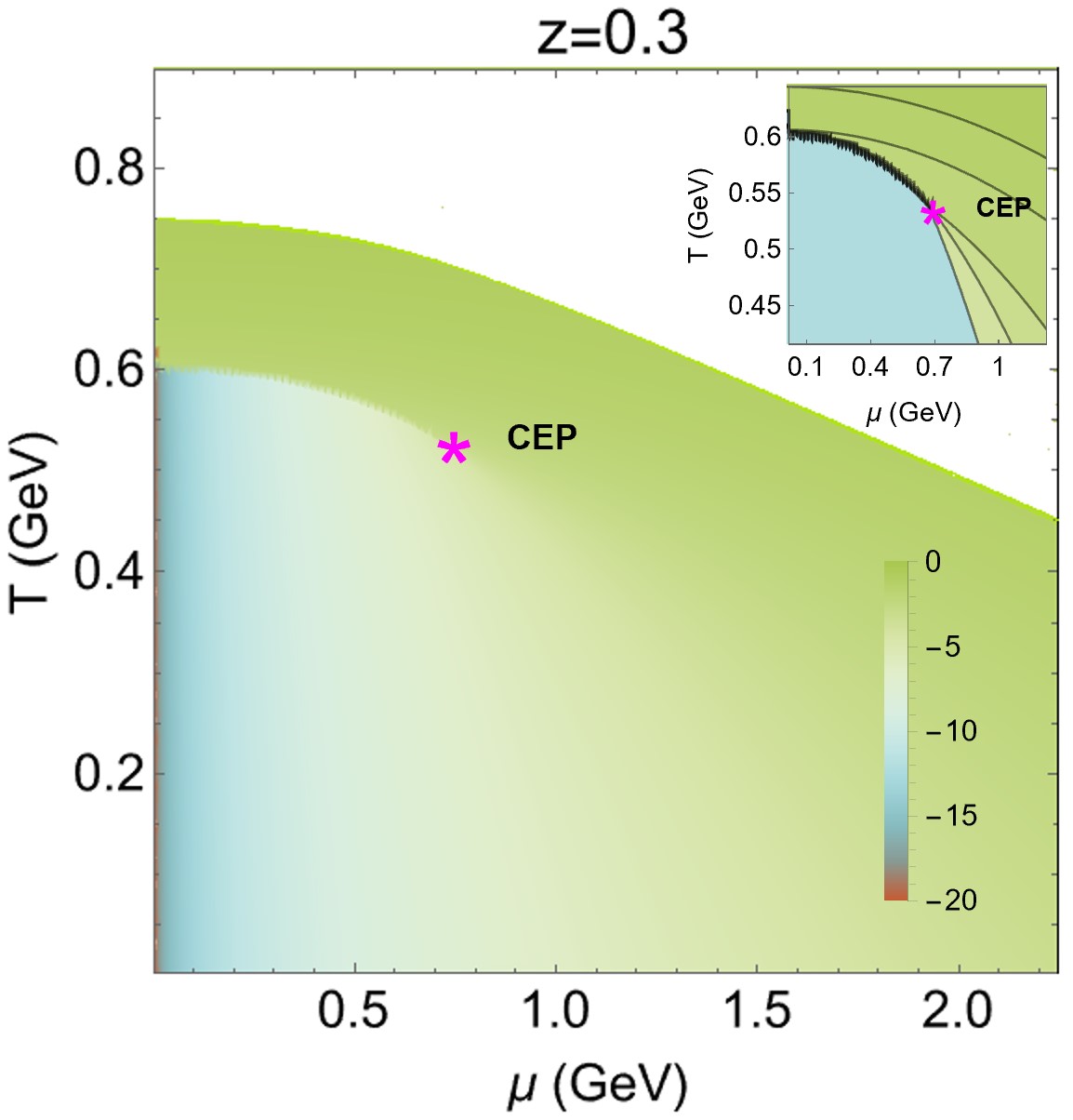} \qquad  \quad
 \includegraphics[scale=0.44]{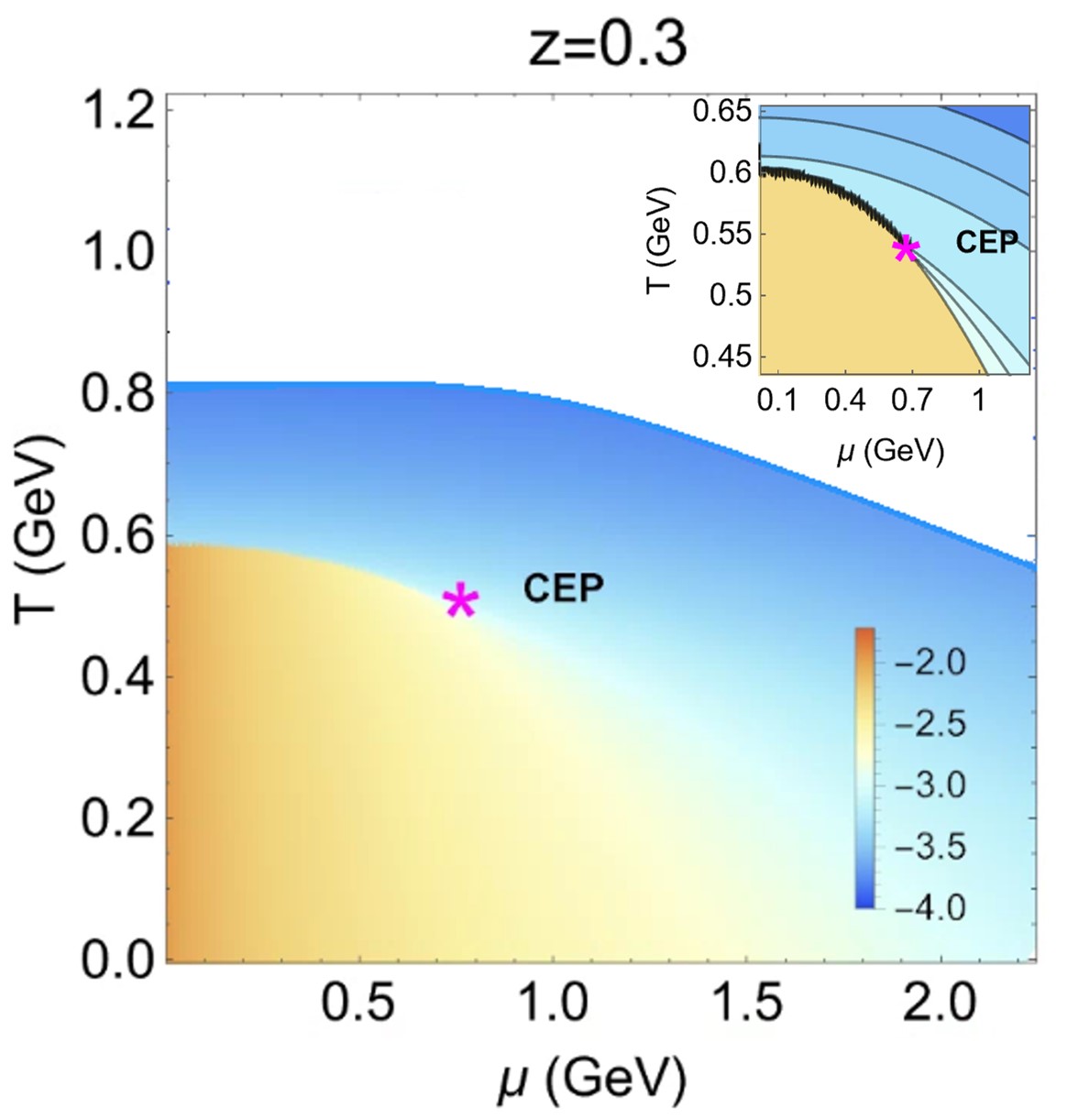}\\
  C\hspace{200pt}D 
 \caption{The density plots of $\log \alpha_{z_h}(z;\mu,T)$  with associated boundary condition
 $z_0=z_h$ (A) and $\log \alpha_{\fz_{_{LQ}}}(z;\mu,T)$ with $z_0=\fz_{LQ}(z_h)$ (B) for light quarks at energy scale $z=1.5$ and 
 $\log \alpha_{z_h}(z;\mu,T)$
 with $z_0=z_h$ (C) and $\log \alpha_{\fz_{_{HQ}}}(z;\mu,T)$ with $z_0=\fz_{HQ}(z_h)$ (D) for heavy quarks at energy scale  $z=0.3$   with zoom areas around CEP's denoted by  magenta stars. Functions $\fz_{LQ}(z_h)$ and $\fz_{HQ}(z_h)$ are given by \eqref{phi-fz-LQ} and \eqref{bceHQ}; $[z]^{-1} =$ GeV.
}
 \label{Fig:logLH}
\end{figure}

Scattering amplitudes of particles formed during heavy ions collisions and  interacting with particles belonging to the strong interactions sector significantly depend on the running QCD coupling constants.
Typical behavior of the running coupling  $\alpha$, in fact the logarithm\footnote{Here $\log\alpha \equiv \log_{e}\alpha$.} of the running coupling  $\log\alpha$, predicted by holographic models, at a certain fixed energy scale, different  temperatures, and chemical potentials is presented in Fig.\,\ref{Fig:logLH} for light  and heavy quarks. A distinctive feature of these diagrams is the indication of coupling constant jumps along the 1st  order phase transition lines. The detailed investigation of these jumps is performed in the main text. These jumps disappear  at CEP marked with  magenta stars in Fig.\,\ref{Fig:logLH}. A more detailed examination of the plots from the first and second columns in Fig.\,\ref{Fig:logLH} shows that there is one significant difference between them.   
Namely, a direction of the coupling constant logarithm increase is oriented from brown to blue in Fig.\,\ref{Fig:logLH}A 
and from blue to brown 
 in Fig.\,\ref{Fig:logLH}B. The same happens in Fig.\,\ref{Fig:logLH}C and Fig.\,\ref{Fig:logLH}D, i.e. the coupling constant logarithm increases from brown to blue and then to green in  Fig.\,\ref{Fig:logLH}C and from blue to brown in Fig.\,\ref{Fig:logLH}D. Such differences are associated with the choice of different boundary conditions for the corresponding holographic models. Let us explain this briefly.
\\

The characteristic feature of the holographic coupling constant $\alpha$ is its dependence on the dilaton field, and it is defined as \cite{Gursoy:2007cb,Gursoy:2007er,Pirner:2009gr,Peet:1998wn}
\be\label{lambda-phi}
\alpha(z)=e^{\varphi(z)} \,.
\ee
It is important to note that, in the holographic approach (gauge/gravity duality) $\alpha$ is 't Hooft coupling\footnote{In most papers and books $\lambda$ is used as 't Hooft coupling.} constant that is defined by Yang-Mills coupling $g_{YM}$ and number of the colors $N_c$ (the rank of the gauge group), i.e. $\alpha=g_{YM}^2\, N_c$.
The dilaton field $\varphi$ itself depends on the chosen boundary condition. We denote the dilaton field with the zero boundary condition at zero holographic coordinate as $\varphi=\varphi_0(z)$, i.e.
\be
\varphi_0(z)\Big|_{z=0}=0.\ee
In general, the dilaton field with the zero boundary condition at the holographic coordinate $z=z_0$ is denoted as
$\varphi_{z_0}(z)$. These two solutions are related by a simple relation
\be
\varphi_{z_0}(z)=\varphi_{0}(z)-\varphi_{0}(z_0),\ee
that gives
\be
\alpha_{z_0}(z)=\alpha_0(z)\,\fG (z_0), \quad \alpha_0(z)=e^{\varphi_{0}(z)},\quad \fG (z_0)=e^{-\varphi_{0}(z_0)}.\ee
As we will see below, $\varphi_{0}(z)$ does not depend on the thermodynamic characteristics of the model both for light and heavy quarks. To obtain the dependence of the running coupling on thermodynamic quantities such as $T$ and $\mu$, one can take $z_0$ related with $z_h$, i.e. $z_0=\fz(z_h)$, where $\fz(z_h)$ is some give function of $z_h$, and get 
\be
\alpha_{\fz}(z;T,\mu)=\alpha_0(z)\,\fG (T,\mu),\quad\mbox{where}\quad\fG (T,\mu)= e^{-\varphi_{0}(\fz(z_h))}.
\ee
One such choice is
$\fz(z_h)=z_h$, the other is given by exponential functions, see \eqref{phi-fz-LQ} and \eqref{bceHQ},
which are different for light and heavy quarks.\\

The different choices of the boundary conditions have phenomenological consequences. The value of string tension $\sigma$ between quarks in the hadronic phase significantly depends on this choice. For example, the option with the exponent \eqref{phi-fz-LQ} reproduces the dependence of $\sigma=\sigma(T)$ on temperature at $\mu=0$ obtained as a result of lattice calculations for light quarks, \cite{Arefeva:2020byn}.
The form \eqref{bceHQ} gives a reasonable dependence of the string tension $\sigma$
on the temperature in the hadronic phase for heavy quarks.
\\

In this paper, see Sect.\,\ref{sec:running}, 
we study the dependence of the running coupling on the energy scale specified by the holographic coordinate $z$, as well as on the thermodynamic parameters $T$ and $\mu$. The plots in Fig.\,\ref{Fig:logLH} show that the running coupling constants
$\alpha_{\fz}(z;T,\mu)$ exhibit the phase transition structure shown in Fig.\,\ref{Fig:PD-LH-HQ-paint}
and the qualitative structure of the running coupling dependence on $(T,\mu)$ is not defined by the form of the function $\fz$ (if
$\fz$ is not an identity constant).
These plots tell us that by studying the scattering amplitudes for particles interacting via QCD, we can obtain information about the 1st order phase transition in QCD containing light or heavy quarks. To obtain the dependence of the running coupling on the energy $E$ and the square of the transferred momenta $Q^2$, it is necessary to present a correspondence between them. We discuss this in Sect.\,\ref{DES-LQ} and Sect.\,\ref{DES-HQ}.
\\

The paper is organized as follows.

In Sect.\,\ref{sec:prelim}
we present  5-dim holographic models for heavy and light quarks and describe  thermodynamic properties
of these models. 

In Sect.\,\ref{sec:running} we describe running coupling constant for light and heavy quarks models.

In Sect.\,\ref{sec:concl}  we review our main
results. 

This work is complemented with Appendix \ref{app:potentials} where we present the behavior of dilaton potentials and gauge kinetic functions for both models, and  Appendix \ref{app:WL} 
where holographic calculations of Wilson loops are reminded.

\newpage

\section{Preliminary}  \label{sec:prelim}
\subsection{Holographic model for light quarks}
\subsubsection{Solutions and background}

We consider Einstein-Maxwell-scalar (EMS) system with the action \cite{Li:2017tdz,Yang:2015aia}
\bea
S&=&\frac{1}{16\pi G_5}\int d^5x\sqrt{- \fg} \left[R-\frac{\ff_0(\varphi)}{4}F^2-\frac{1}{2}\partial_{\mu}\varphi\partial^{\mu}\varphi-\cV(\varphi)\right],\label{action}
\eea
where $G_5$ is the 5-dimensional Newtonian  gravitational constant,  $g_{\mu\nu}$ is the metric tensor, $\fg=det\,g_{\mu\nu}$ is the determinant of the metric tensor, $F_{\mu\nu}$ is the electromagnetic tensor of the gauge Maxwell field $A_{\mu}$, $F_{\mu\nu}=\partial_{\mu}A_{\nu}-\partial_{\nu}A_{\mu}$, $\varphi$ is the dilaton field, $\ff_0(\varphi)$ is the gauge kinetic function associated to the Maxwell field, and $\cV(\varphi)$ is the potential of the dilaton field $\varphi$.\\

Corresponding to (\ref{action}) equations of motion (EOM) in general take the following form:
\bea
\nabla^2\varphi&=&\frac{\partial \cV}{\partial \varphi}+ \frac{F^2}{4}\frac{\partial \ff_0}{\partial \varphi} ~,\quad \partial_{\mu}\left[\sqrt{-\fg}\, \ff_0(\varphi)F^{\mu\nu}\right]=0 ~,\\
R_{\mu\nu}-\frac{1}{2}g_{\mu\nu}R&=&\frac{\ff_0(\varphi)}{2}\left(F_{\mu\rho}F^{\rho}_{\nu}-\frac{1}{4}g_{\mu\nu}F^2\right)+\frac{1}{2}\left[\partial_{\mu}\varphi\partial_{\nu}\varphi-\frac{1}{2}g_{\mu\nu}(\partial\varphi)^2-g_{\mu\nu}\cV(\varphi)\right].\nonumber\\
\eea
To solve this system of equations, we consider the ansatz for the metric, scalar field and Maxwell field as \cite{Li:2017tdz,Yang:2015aia}
\bea
ds^2=B^2(z)\left[-g(z)dt^2+d\Vec{x}^2+\frac{dz^2}{g(z)}\right],\label{metric}\\
 \quad  \varphi=\varphi(z),\quad \label{warp-factor} A_{\mu}=\left(A_t(z)~,\Vec{0},0\right)~,
\eea
where 
\be
\label{warp-factor} B(z)=\frac{L\,e^{A(z)}}{z}
\ee is the warp factor, $g(z)$ is the blackening function, $\Vec{x}=(x_1,x_2,x_3)$, $A(z)$ is a scale  factor that has different functionality associated to the light and heavy quarks and $L$ is the AdS radius that we set $L=1$.
Then, EOM are 

\cite{Li:2017tdz,Yang:2015aia,
  Arefeva:2018hyo, Arefeva:2022avn, Arefeva:2022bhx, Arefeva:2020vae,Arefeva:2020byn,Arefeva:2021mag,Arefeva:2023ter} 
\bea
\label{phi2prime}
\varphi''+\left(\frac{g'}{g}+3A'-\frac{3}{z}\right)\varphi'+\left(\frac{z^2e^{-2A}A'_tf_{0,\varphi}}{2g}-\frac{e^{2A}V_{\varphi}}{z^2g}\right)=0,\\\label{At2prime}
A''_t+\left(\frac{f'_{0}}{f_{0}}+A'-\frac{1}{z}\right)A'_t=0,\\
\label{phiprime}
A''-A'^2+\frac{2}{z}A'+\frac{\varphi'^2}{6}=0,\\\label{g2prime}
g''+\left(3A'-\frac{3}{z}\right)g'-e^{-2A}z^2f_0 A_t'^2=0,\\\label{A2primes}
A''+3A'^2+\left(\frac{3g'}{2g}-\frac{6}{z}\right)A'-\frac{1}{z}\left(\frac{3g'}{2g}-\frac{4}{z}\right)+\frac{g''}{6g}+\frac{e^{2A}V}{3z^2g}=0.
\eea
here all functions  depend on the omitted holographic coordinate $z$   and $V(z)=\cV(\varphi(z))$,  
$V_\varphi(z)=\cV_\varphi(\varphi(z))$ and $f_0(z)=\ff_0(\varphi(z))$. 
\\

Therefore, in isotropic case we have five nontrivial equations: three the Einstein equations, one equation for the scalar field $\varphi$, one equation for the component of vector potential $A_t$. The \eqref{phi2prime} is the consequence of \eqref{At2prime}--\eqref{A2primes} due to the Bianchi identity. Also we have six nonzero functions: two nonzero diagonal components of the metric $g_{MN}$, one  scalar field $\varphi$,  one component of vector potential $A_t$, the potential of scalar field $V$   and one coupling function for the Maxwell tensor $\ff_0(\varphi)$.   Therefore, we should fix two functions $f_0(z)$ and $B(z)$ to solve the system of EOM. In this case we obtain four  equations \eqref{At2prime}--\eqref{A2primes} for four unknown functions ($\varphi$, $A_t$, $g(z)$, $V(z)$). 
\\

To solve the EOM (\ref{phi2prime})--(\ref{A2primes}) the usual boundary conditions are used
\begin{gather}
  A_t(0) = \mu, \quad A_t(z_h) = 0, \label{eq:4.24} \\
 \quad  g(0) = 1, \qquad g(z_h) = 0. \label{eq:4.25} 
\end{gather} 
For dilaton field one can use different  boundary conditions.

We denote these different fields as $\varphi(z,z_0)$, so that 
\be\label{phi-z0}
\varphi(z,z_0)\Big|_{z=z_0}=0 \,.
\ee
We will consider three different types of boundary conditions different by choice of $z_0$:
\bea \label{bc0}
z_0&=&0,\\
\label{bch}
z_0&=&z_h,\\ \label{bce}
z_0&=&\fz(z_h),
\eea
where $\fz (z_h)$ is a smooth function of $z_h$.
We denote the dilaton field with zero boundary condition \eqref{bc0}  as   $\varphi_0(z)$, i.e $\varphi(z,0)=\varphi_0(z)$,
with the first boundary condition \eqref{bch} as $\varphi_{z_h}(z)$ and with the second boundary condition \eqref{bce} as $\varphi_\fz(z)$, i.e.
\be
\label{phi-z0-gen}
\varphi_\fz(z)=\varphi(z,\fz(z_h)),
\qquad\mbox{i.e.}\qquad
\varphi(\fz(z_h),\fz(z_h))=0
\ee

For light quarks\footnote{For heavy quarks model we need to consider $\fz_{\,_{HQ}}(z_h)$ that is a different function.} we will take
\be\label{phi-fz-LQ}
z_0=\fz_{\,_{LQ}}(z_h)=10 \, e^{(-\frac{z_h}{4})}+0.1 \,.\ee


At zero temperature and zero chemical potential, i.e. $T=\mu=0$, the vector meson spectrum should satisfy the linear Regge trajectories, \cite{Karch:2006pv,Li:2017tdz,Yang:2015aia}.  
Assuming that the Maxwell field $A_\mu$ also provides
linear Regge trajectories, it is necessary to choose a gauge kinetic function in the form  \cite{Li:2017tdz}
\bea \label{wfLc}
f_0(z)=e^{-c \, z^2-A(z)},
\eea
where the  scale factor $A(z)$ for the light quarks has the form 
\be\label{wfL}
A(z)=-a\log(bz^2+1)~,
\ee
 and $a$, $b$ and $c$ are parameters that can be fitted with experimental data as $a = 4.046$, $b = 0.01613$ GeV${}^2$ and $c=0.227$   GeV${}^2$ \cite{Li:2017tdz}.  
 Note, that  if we change boundary condition for $\varphi$ the Regge spectrum will not change. Indeed, obtaining linear Regge trajectory one  solves Schrodinger equation, i.e. (Eq. 10  in \cite{Karch:2006pv}), that has no direct dependence on dilaton and just depends on its first and second derivatives.
\\

 Under boundary conditions \eqref{eq:4.24}, \eqref{eq:4.25}, analytical solutions of the system of EOM
(\ref{phi2prime})--(\ref{A2primes}) are
\bea
\label{phiprime} 
\varphi'(z)&=&\sqrt{-6\Bigg(A''-A'^2+\frac{2}{z}A'\Bigg)},\label{spsol}\\
A_t(z)&=&\mu\frac{e^{cz^2}-e^{cz^2_h}}{1-e^{cz^2_h}},\label{atexp}
\\
g(z)&=&1-\frac{1}{\int_0^{z_h}y^3e^{-3A}dy}\Bigg[\int_0^zy^3 e^{-3A}dy-\frac{2c\mu^2}{(1-e^{cz_h^2})^2}\Bigg|\begin{matrix}
\int_0^{z_h}y^3e^{-3A}dy & \int_0^{z_h}y^3e^{-3A}e^{cy^2}dy \\
\int_{z_h}^z y^3e^{-3A}dy & \int_{z_h}^{z}y^3e^{-3A}e^{cy^2}dy
\end{matrix}\Bigg|\Bigg],\nonumber\\
\label{g}
\\\nn
\,
\\
V(z)&=&-3z^2g e^{-2A}\Bigg[A''+3A'^2+\Bigg(\frac{3g'}{2g}-\frac{6}{z}\Bigg)A'-\frac{1}{z}\Bigg(\frac{3g'}{2g}-\frac{4}{z}\Bigg)+\frac{g''}{6g}\Bigg].\label{Vsol}
\eea

\subsubsection{Phase structure for light quarks model} \label{phaseLQ}

Using the metric \eqref{metric} 
temperature can be written as:
\begin{gather}
  \begin{split}
    T &= \cfrac{|g'|}{4 \pi} \, \Bigl|_{z=z_h}
  \end{split}\, , \label{eq:2.03}
\end{gather}
and the entropy can be obtained as
\begin{gather}
  s = \frac{B^{3}(z_h)}{4}\,,\label{eq:2.04}
\end{gather}
here we set $G_5=1$. The entropy monotonically decreases with horizon growth.

The BH-BH (black hole - black hole, or Hawking-Page-like) phase transition caused by non-monotonicity of the temperature function
produces a  jump of density $\rho$, that is a coefficient in $A_t(z)$
expansion.
In other words, one can obtain the near boundary expansion of the time component of the vector field $A_t(z)$, i.e. \eqref{atexp} for light quark model\footnote{The similar expansion can be obtained for heavy quark model.}:
\be
  A_t(z) = \mu - \rho z^2 + \dots 
  = \mu - \cfrac{c \mu z^2}{1 - e^{c z_h^2}} + \dots, \qquad
  \rho = - \, \cfrac{c \mu}{1 - e^{c z_h^2}} \,,  \label{eq:2.041}
\ee
here $\mu$ is the baryon number chemical potential and $\rho$ is the quark number density.

To get BH-BH phase transition line we
need to calculate free energy as a function of temperature:
\begin{gather}
  F =  \int_{z_h}^{z_{h_2}} s \, T' dz. \label{eq:2.05}
\end{gather}
While $T \ge 0$, i.e. for small chemical potentials, we integrate to
$z_{h_2} = \infty$. When second horizon at $T = 0$ appears, one
should integrate to it's value, i.e. to $z_{h_2} = 4.609$ GeV${}^{-1}$. These conditions determine the end-point of the phase diagram, i.e. maximum permissible chemical potential $\mu_{max}$.

\begin{figure}[h!]
  \centering
  \includegraphics[scale=0.49]{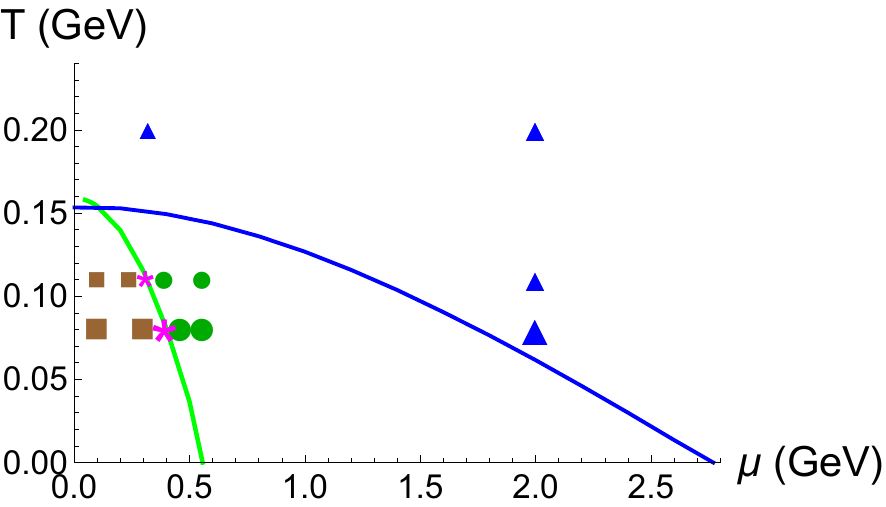} 
  \includegraphics[scale=0.52]{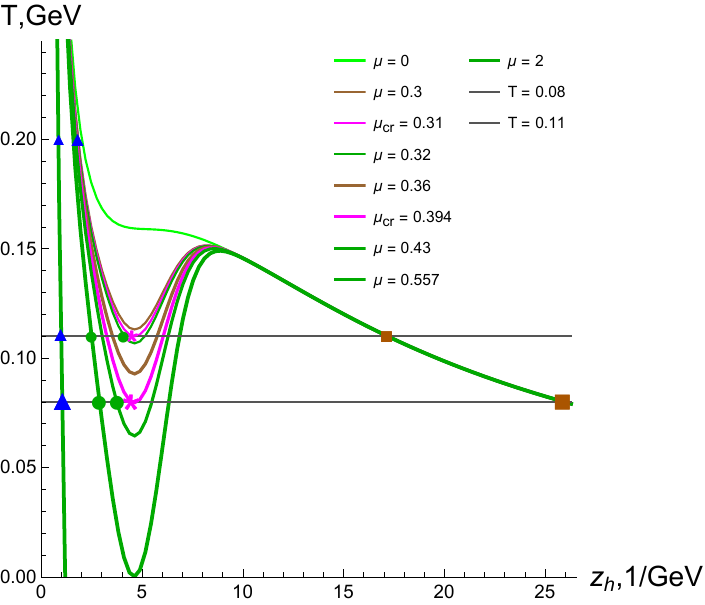} \\
  A\hspace{225pt}B\\
\caption{Phase diagram $(\mu,T)$-plane (A) and the temperature $T(z_h)$ (B) of the light quarks model indicating the points at $T=0.08, 0.11$ and $0.2$. Confinement/deconfinement is depicted by blue line and 1st order phase transition is depicted by green line (including magenta stars).
The hadronic, quarkyonic and QGP phases are represented by brown squares, green disks and blue triangles, respectively. $[\mu]=[T]=[z_h]^{-1}=$ GeV.
}
 \label{Fig:T-zh-mu-LQ-A}
\end{figure}

In Fig.\,\ref{Fig:T-zh-mu-LQ-A}A the phase diagram of the light quarks model describes two different types of phase transitions. It is very important to note the green line that corresponds to the 1st order phase transition is obtained using thermodynamics of the theory, i.e. free energy calculations and the blue line corresponds to the confinement/deconfinement phase transition is obtained via Wilson loop calculations  \cite{Li:2017tdz}. In fact, the phase diagram shows three different phases, i.e. hadronic, quarkyonic and QGP phases are indicated by brown squares, green disks and blue triangles, respectively.
We have concentrated on three different fixed temperatures, i.e. $T=0.08, 0.11$ and $0.2$ (GeV) and different values of chemical potential $\mu= 0.30, 0.31, 0.32, 0.36, 0.394, 0.43, 0.557$ and $2$ (GeV) to investigate the running coupling constant in different phases of the theory. 

ssss

In Fig.\,\ref{Fig:T-zh-mu-LQ-A}B, the dots in correspondence with different phases of light quarks phase diagram are depicted in $(z_h,T)$-plane.
We see that in this figure there are fewer brown squares, only two of them related to different sizes. This is due to the fact that for very small values of chemical potential the transition temperature occurs at values higher than $T=0.08, 0.11$ and $0.2$ (GeV). 
Using this plot, we can obtain the proper $z_h$ values associated to definite $\mu$ and $T$ in such a way that we can obtain suitable boundary condition for dilaton field, i.e. (\ref{phi-z0}) and (\ref{bch})  for light quarks model. Then, we can calculate the running coupling constant  for light quarks in the next sections using this boundary condition.

\begin{figure}[h!]
\begin{minipage}{10cm}
     \centering
\includegraphics[scale=0.30]{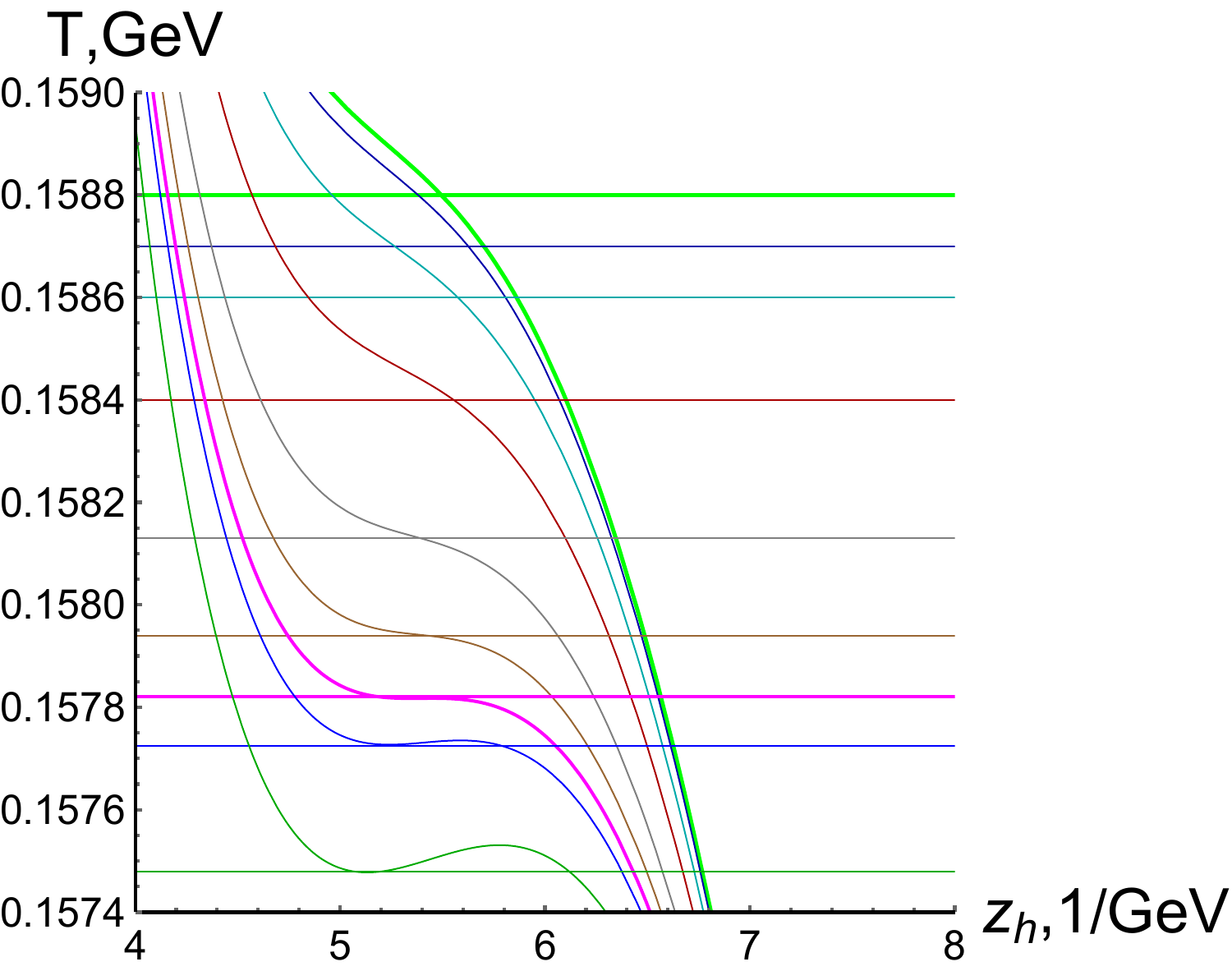}
\end{minipage}
 \begin{minipage}{3cm}
    \includegraphics[scale=0.25]{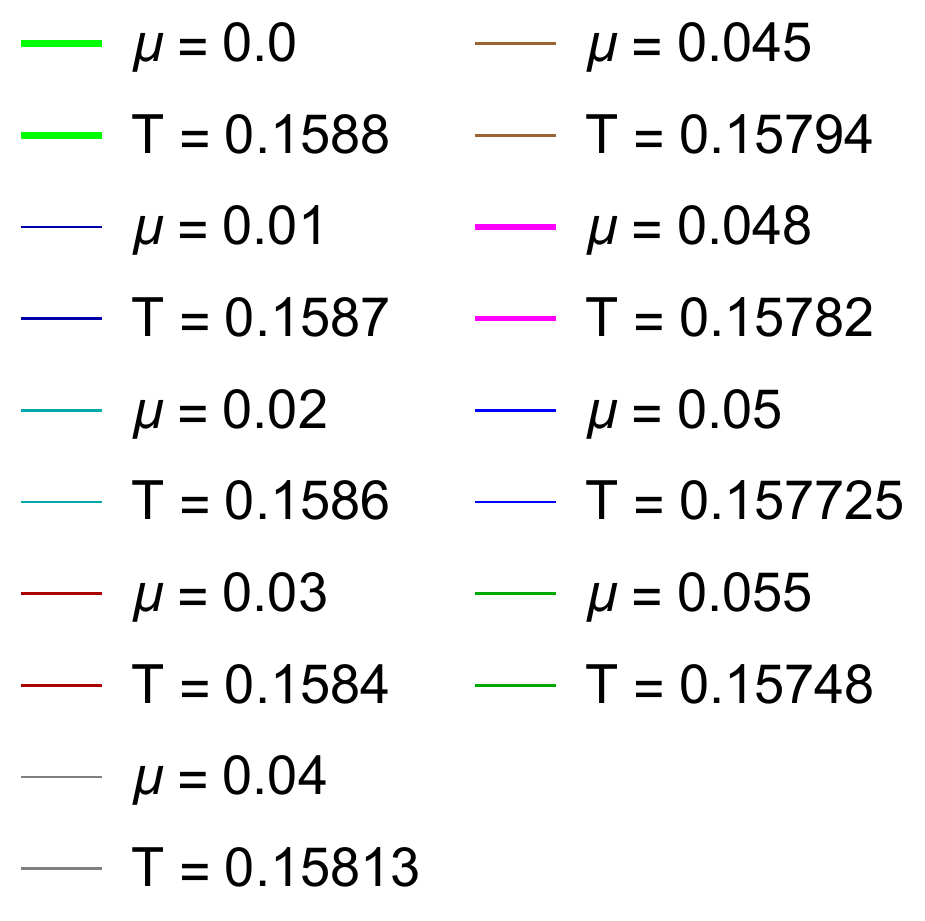} 
 \end{minipage}
\caption{Temperature $T(z_h)$ for light quarks model. The plot shows    that  there is no 1st order phase transition for small chemical potential $\mu$. $[\mu]=[T]=[z_h]^{-1}=$ GeV.
}\label{fig:No1PT-LQ}
\end{figure}

The phase transition for small chemical potential in details is presented in Fig.\,\ref{fig:No1PT-LQ}. There is no phase transition for $\mu < 0.048$ for light quarks model.

Fig.\,\ref{Fig:T-zh-mu-LQ-A}B shows that in isotropic case for $\mu = 0$
temperature is a monotonically decreasing function of
horizon. Increasing chemical potential one
local minimum for temperature as a function of $z_h$ appears. As we will
see below, this is directly related to the Hawking-page-like phase transition. 
Indeed, in the isotropic case 1st  order phase transition for light quarks shouldn't exist
near zero chemical potential and we should see a crossover,
Fig.\,\ref{fig:No1PT-LQ}. The larger chemical potential is the lesser
temperature value at this local minimum becomes. For $\mu \approx
0.557$ GeV local minimum temperature $T_{min} = 0$ and second horizon
appears.

\begin{figure}[h!]
  \centering
  \includegraphics[scale=0.34]{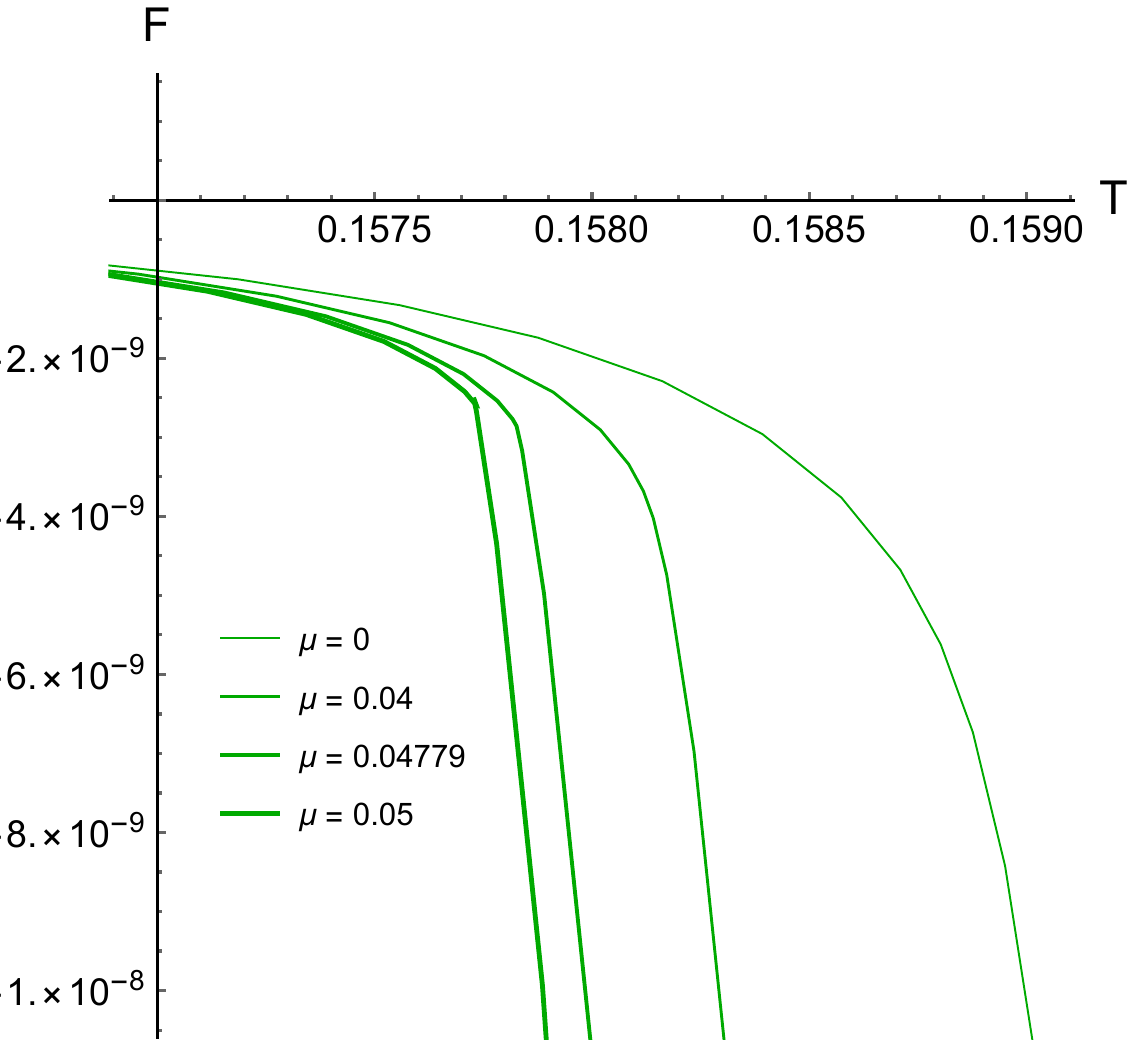}\qquad \qquad \includegraphics[scale=0.34]{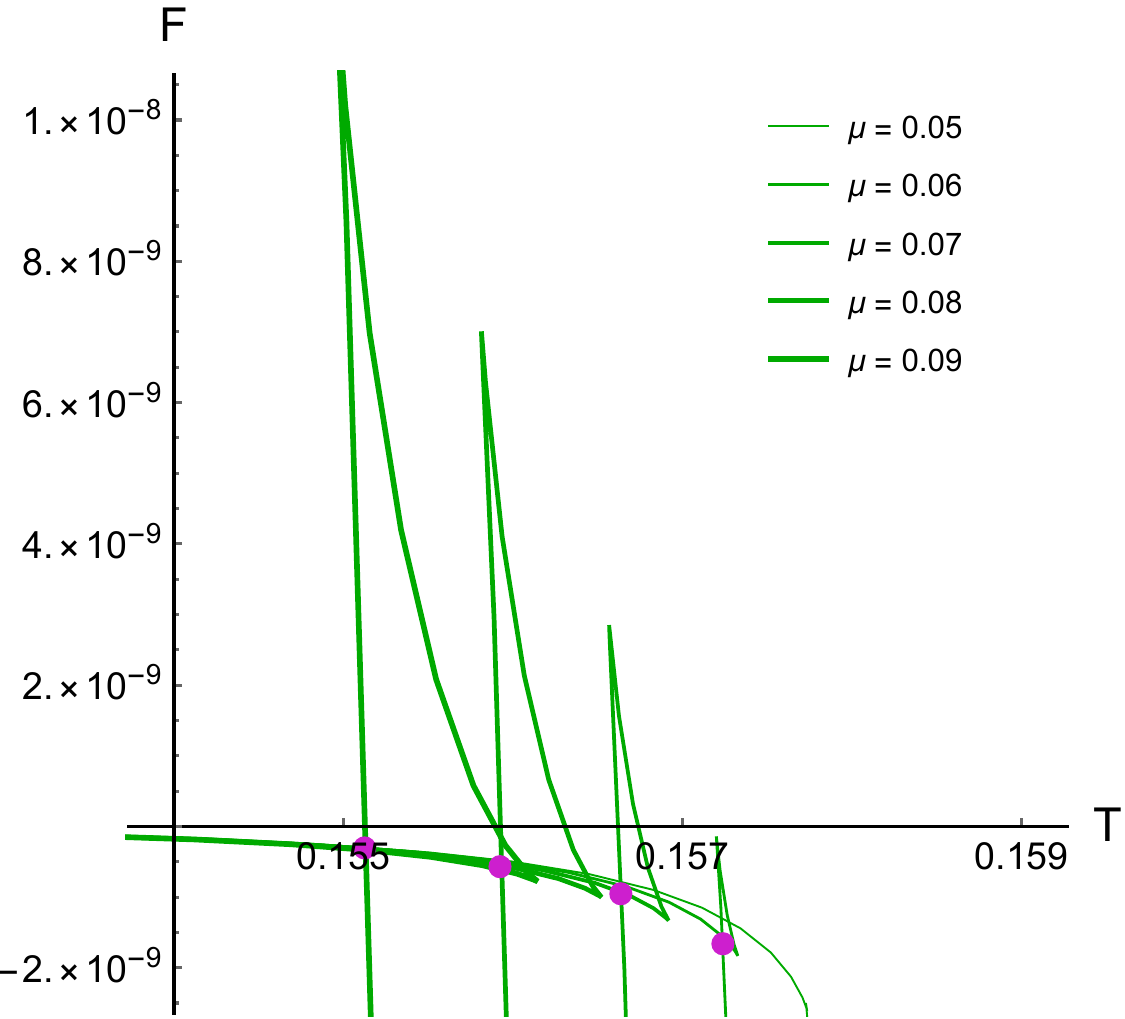}\\
  A \qquad \qquad \qquad \qquad \qquad \qquad B
  \caption{Free energy $F(T)$ in isotropic light quarks model for different $\mu$ before phase transition (A) and after phase transition (B). Magenta points denote the Hawking-Page-like phase transition temperature; $[F]=[\mu]=[T]=$ GeV. 
  \label{Fig:FT}}
\end{figure}

In Fig.\,\ref{Fig:FT} the free energy $F(T)$ in isotropic  light quarks model for different $\mu$ is depicted. The BH-BH phase transition starts from a critical point $\mu_{c} = 0.04779$ GeV, $T_{c} =
0.1578$ GeV Fig.\,\ref{Fig:FT}A. For the Hawking-Page-like phase transition the free energy should be a multi-valued function of temperature. Graphically it is displayed as a
``swallow-tail''. The point where the free energy curve intersects itself determines the Hawking-Page-like phase transition temperature.  The larger $\mu$ becomes the more pronounced the ``swallow-tail'' is (Fig.\,\ref{Fig:FT}B).

\begin{figure}[t!]
  \centering
\includegraphics[scale=0.58]{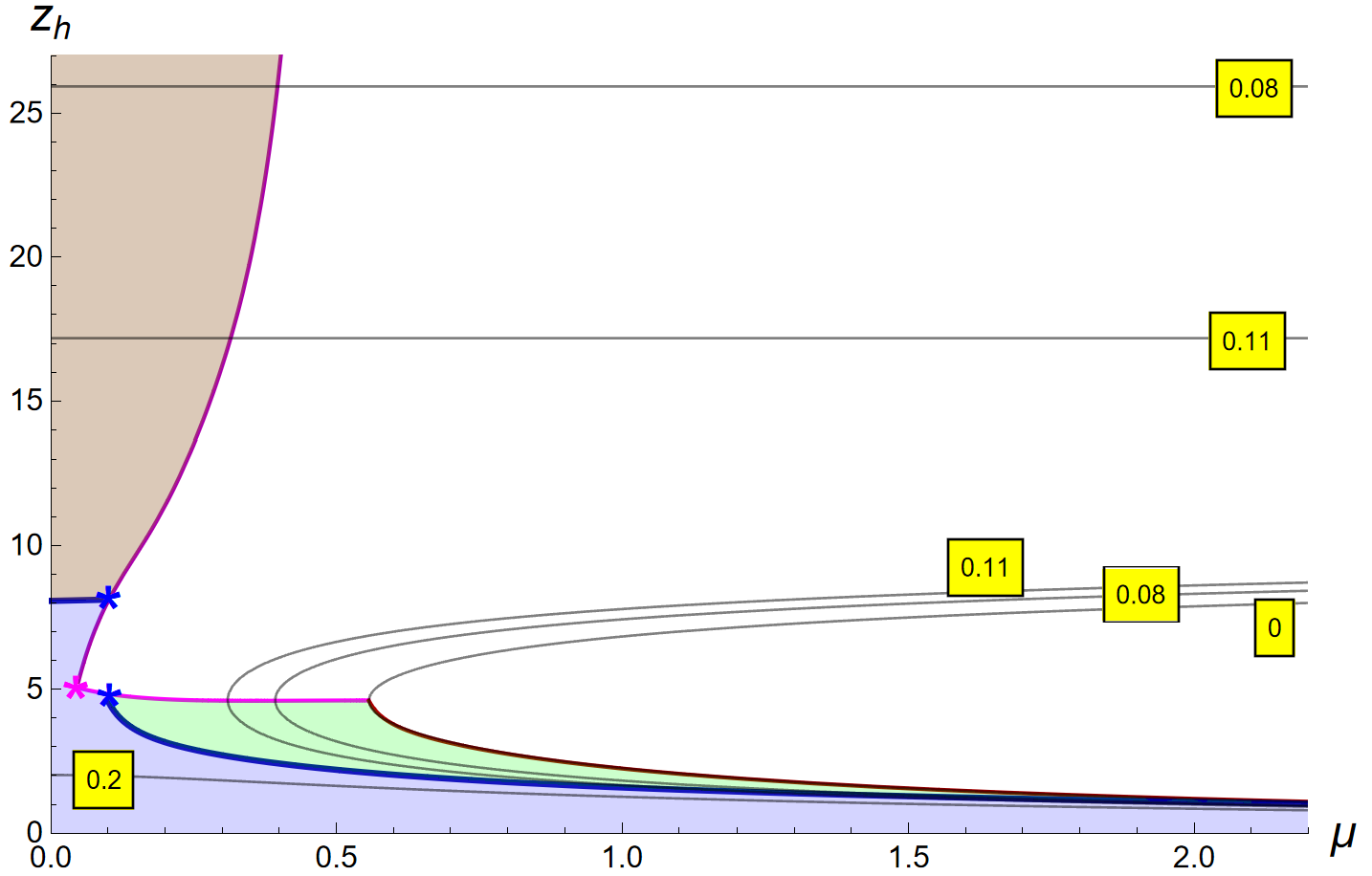}
\caption{The light quarks model. 2D plot in $(\mu,z_h)$-plane with different phases, i.e. QGP, quarkyonic and hadronic corresponding to blue, green and brown regions, respectively.  Solid black lines show the temperature indicated in yellow squares. The intersection of the confinement/deconfinement and 1st order phase transition lines is denoted by the blue stars. The magenta star indicates the CEP; $[\mu]=[T]=[z_h]^{-1} =$ GeV. 
}
\label{Fig:PhL2D}
\end{figure}

\begin{figure}[t!]
\centering
\includegraphics[scale=0.26]{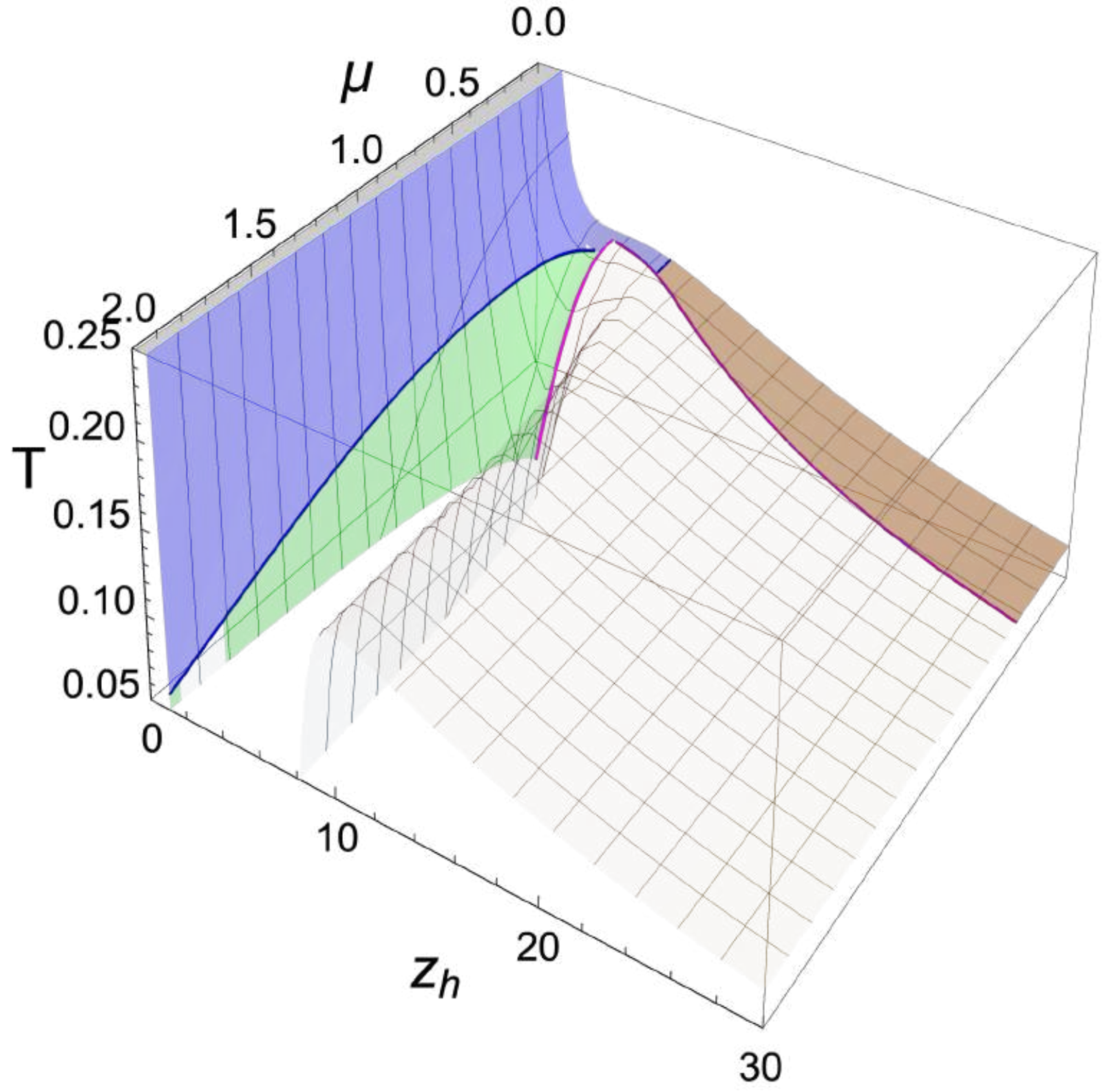}\quad
\includegraphics[scale=0.40]{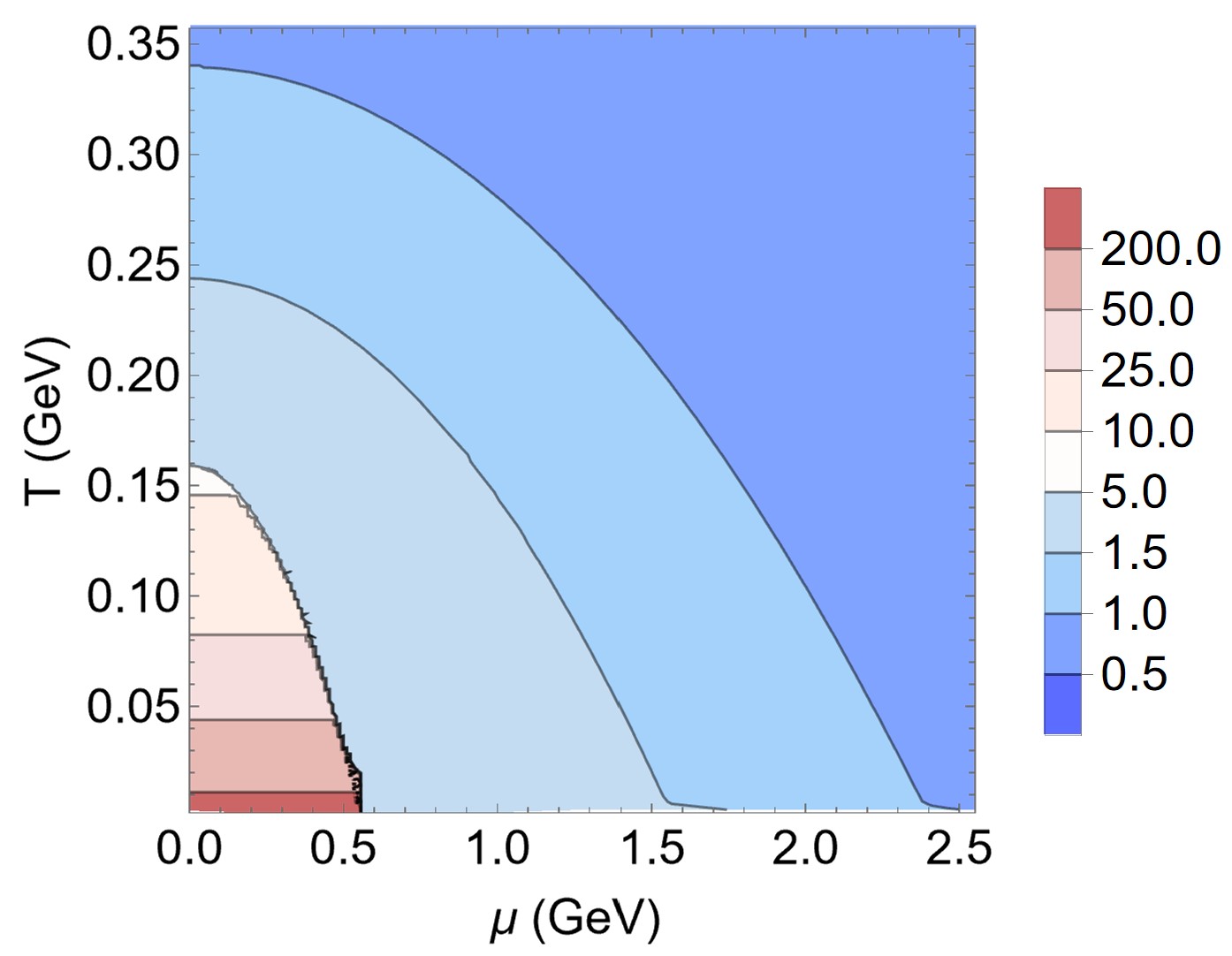}\\
A\hspace{210pt}B
\caption{Light quarks model. A) The 3D plot of temperature as function of $\mu$ and $z_h$, $T=T(\mu,z_h)$. The   brown part of the surface corresponds to the hadronic phases, the blue one corresponds to  the quark-gluon plasma and the green one to the quarkyonic phase, compare with  2D plot in Fig.\,\ref{Fig:PhL2D}. B) Density plot (with contours) for the horizon  as function of $\mu$ and $T$,  $z_h=z_h(\mu,T)$. This contour plot shows that $z_h$ in the hadronic phase depends mainly on $T$ and does not depend on the chemical potential $\mu$, meanwhile it depends on both $T$ and $\mu$ in the quark-gluon and quarkyonic phases; $[T]=[\mu]=[z_h]^{-1} =$ GeV. \\
}
 \label{Fig:PhL3d}
\end{figure}

To describe in detail the procedure that is used to construct the 2D $(\mu,z_h)$-plane phase transition diagram, Fig.\,\ref{Fig:PhL2D}, for light quarks model\footnote{This procedure to obtain 2D  $(\mu,z_h)$-plane is also the same as for heavy quarks model.} we consider the  following steps:

\begin{itemize}
\item We extract coordinates $(\mu,T)$ of the points corresponding to the 1st order and  confinement/deconfinement phase transition lines from the original phase diagram, i.e. Fig.\,\ref{Fig:T-zh-mu-LQ-A}A.
\item For each chosen point $(\mu,T)$ on the phase transition lines, we produce $T$ as a function of $z_h$, i.e. $T(z_h)$ plot, for  fixed but different $\mu$  and a line of corresponding constant $T$, Fig.\,\ref{Fig:T-zh-mu-LQ-A}B.
 \item For each point $(\mu,T)$ of original phase diagram Fig.\,\ref {Fig:T-zh-mu-LQ-A}A, we extract suitable coordinate $z_h$
 corresponding to the intersection point between constant $T$ and graph of $T(z_h)$ at fixed $\mu$ in this way:
 \begin{itemize}
 \item for the confinement/deconfinement  transition - We consider the first intersection at the stable branch of plot $T(z_h)$ associated with the large black hole. Corresponding result in Fig.\,\ref{Fig:PhL2D} are blue lines.
 \item for the 1st order phase transition - We consider the first intersection at the stable branch of plot $T(z_h)$ associated with the large black hole and the last intersection at the unstable branch of plot $T(z_h)$ associated with the small black hole that produce the light magenta line and dark magenta line in Fig.\,\ref{Fig:PhL2D}, respectively.
 \item for $T=0$ (the 2nd horizon) - We consider different fixed $\mu$ and extract corresponding coordinates $z_h$ where the plots $T(z_h)$ cross $T=0$ line. Corresponding result in Fig.\,\ref{Fig:PhL2D} is the red line.
\end{itemize}
\item Finally, we use the coordinates of the resulted points $(\mu,z_h)$ to find approximations of functions $z_h(\mu)$ for confinement/deconfinement transition line, 1st order phase transition line and the 2nd horizon to plot Fig.\,\ref{Fig:PhL2D}. Also, we can produce 3D plot $(\mu,z_h,T)$ as in Fig.\,\ref{Fig:PhL3d}.
\end{itemize}

The phase structure of light quarks model in 2D $(\mu,z_h)$-plane in Fig.\,\ref{Fig:PhL2D} shows different domains of phases, i.e. QGP, quarkyonic and hadronic correspond to blue, green and brown regions, respectively.
The solid blue lines in Fig.\,\ref{Fig:PhL2D} correspond to the confinement/deconfinement phase transition line in Fig.\,\ref{Fig:T-zh-mu-LQ-A}A obtained via Wilson loop calculations \cite{Li:2017tdz}.
The solid magenta lines in Fig.\,\ref{Fig:PhL2D} correspond to the 1st order phase transition line in Fig.\,\ref{Fig:T-zh-mu-LQ-A}A obtained via free energy calculations \cite{Li:2017tdz} and the solid red line corresponds to the second horizon where $T=0$. All these lines have been obtained via approximation procedure. We see that for $\mu<0.048$ there is crossover region and at $T=0.154$ GeV  the phase transition (without any jump) occurs between hadronic and QGP phases. But for $0.0101<\mu<0.560$ when we change the temperature there is a 1st order phase transition between hadronic and quarkyonic phases with a jump that is clear in Fig.\,\ref{Fig:PhL2D}. Also, for $\mu>0.101$ by changing the temperature there is the confinement/deconfinement transition between quarkyonic and QGP phases without any jump.
Black solid lines show the temperature indicated in yellow squares and the red square corresponds to the intersection of the confinement/deconfinement transition line and 1st order phase transition which is denoted by the blue stars. The magenta star indicates the end of the 1st  order phase transition.
 
To have a better intuition of phase structure of light quarks in our model Fig.\,\ref{Fig:PhL2D}, the 3D plot $(\mu, z_h, T)$ is depicted in Fig.\,\ref{Fig:PhL3d}A. 
The density plot with contours of the horizon $z_h(\mu,T)$ for light quarks model is shown in Fig.\,\ref{Fig:PhL3d}B. Each contour corresponds to fixed value of $z_h$. This complementary plot describes the phase structure of the light quarks model as a contour plot for $z_h$ and completes Fig.\,\ref{Fig:PhL2D} and Fig.\,\ref{Fig:PhL3d}A. It is important to note that we used a consistent color scheme to indicate different phases on phase
diagrams, i.e. blue for the QGP phase, green for the quarkyonic phase, and brown
for the hadronic phase. However, for the density plot (with contours), this color
scheme is not strictly followed.

\subsubsection{Boundary conditions and Cornell potential for light quarks model}
\label{BC-CP-LQ}
In this subsection we show that confinement/deconfinement  phase transitions structure does not depend on the boundary condition for dilaton field $\varphi$. Although, the Cornell potential \cite{Andreev:2006ct,Yang:2015aia,Arefeva:2019yzy,Asadi:2021nbd,Slepov:2021gvl,He:2010ye} and its asymptotics strongly depend on the boundary condition for the dilaton field.

$$\,$$
{\bf 2.1.3.a. $z_0=0$  boundary condition}\\

\begin{figure}[h!]
\centering
\includegraphics[scale=0.42]{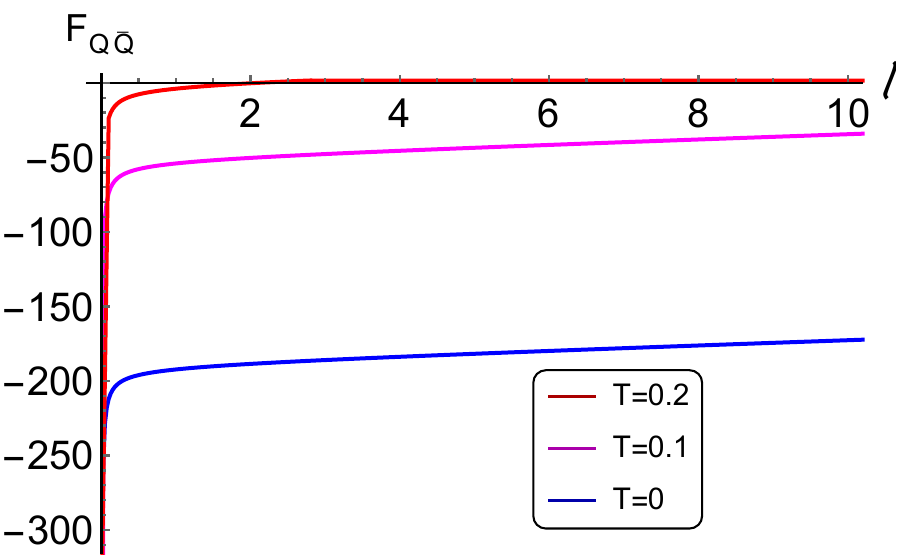}
  \quad \includegraphics[scale=0.42]{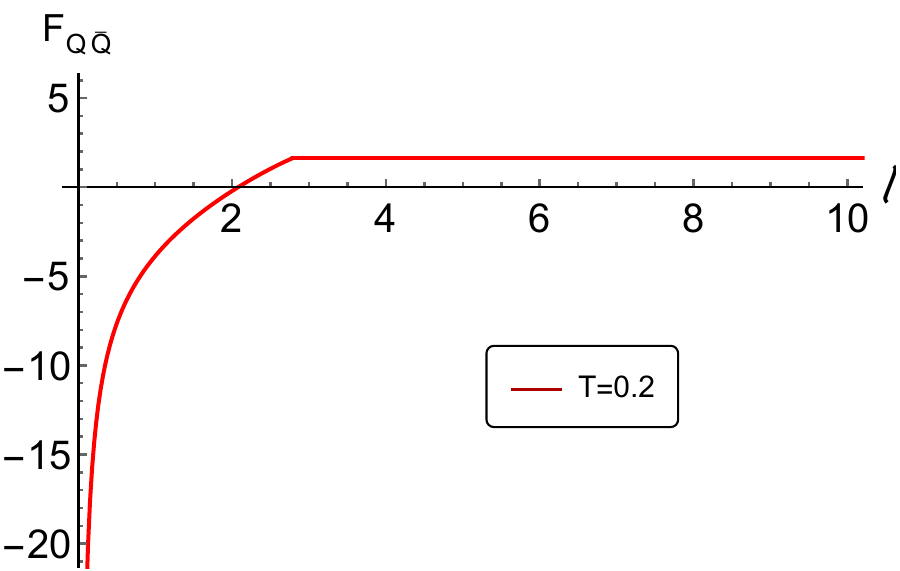}\\
 A\hspace{175pt}B
 \caption{$F_{Q\bar{Q}}$ as a function of $\ell$ for light quarks model with boundary condition $z_0=0$ with different temperatures $T=0$, $T=0.1$, $T=0.2$ (A) and zoom of $T=0.2$ (B). We set chemical potential $\mu=0$; $[F_{Q\bar{Q}}]=[\mu]=[T]=[\ell]^{-1} =$ GeV.
}
 \label{Fig:sigma-LQ-newbc0}
\end{figure}

This boundary condition was considered in \cite{Li:2017tdz}. We obtained the energy configuration of quark and anti-quark $F_{Q\bar{Q}}$  as a function of interquark distance $\ell$ to plot the Cornell potential (\ref{cornel1}) for light quarks model with different temperatures $T=0, 0.1$ and $0.2$ (GeV)  in Fig.\,\ref{Fig:sigma-LQ-newbc0}. For the temperature less than critical temperature, i.e. $T_c=0.15$ GeV the linear part of Cornell potential exists and one can obtain the QCD string tension $\sigma$, see Fig.\,\ref{Fig:sigma-LQ-newbc0}A. But, for temperatures greater than $T_c$ since we have phase transition from confinement to deconfinement, then there is no $\sigma$, see Fig.\,\ref{Fig:sigma-LQ-newbc0}B.

\begin{figure}[h]
\centering
 \quad \includegraphics[scale=0.5]{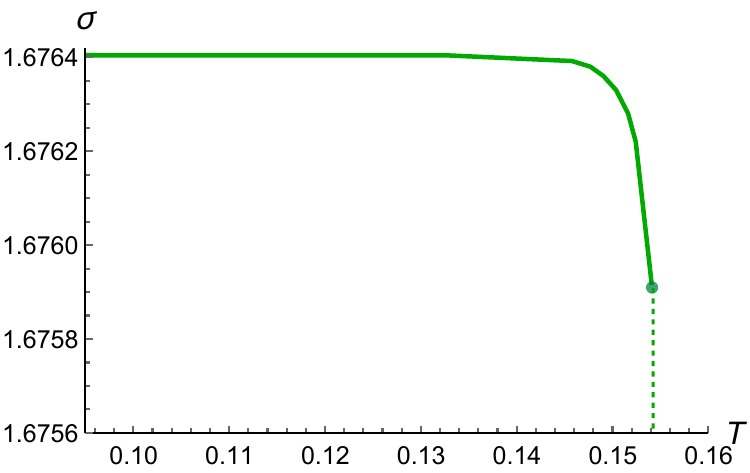}
 \caption{The string tension $\sigma(T)$ for light quarks model with boundary condition $z_0=0$ and  chemical potential $\mu=0$ \cite{Arefeva:2020byn}; $[\sigma]^{\frac{1}{2}}=[\mu]=[T] =$ GeV.
}
 \label{Fig:sigma-LQ-ARS}
\end{figure}

Using the boundary condition $z_0=0$, the temperature dependence of the QCD string tension $\sigma(T)$ can be obtained in Fig.\,\ref{Fig:sigma-LQ-ARS}. It shows that it is almost constant for $\mu=0$ and different temperatures and close to $T_{c}$ the value of $\sigma$ decreases and after phase transition gets zero. The behavior of $\sigma(T)$ as a decreasing function 
physically is acceptable and is confirmed by lattice results \cite{Cardoso:2011hh}. Although, the boundary condition $z_0=0$ cannot produce the proper results for coupling constant as a function of thermodynamic parameters in our holographic light quarks model.

$$\,$$

{\bf 2.1.3.b. $z_0=z_h$  boundary condition}\\

\begin{figure}[h!]
\centering
\includegraphics[scale=0.5]{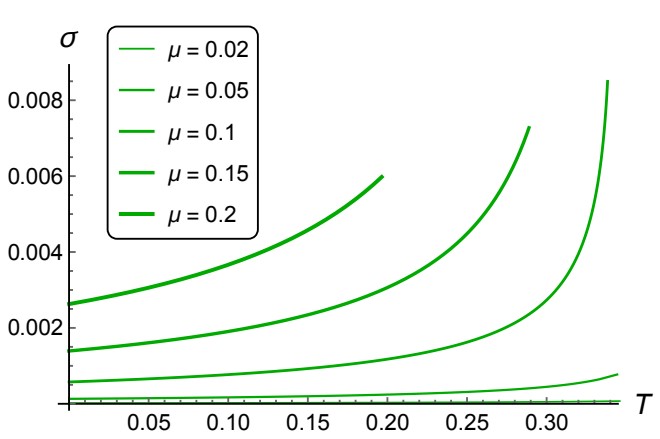} 
 \caption{The string tension $\sigma(T)$ for light quarks model with boundary condition $z_0=z_h$ and different chemical potentials \cite{Arefeva:2019yzy}; $[\sigma]^{\frac{1}{2}}=[\mu]=[T] =$ GeV.\\
}
 \label{Fig:sigma-LQ-nbczh}
\end{figure}

Considering the second boundary condition $z_0=z_h$ that was utilized in the paper \cite{Arefeva:2018hyo}, the string tension $\sigma(T)$ for light quarks model with different chemical potentials is plotted in Fig.\,\ref{Fig:sigma-LQ-nbczh}. String tension increases with enhancing temperature and this result is in contrast with lattice results \cite{Cardoso:2011hh}. Therefore, it shows that we cannot trust to the boundary condition $z_0=z_h$ to produce correct results for running coupling constant. 

$$\,$$

{\bf   $z_0=\fz_{LQ}(z_h)$ boundary condition}\\

In this subsection we use a new boundary condition \eqref{phi-fz-LQ}, $z_0=\fz_{\,_{LQ}}(z_h)=10 \exp(-0.25 z_h)+0.1 $ for light quarks model:
to cover the proper behavior for string tension as a function of temperature that has been introduced in \cite{Arefeva:2020byn}. 

\begin{figure}[h]
\centering
\includegraphics[scale=0.4]{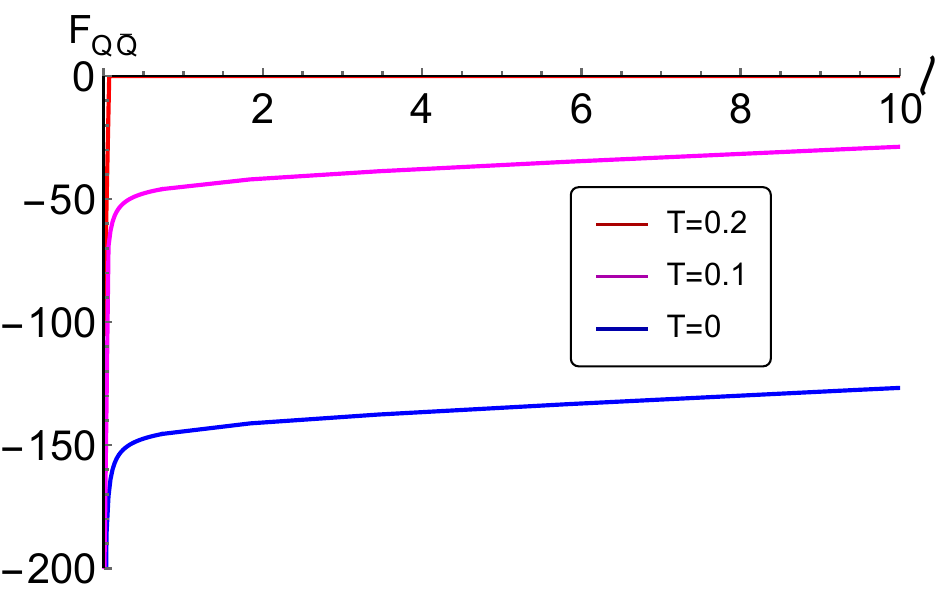}\qquad \includegraphics[scale=0.4]{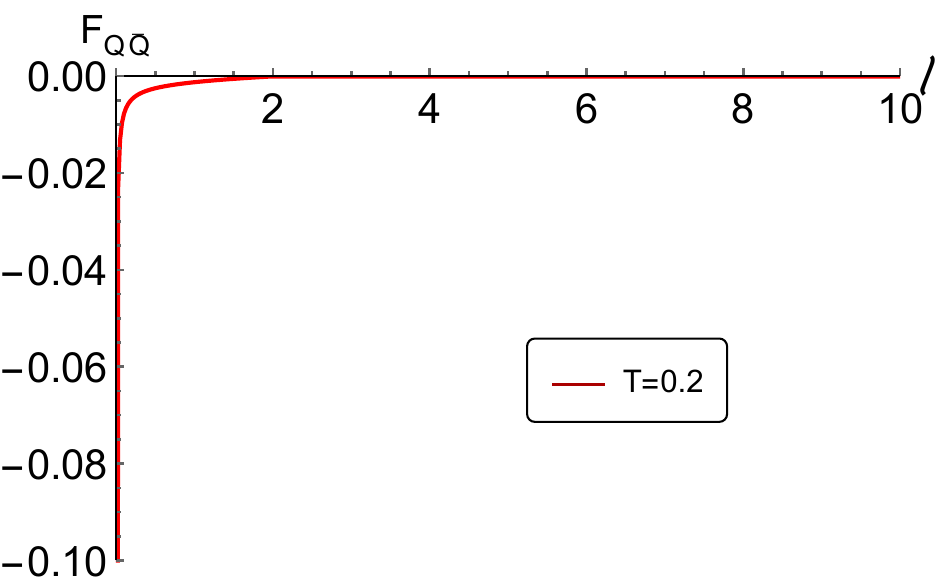}\\
\qquad\qquad A\hspace{200pt}B
 \caption{$F_{Q\bar{Q}}$ as a function of $\ell$ for light quarks model with the boundary condition \eqref{phi-fz-LQ} with different temperatures $T=0$, $T=0.1$, $T=0.2$ (A) and zoom of $T=0.2$ (B). We set chemical potential $\mu=0$; $[F_{Q\bar{Q}}]=[\mu]=[T]=[\ell]^{-1} =$ GeV.\\
 }
 \label{Fig:sigma-LQ-nbc2}
\end{figure}

In Fig.\,\ref{Fig:sigma-LQ-nbc2}, the energy configuration of quark and anti-quark $F_{Q\bar{Q}}$  as a function of interquark distance $\ell$ for light quarks model with boundary condition \eqref{phi-fz-LQ} with different temperatures $T=0, 0.1$ and $0.2$ (GeV) (A) and zoom of $T=0.2$ GeV (B) with $\mu=0$ is plotted. The linear part of the Cornell potential (\ref{cornel1}), i.e. $\sigma$ exists for the temperature less than the critical temperature, i.e. $T_c=0.15$ GeV Fig.\,\ref{Fig:sigma-LQ-nbc2}A and there is no $\sigma$ for $T>T_c$ because of confinement/deconfinement phase transition Fig.\,\ref{Fig:sigma-LQ-nbc2}B.

\begin{figure}[h!]
\centering
\includegraphics[scale=0.43]{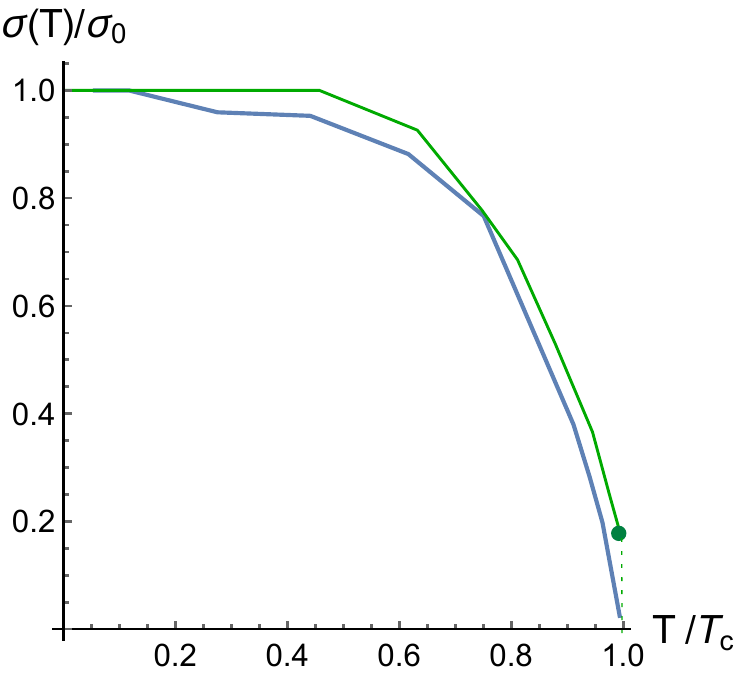}\qquad\includegraphics[scale=0.30]{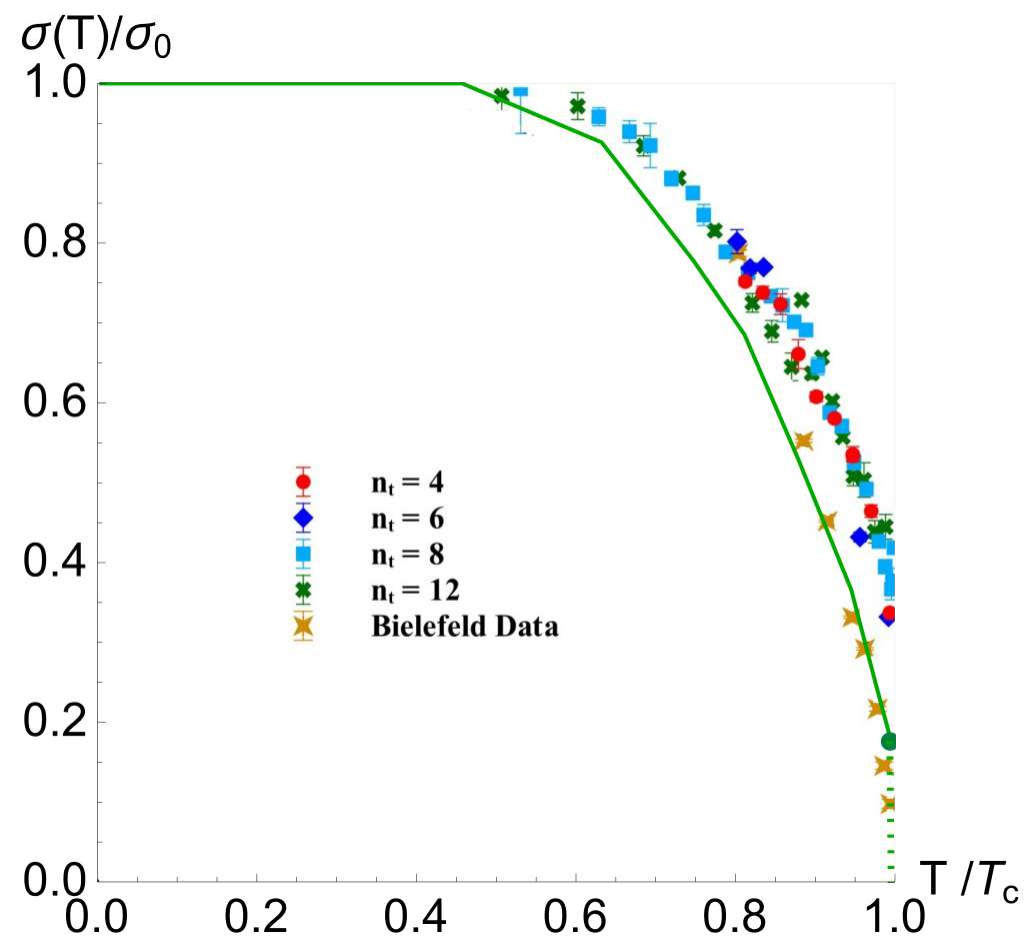}\\
 A\hspace{175pt}B
 \caption{Comparison of the $\sigma(T)$ for light quarks at $\mu=0$ with the boundary condition \eqref{phi-fz-LQ} between analytical holographic calculations using effective potential (green curve) and numerical calculations of Cornell potential (blue curve) (A) and $\sigma(T)$ found in the lattice (dots) (B). $\sigma_0$ is the string tension at $T=0$ and $T_c=0.15$ is the confinement/deconfinement phase transition temperature; $[\sigma]^{\frac{1}{2}}=[\mu]=[T] =$ GeV.\\
}
 \label{Fig:sigma-LQ-gbc5}
\end{figure}

The comparison of the $\sigma(T)$ for light quarks at $\mu=0$ considering the boundary condition \eqref{phi-fz-LQ} between analytical holographic calculations using effective potential (green curve) and numerical calculations of Cornell potential (blue curve) Fig.\,\ref{Fig:sigma-LQ-gbc5}A and $\sigma(T)$ found in the lattice (dots) Fig.\,\ref{Fig:sigma-LQ-gbc5}B are depicted. $\sigma_0$ is the string tension at $T=0$ and $T_c=0.15$ GeV is the confinement/deconfinement phase transition temperature. With this boundary condition \eqref{phi-fz-LQ} the $\sigma(T)$ fits the known lattice data Fig.\,\ref{Fig:sigma-LQ-gbc5}B, see \cite{Cardoso:2011hh}.
Therefore, we will utilize this boundary condition in SubSect.\,\ref{gbc10} to calculate running coupling constant that is compatible with lattice results.

\newpage
\subsubsection{Dependence of energy scale $E$ on the holographic $z$ coordinate}\label{DES-LQ}
To obtain the dependence of the running coupling on the energy scale $E$ or the square of the transferred momenta $Q^2$, it is necessary to relate the holographic coordinate $z$ to the $E$ or $Q^2$ \cite{Galow:2009kw}.
The energy scale $E$ in the boundary field theory can be identified with the warp factor in the metric \eqref{metric} \cite{Galow:2009kw}. For the light quark model we have: 
\be \label{BBL}
E = B(z) = \frac{1}{z\, \left(1+ b z^2 \right)^{a}}\, ,
\ee
where $a$ and $b$ are introduced in \eqref{wfL}. 
This relation shows that we can cover UV limit of QFT for small values of $z$ and IR limit considering large $z$. Therefore, the bulk theory can describe all possible energy scales of the boundary field theory.

\begin{figure}[!ht]
\begin{center}
\includegraphics[scale=0.3]    {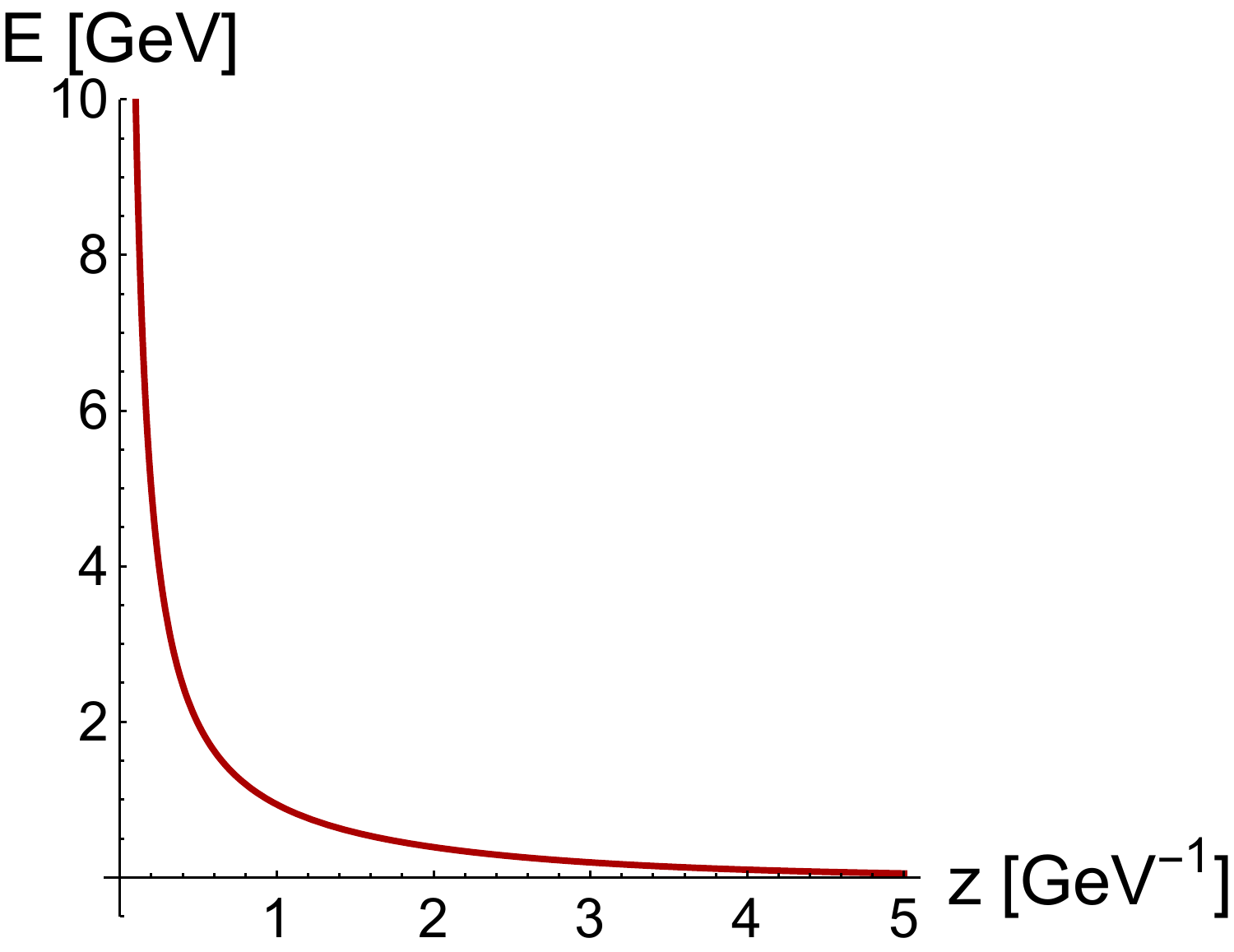} 
\end{center}
\caption{The Energy scale \( E \) (GeV) in the boundary field theory as a function of the holographic coordinate \( z \) (GeV${}^{-1}$) for light quark model.}
\label{E-of-z-LQ}
\end{figure}

In Fig.\,\ref{E-of-z-LQ}, for light quark model, the energy scale \( E \) (GeV) in the boundary field theory as a function of the holographic coordinate $z$ (GeV${}^{-1}$), corresponding to the warp factor $B(z)$ given by (\ref{BBL}) is presented. We see that small $z$ corresponds to large $E$ and large $z$ corresponds to small $E$, as it should be.
The Fig.\,\ref{E-of-z-LQ} shows that $E(z)$ is a  monotonic function and there is a one-to-one correspondence between energy scale $E$ and holographic coordinate $z$ (or energy scale $z$).
We emphasize that the energy scale $E$ in QFT can
get a wide range and differs from the energy scale of QCD, i.e. $\Lambda_{QCD}$, which is fixed at the confinement scale, specially 
$\Lambda_{QCD}=264$ MeV
\cite{Galow:2009kw}.

 \newpage

\subsection{Holographic model for heavy quarks}

\subsubsection{Solutions and background} \label{HQpreli}
The holographic model describing heavy quarks is described by the same action \eqref{action} \cite{Yang:2015aia,Arefeva:2023jjh}, but the warp factor that is specified by the scale factor $A(z)$ is different from \eqref{wfL} and is given by \cite{Yang:2015aia}
\bea \label{scaleHQ}
A(z)=-\frac{\fc}{3}z^2- p\, z^4~,
\eea
$\fc$ and $p$ are parameters that can be fitted with the experimental data as $\fc$= 1.16 GeV${}^2$ and $p = 0.273$ GeV${}^4$. 
To respect the linear Regge trajectories the gauge kinetic function for heavy  quarks model is chosen in the form \cite{Yang:2015aia}
\bea
 ~~~~f_0(z)=e^{-\fc \, z^2-A(z)},
\eea
 compare with \eqref{wfLc}.
As in the case of light quarks, the analytic solution for heavy quarks model is obtained by solving the system of EOM (\ref{phi2prime})-(\ref{A2primes})  with 
 the boundary conditions (\ref{eq:4.24}) and (\ref{eq:4.25}). 
The dilaton field $\varphi(z)$
we consider  the boundary condition \eqref{phi-z0}.
Also, the same as light quarks we will consider  for heavy quarks zero, first and second boundary conditions correspond to (\ref{bc0}), (\ref{bch}) and (\ref{bce}), respectively. In the second boundary condition  for heavy quarks model we propose a new function
\bea \label{bceHQ}
z_0&=& \fz_{\,_{HQ}}(z_h) =e^{(-\frac{z_h}{4})} + 0.1 \,.
\eea
Therefore, analytical solutions for heavy quarks model are functionally the same, i.e. are given by equations \eqref{phiprime}-\eqref{Vsol}, only the scale factor $A(z)$ is different and  constant parameter $c$ should be replaced by $\fc$ as a new constant parameter for heavy quarks.

\subsubsection{Phase structure for heavy quarks model} \label{phaseHQ}

\begin{figure}[h!]
\centering\includegraphics[scale=0.57]{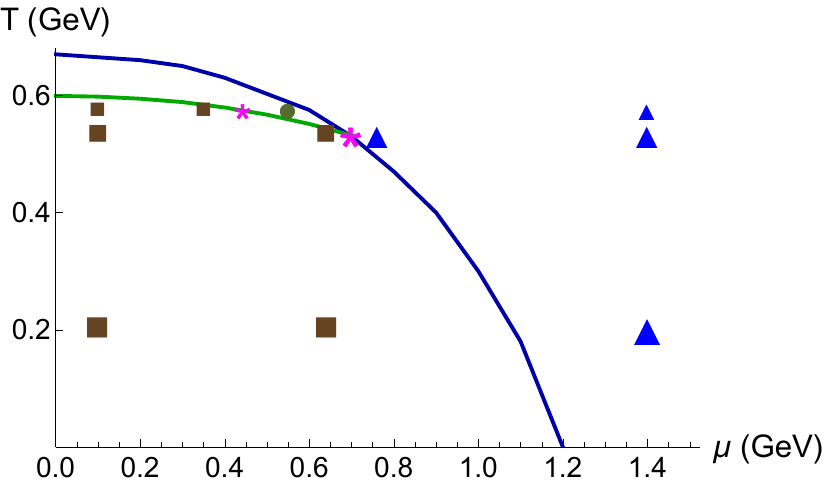} \includegraphics[scale=0.33]{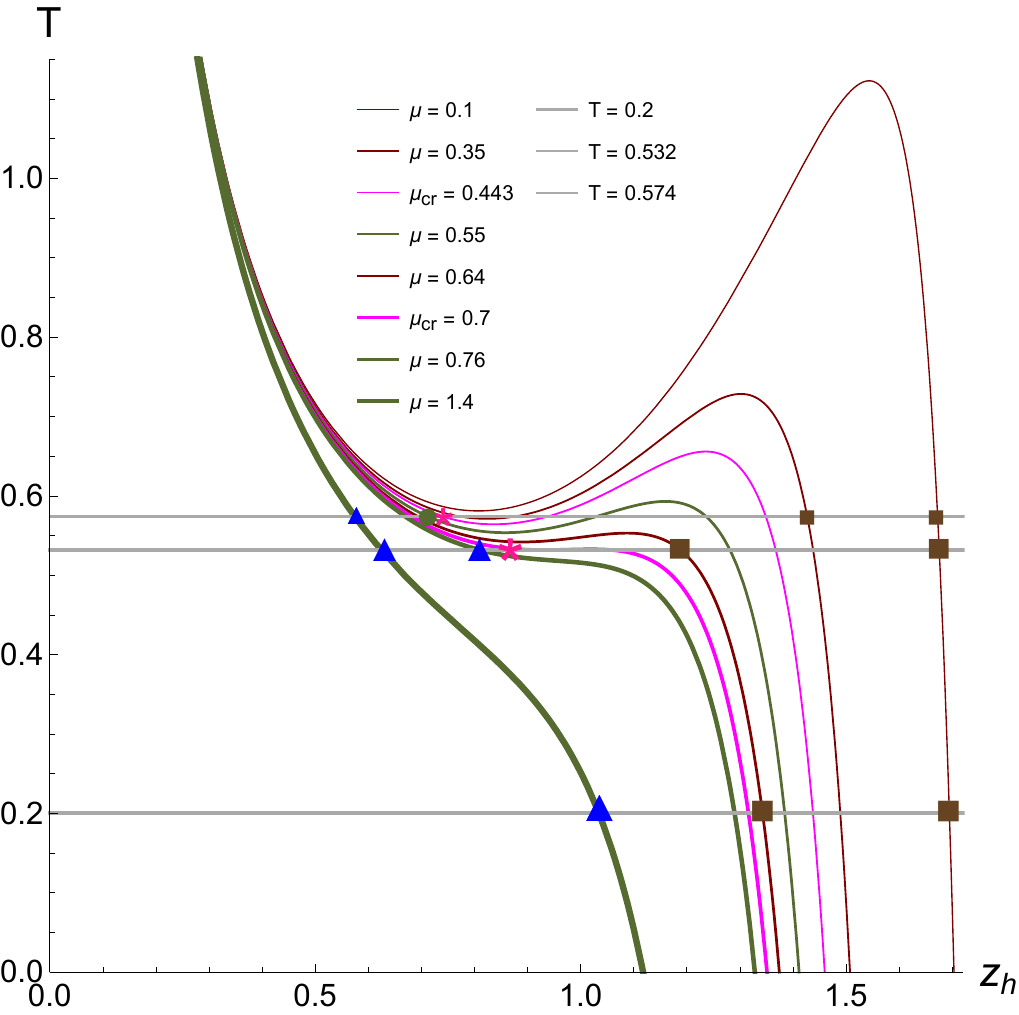}\\
A\hspace{210pt}B
\caption{Phase diagram $(\mu,T)$-plane (A) and the temperature $T(z_h)$ (B) of the heavy quarks model indicating the points at  $T=0.2, 0.532$ and $0.574$. Confinement/deconfinement is depicted by blue line and 1st order phase transition is depicted by green line (including magenta stars). The hadronic, quarkyonic and QGP phases are represented by brown squares, green disk and blue triangles, respectively; $[z_h]^{-1} =$ GeV.\\
}
 \label{Fig:Ph-TmuHQ}
\end{figure}

The phase diagram of the heavy quarks model describes two different types of phase transitions in Fig.\,\ref{Fig:Ph-TmuHQ}A. It is very important to note that the green line that corresponds to the 1st order phase transition is obtained using thermodynamics of the theory, i.e. free energy calculations and the blue line corresponds to the confinement/deconfinement phase transition is obtained via Wilson loop calculations \cite{Yang:2015aia,Arefeva:2023jjh}. As a matter of fact, the phase diagram Fig.\,\ref{Fig:Ph-TmuHQ}A shows three different phases, i.e. hadronic, quarkyonic and QGP phases are indicated by brown squares, green disk and blue triangles, respectively.
In the phase diagram we concentrated on three  fixed temperatures, i.e. $T=0.2, 0.532$ and $0.574$ (GeV)  and different values of chemical potential $\mu= 0.10, 0.35, 0.443, 0.55, 0.64, 0.70, 0.76$ and $1.5$ (GeV).
The points on the 1st order phase transition line are indicated by magenta stars. 

The behavior of the temperature $T$ as a function of $z_h$ for heavy quark model for different chemical potentials $\mu$ including different points correspond to different phases is shown in Fig.\,\ref{Fig:Ph-TmuHQ}B. For $\mu\geq 0.76$ there is no phase transition. Using this plot we can obtain the $z_h$ values associated to definite $\mu$ and $T$ in such a way that we can obtain suitable for dilaton field, i.e. (\ref{phi-z0}) and (\ref{bch})  for heavy quarks model. Then we can calculate the running coupling constant for heavy quarks in the next sections using this boundary condition. 

\begin{figure}[t!]
\begin{center}
\includegraphics[scale=0.30]{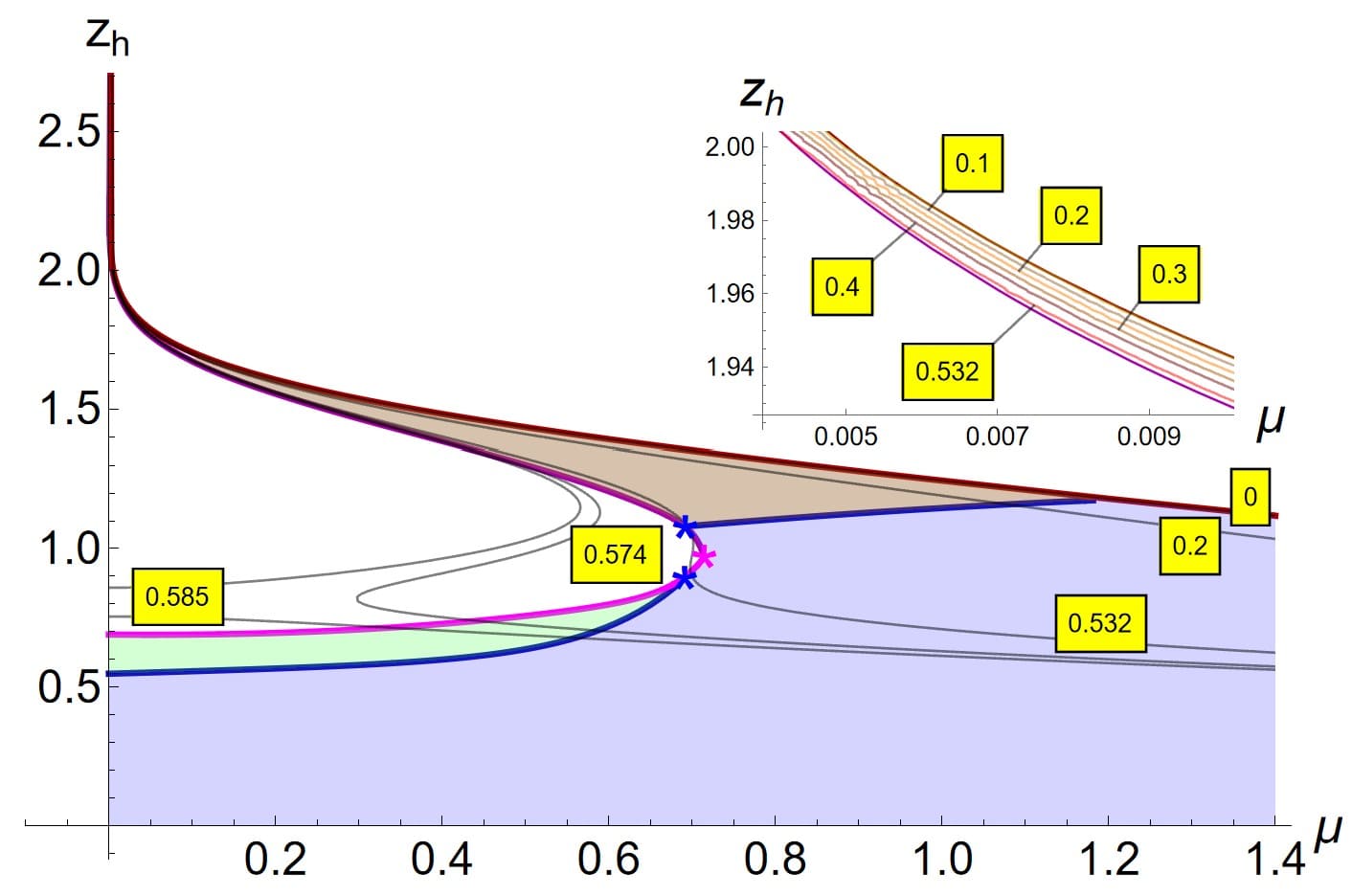}
\end{center}
 \caption{ 2D plot $(\mu,z_h)$-plane for heavy quarks model with different phases, i.e. hadronic, quarkyonic and QGP in correspondence with brown, green and blue regions, respectively. The zoom of the main plot is at $\mu\in[0.0038, 0.008]$, $ z_h\in[1.95, 2]$. Different temperatures are shown in yellow squares with associated lines. The intersection of the confinement/deconfinement transition line and 1st order phase transition is denoted by the blue stars. The magenta star indicates CEP; $[\mu]=[z_h]^{-1} =$ GeV. 
 }
 \label{Fig:HQphase-colors-Е}
\end{figure}

Using the same procedure that we described in SubSect.\,\ref{phaseLQ}, the phase structure of heavy quarks model as 2D $(\mu,z_h)$-plane is plotted in Fig.\,\ref{Fig:HQphase-colors-Е} with different domains of phases, i.e. hadronic, quarkyonic and QGP in correspondence with brown, green and blue regions, respectively. The solid thick blue lines in Fig.\,\ref{Fig:HQphase-colors-Е} correspond to the confinement/deconfinement phase transition in Fig.\,\ref{Fig:Ph-TmuHQ} obtained via Wilson loop calculations \cite{Yang:2015aia,Arefeva:2023jjh}. 
The solid thick magenta lines in Fig.\,\ref{Fig:HQphase-colors-Е} correspond to the 1st order phase transition in 
Fig.\,\ref{Fig:Ph-TmuHQ} obtained via free energy calculations \cite{Yang:2015aia,Arefeva:2023jjh} and the solid thick red line corresponds to the second horizon where $T=0$. All these lines have been obtained via approximation procedure.
For $0<\mu<0.694$ when we change the temperature in Fig.\,\ref{Fig:HQphase-colors-Е} there is a jump between hadronic and quarkyonic phases that represents the 1st order phase transition, while there is the confinement/deconfinement phase transition between quarkyonic and QGP phases without any jump. Also, for $0.694<\mu<1.181$  there is a confinement/deconfinement phase transition between hadronic and QGP without any jump.
The temperature is shown by black solid lines indicated in yellow squares. The intersection of the confinement/deconfinement and 1st order phase transition lines is denoted by the blue stars. The magenta star indicates  CEP that is the end of the 1st order phase transition. 

\begin{figure}[h!]
  \centering
 \includegraphics[scale=0.45]{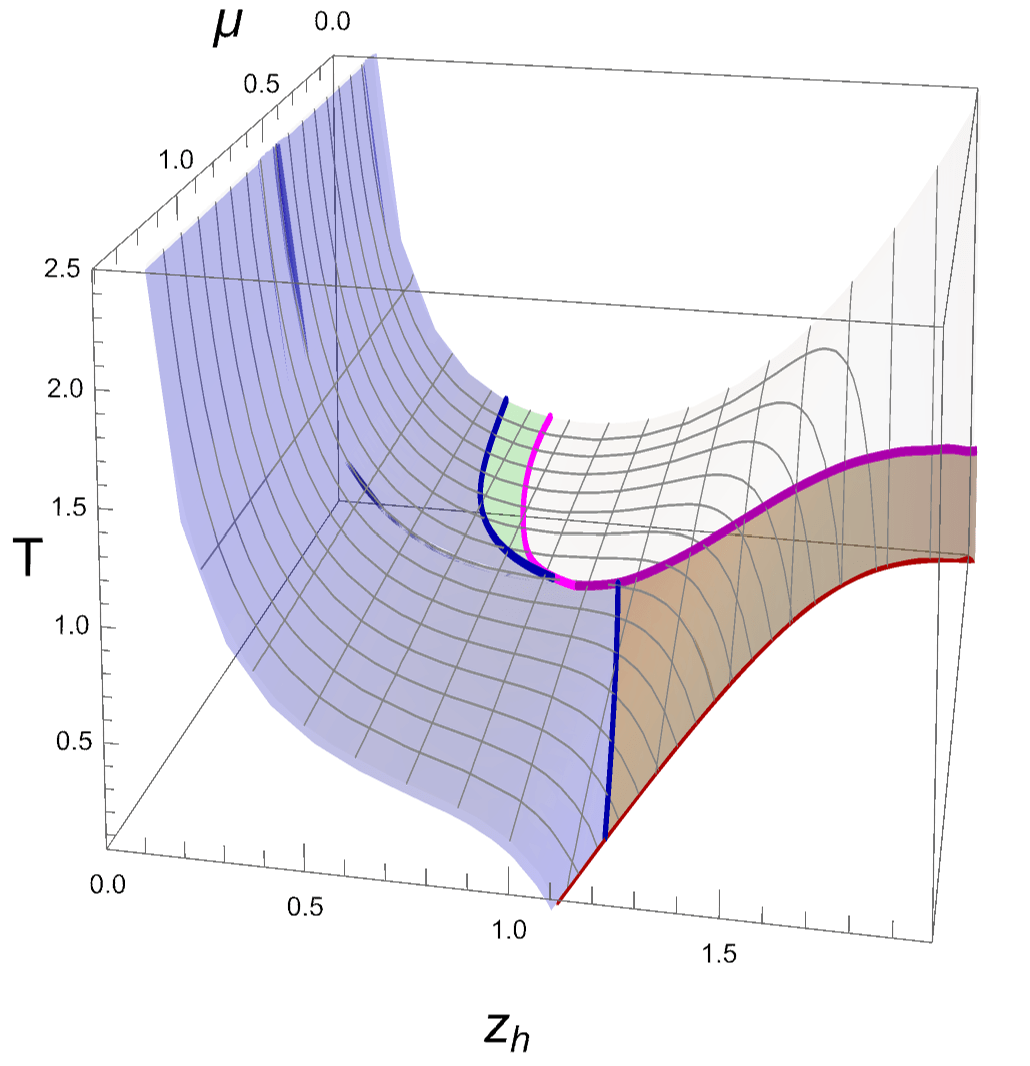}\\
 A\\\,\\
 \includegraphics[scale=0.43]{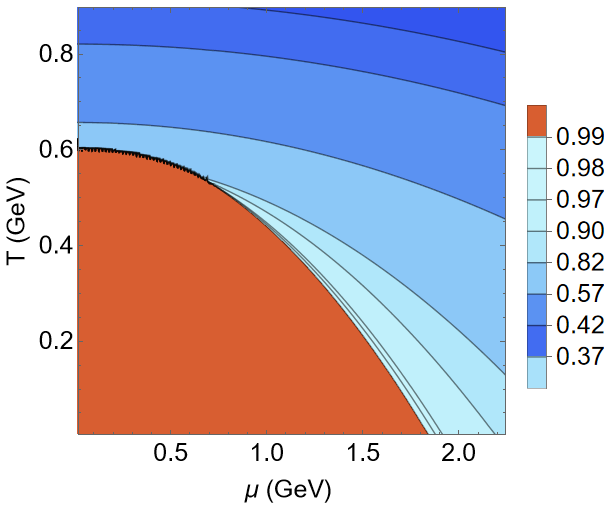}
 \quad
\includegraphics[scale=0.43]{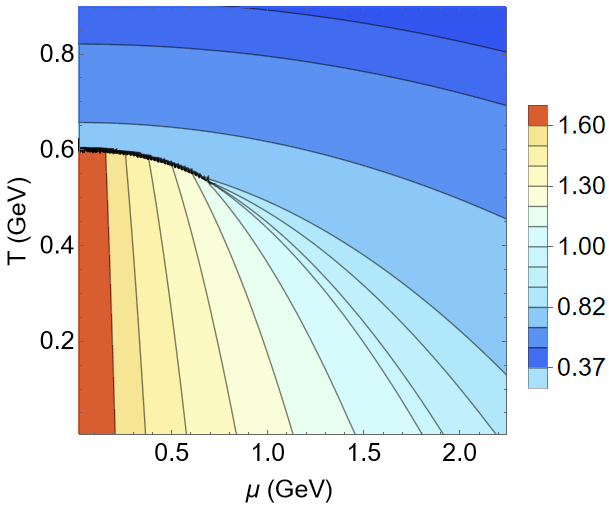}\\
 B\hspace{210pt}C
\caption{Heavy quarks model. A) The 3D plot $T=T(\mu,z_h)$  explaining different phase transitions mentioned in 2D plot in Fig.\,\ref{Fig:HQphase-colors-Е}. Here $[T]=[\mu]=[z_h]^{-1} =$ GeV. 
B) Density plot with contours for the horizon $z_h=z_h(\mu,T)$. We see that the contours with $z_h=0.99,\,0.98,\,0.97,\,0.90,\, 0.82$ joint at CEP.  C) The same as B) with  more contours with fixed $z_h$ in the hadronic region. 
The contours in plot C show that $z_h$ in the hadronic phase depends mainly on $\mu$ and does not essentially depend on $T$ (the reason for this behavior is the almost vertical orientation of the brown part of the surface $T=T(\mu,z_h)$, presented in panels A and B), meanwhile it depends on both $T$ and $\mu$ in the quark-gluon and quarkyonic phases.
\\ 
}
\label{Fig:PhH}
 \end{figure}

To have a better intuition the 3D phase structure $(\mu, z_h, T)$ of heavy quarks model is depicted in Fig.\,\ref{Fig:PhH}A.
The density plot with contours of the horizon $z_h(\mu,T)$ for heavy quarks model is depicted in Fig.\,\ref{Fig:PhH}B and more contours in hadronic region in Fig.\,\ref{Fig:PhH}C . Each contour corresponds to fixed value of $z_h$. This complementary plot describes the phase structure of the heavy quarks model as a contour plot for $z_h$ and completes Fig.\,\ref{Fig:HQphase-colors-Е} and Fig.\,\ref{Fig:PhH}A. Also, the contours in Fig.\,\ref{Fig:PhH}C  show that $z_h$ in the hadronic phase depends mainly on $\mu$ and does not essentially depend on $T$.

\begin{figure}[h!]
  \centering
 \includegraphics[scale=0.3]{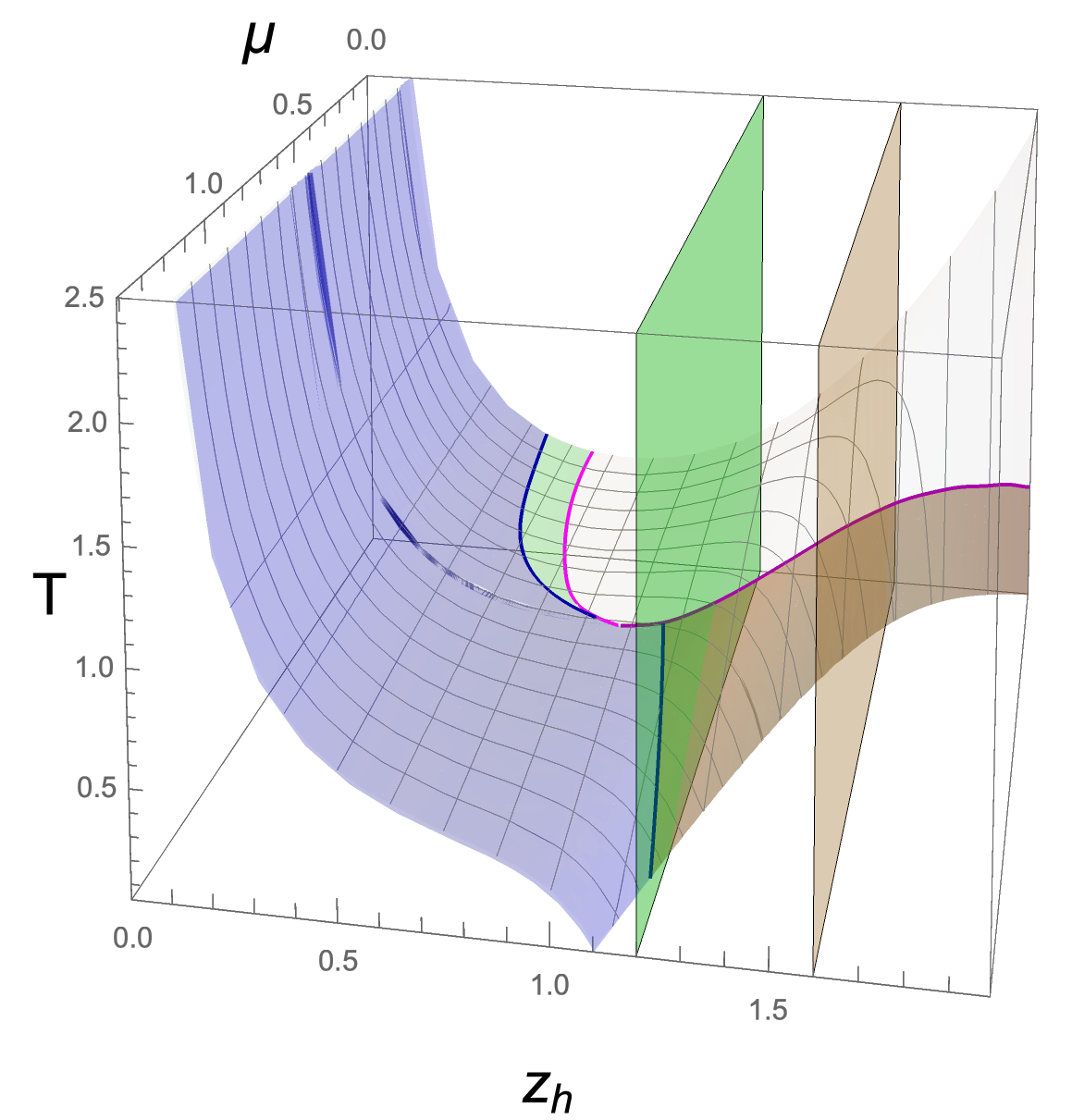}
\includegraphics[scale=0.32]{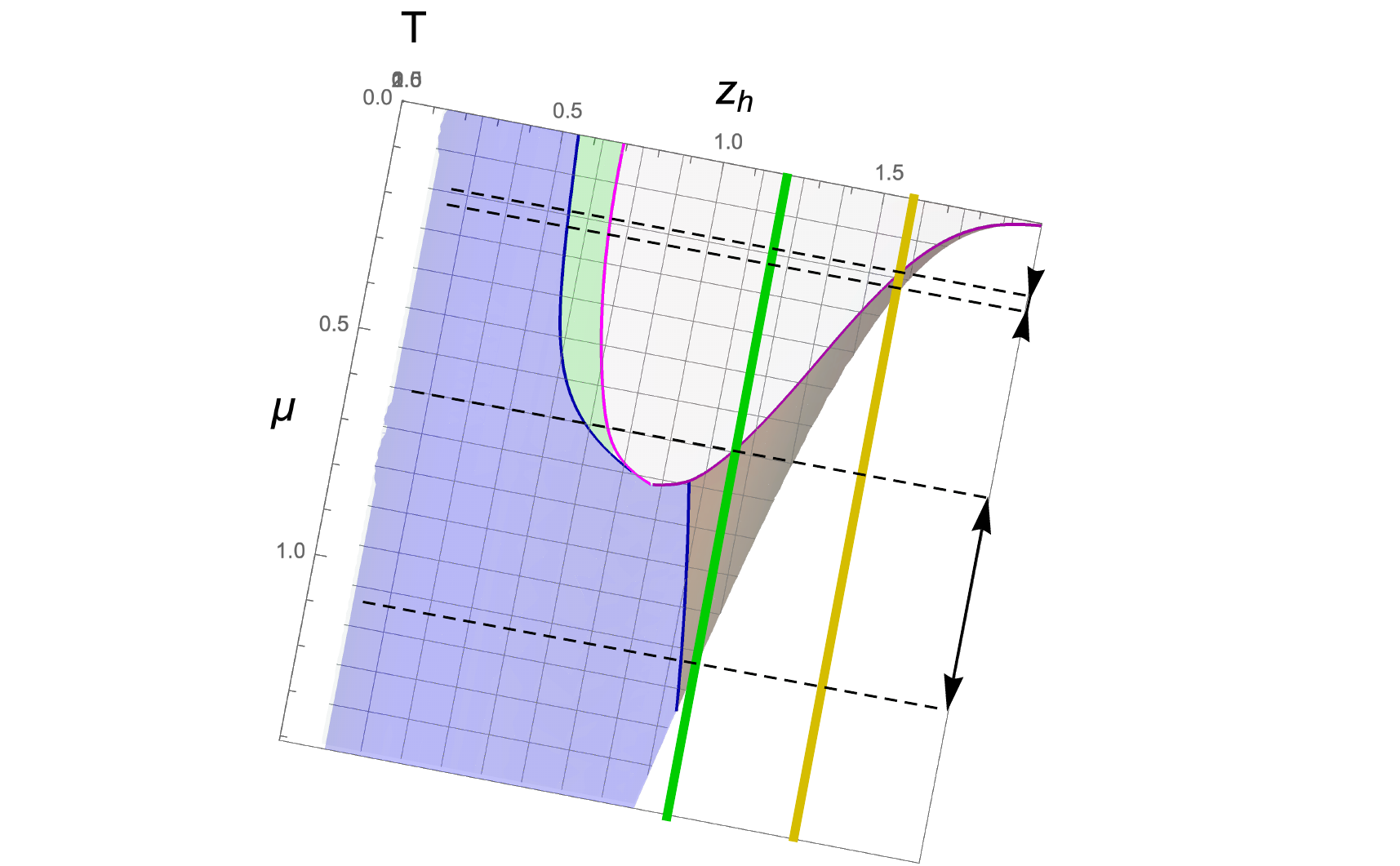}\\
 A\hspace{190pt}B\\\,
\caption{Heavy quarks model. A) The same as Fig.\,\ref{Fig:PhH}A with planes at fixed $z_h$:  $z_h=1.2$ (green), $z_h=1.6$ (beige). B) The orthographic top view of graph A). We see how the $\mu$ coordinate changes at the intersection of the green plane with the hadronic (brown) part of the 3D graph. This option is indicated by a segment with arrows. But the coordinate $\mu$ practically does not change when the beige plane intersects the hadronic part of the 3D graph. This slight variation is indicated by reverse arrows; $[T]=[\mu]=[z_h]^{-1} =$ GeV.}
 \label{Fig:PhH-ad}
\end{figure}

In Fig.\,\ref{Fig:PhH-ad}  complementary plots of Fig.\,\ref{Fig:PhH}A with two surfaces at fixed $z_h$, i.e. $z_h=1.2$ GeV${}^{-1}$ (green), $z_h=1.6$ GeV${}^{-1}$ (beige) for heavy quarks model to understand the behavior of contours in hadronic region in Fig.\,\ref{Fig:PhH}C,  are depicted. 
In Fig.\,\ref{Fig:PhH-ad}B the orthographic top view point of panel A is plotted. We see how the $\mu$ coordinate changes at the intersection of the green plane with the hadronic (brown) part of the 3D plot, i.e. Fig.\,\ref{Fig:PhH-ad}A. This option is indicated by a segment with arrows. But the coordinate $\mu$ practically does not change when the beige surface intersects the hadronic part of the 3D plot. This slight variation is indicated by reverse arrows. It shows that at larger fixed values of $z_h$ ($z_h=1.6$ GeV${}^{-1}$) there is a crucial change in $T$ while there is a small change in $\mu$. Utilizing this point, we can see that in the hadronic region of the Fig.\,\ref{Fig:PhH}C, at large fixed value of $z_h$ there is sharp change in $T$ and not crucial change on $\mu$. It is important to note that the Fig.\,\ref{Fig:PhH-ad} obtained from thermodynamics of the model for heavy quarks.
In addition, since the running coupling $\alpha(z;\mu,T)$ depends on the $z_h$, i.e. increases by increasing $z_h$ and decreases by decreasing $z_h$, the same behavior has been obtained for running coupling of heavy quarks model in Sect.\,\ref{HQ-running10}.

$$\,$$

\newpage
$$\,$$

\subsubsection{Boundary condition and Cornell potential  for heavy quarks model}

To obtain the physical results for heavy quarks model the new boundary condition \eqref{bceHQ} is proposed. To check this boundary condition, we study the Cornell potential \eqref{cornel1} and associated QCD string tension, $\sigma$, as a function of temperature.

\begin{figure}[h!]
  \centering
\includegraphics[scale=0.47]{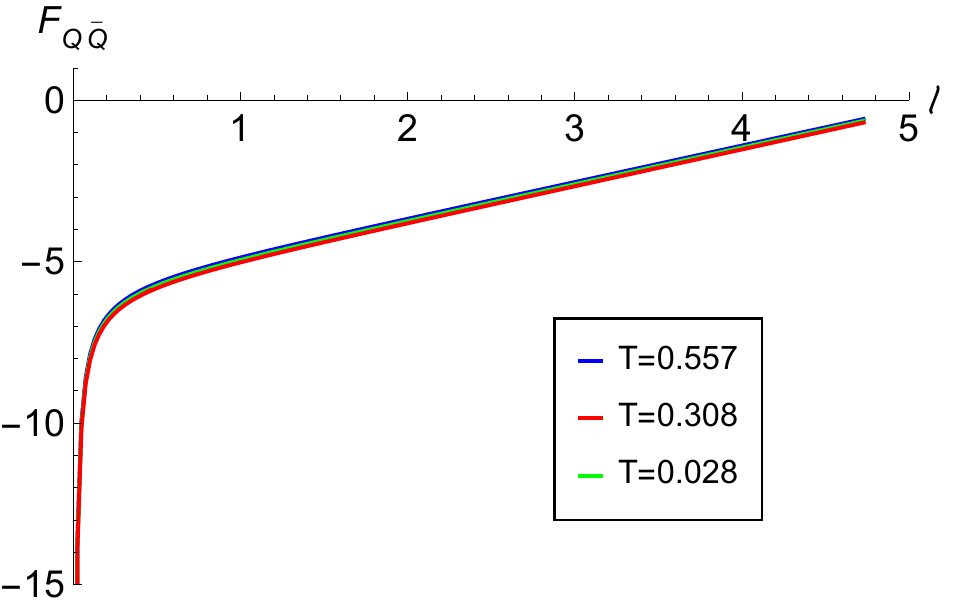}\qquad
 \includegraphics[scale=0.36]{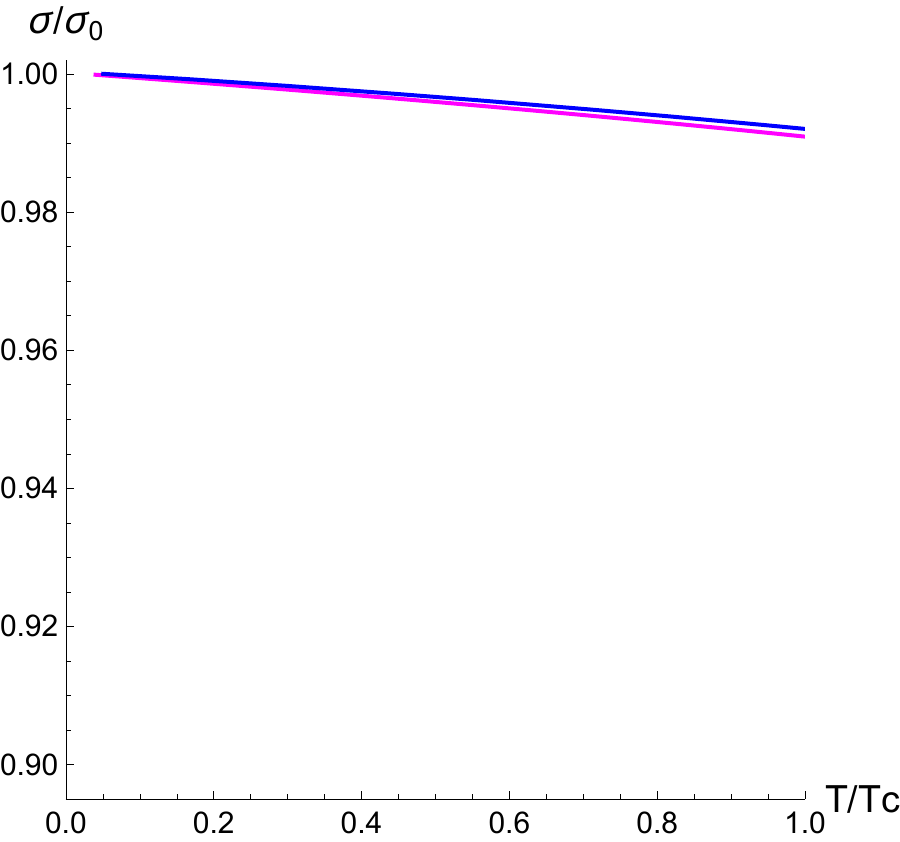}\\
 A\hspace{200pt}B
\caption{$F_{Q\bar{Q}}$ as a function of $\ell$ for heavy quarks model with the second boundary condition \eqref{bceHQ} with different $T$ and fixed $\mu=0.01$ (A) The $\sigma(T)$ for $\mu=0.01$ and $T_c=0.56$. Blue line is obtained from linear part of Cornell potential and Magenta line is obtained from asymptotic of effective potential (B); $[F_{Q\bar{Q}}]=[\sigma]^{\frac{1}{2}}=[\mu]=[T]=[\ell]^{-1} =$ GeV. \\
}
 \label{Fig:cornel100}
\end{figure}

The dependence of energy configuration of quark and anti-quark $F_{Q\bar{Q}}$ on the interquark distance $\ell$ for  different temperatures $T=0.028, 0.308$ and $0.557$ (GeV) and fixed $\mu=0.01$ GeV is depicted in Fig.\,\ref{Fig:cornel100}A. 
Then we obtained the $\sigma (T)$ using two methods, i.e. linear part of the Cornell potential (\ref{cornel1}) in blue color and asymptotic of effective potential for $\ell\to \infty$ (\ref{ten}) in magenta color for $\mu=0.01$ GeV and $T_c=0.56$ GeV in Fig.\,\ref{Fig:cornel100}B.
We see that the $\sigma$ decreases for the boundary condition \eqref{bceHQ}, that confirms the physical behavior of QCD string tension versus temperature \cite{Petreczky:2018xuh,Rothkopf:2019ipj}. Therefore, this shows that the new boundary condition \eqref{bceHQ} is proper one and can be used to obtain the correct behavior of running coupling constant as a function of parameters for the heavy quarks model.

\newpage

\subsubsection{Dependence of energy scale $E$ on the holographic $z$ coordinate}\label{DES-HQ}

\begin{figure}[!ht]
  \begin{center}
  \includegraphics[scale=0.25]    {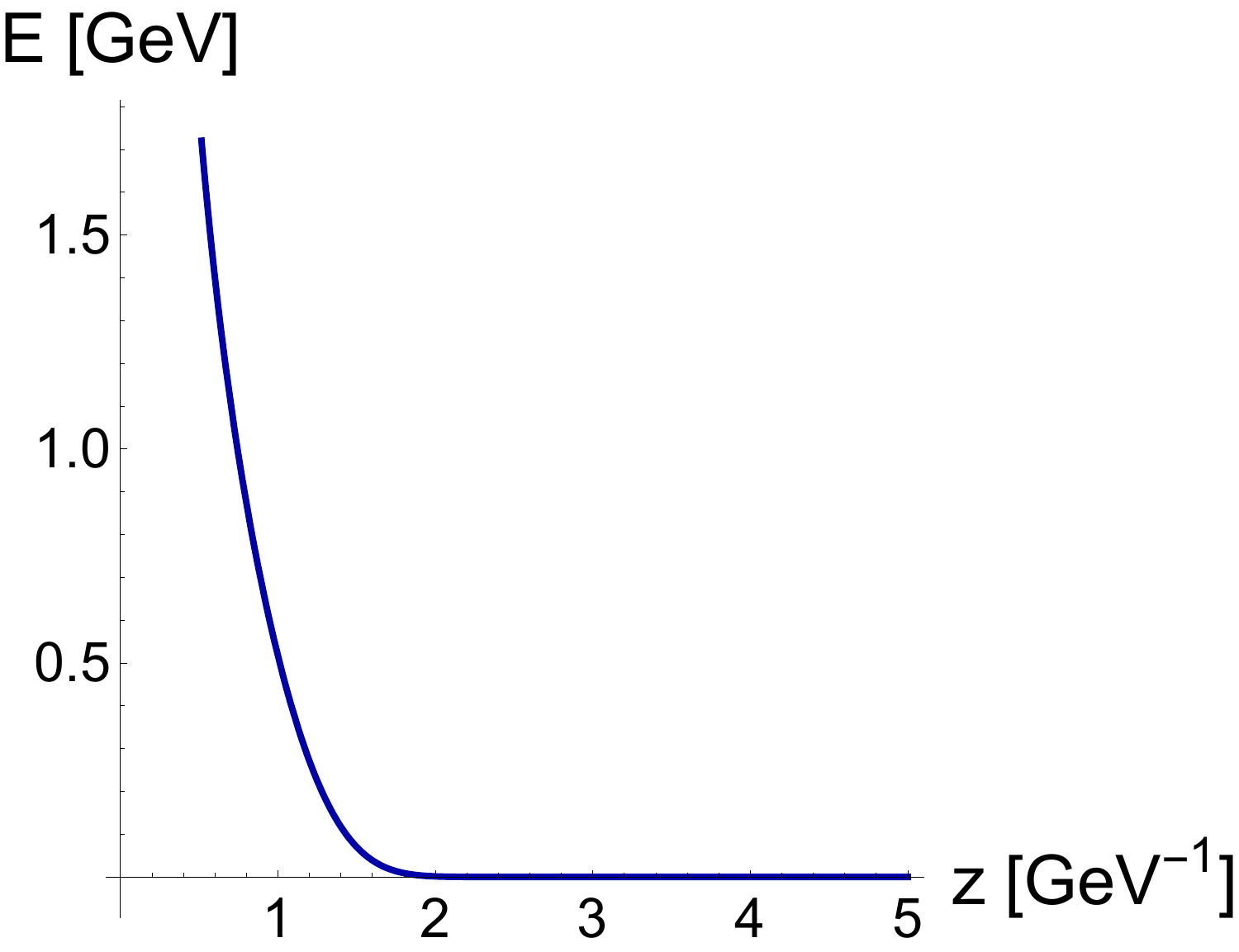} \qquad\includegraphics[scale=0.25]    {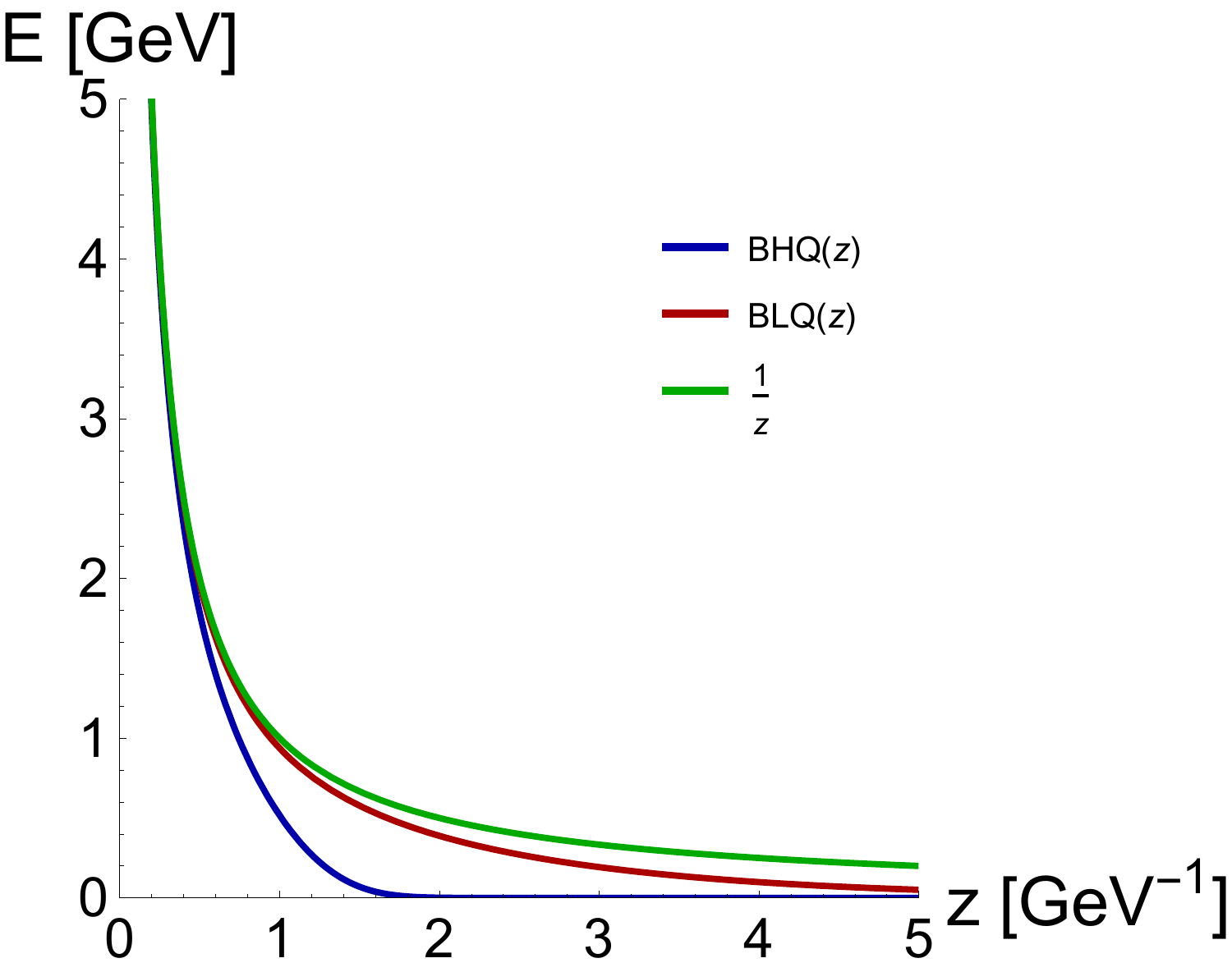}\\
  A\hspace{185pt}B
  \end{center}
  \caption{A) The Energy scale \( E \) (GeV) in the boundary field theory as a function of the holographic coordinate \( z \) (GeV${}^{-1}$) for heavy quark model.
  B) The comparison of warp factors for heavy (darker blue (BHQ)), light (darker red (BLQ)) quarks and AdS (green) cases. We see that for small $z$ these three factors are almost coincide.}
\label{E-of-z-HQ}
\end{figure}

Relating the holographic coordinate $z$ to the energy scale $E$ or the square of the transferred momenta $Q^2$, one can express the dependence of running coupling  on the $E$ or $Q^2$.
The warp factor in the metric \eqref{metric} can be identified with the energy scale $E$ in the boundary field theory \cite{Galow:2009kw}. For the heavy quark model we have: 
\be \label{BBH}
E=B(z)=\frac{1}{z\,\, e^{\frac{\fc}{3}z^2+ p\, z^4}}\, ,
\ee
where $\fc$ and $p$ are introduced in \eqref{scaleHQ}.
The Energy scale \( E \) (GeV) in the boundary field theory as a function of the holographic coordinate \( z \) (GeV${}^{-1}$) for heavy quark model is presented in Fig.\,\ref{E-of-z-HQ}. The energy scale $E$ is in correspondence with the warp factor $B(z)$ given by  (\ref{BBH}). We can see that small $z$ corresponds to large $E$ and large $z$ corresponds to small $E$, as it should be.

The dependence of energy scale $E$ on the energy scale $z$ is monotonically decreasing function. By considering the asymptotics of $E(z)$ when $z \to 0$  the plots for $E(z)$ for light and heavy quarks models and AdS case correspond to each other. In addition, for large values of $z$ there is no qualitative difference between the light and heavy quarks models and AdS case.

\newpage

\section{Running coupling constant} 
\label{sec:running}

The running coupling constant in holography was defined in as
\eqref{lambda-phi} $\alpha(z)=\exp (\varphi(z))$. Here, $z$ is the holographic coordinate that represents the energy scale of the theory and $\varphi$ is the dilaton field that solves the system of equations  (\ref{phi2prime})--(\ref{A2primes}). To obtain the running coupling for light and heavy  quarks model we have to fix the boundary conditions \eqref{phi-z0} for the dilaton field. Let us denote the solution with boundary condition \eqref{bc0} as $\varphi_0(z)$
and corresponding $\alpha(z)$ as $\alpha_0(z)$, i.e. $\alpha_0(z)=e^{\varphi_0(z)}$. Changing the dilaton boundary condition we make just a
rescaling of the coupling constant 
\bea 
\alpha_0(z)&\to &\alpha(z,z_0)= \alpha_0(z)\,\fG(z_0), \quad\mbox{where}
,\quad \fG (z_0)=e^{-\varphi_{0}(z_0)},
\eea
that does not change the dependence of the running coupling on holographic coordinate $z$. By choosing a boundary condition at point $z_0$, depending on the position of the horizon, i.e. $z_0=\fz(z_h)$, we get different behavior of the running coupling constant depending on the thermodynamic characteristics.
As it is mentioned in Sect.\,\ref{BC-CP-LQ} we considered zero, first and second boundary conditions correspond to \eqref{bc0}, \eqref{bch}  and \eqref{bce}, respectively.
 But we will see from the calculations below, that only one of them, namely \eqref{bce}, gives physically reasonable results.

\subsection{Running coupling for light quarks model}

\subsubsection{Running coupling for light quarks model with boundary condition  $z_0=0$}

\begin{figure}[h!]
    \centering
    \includegraphics[scale=0.5]{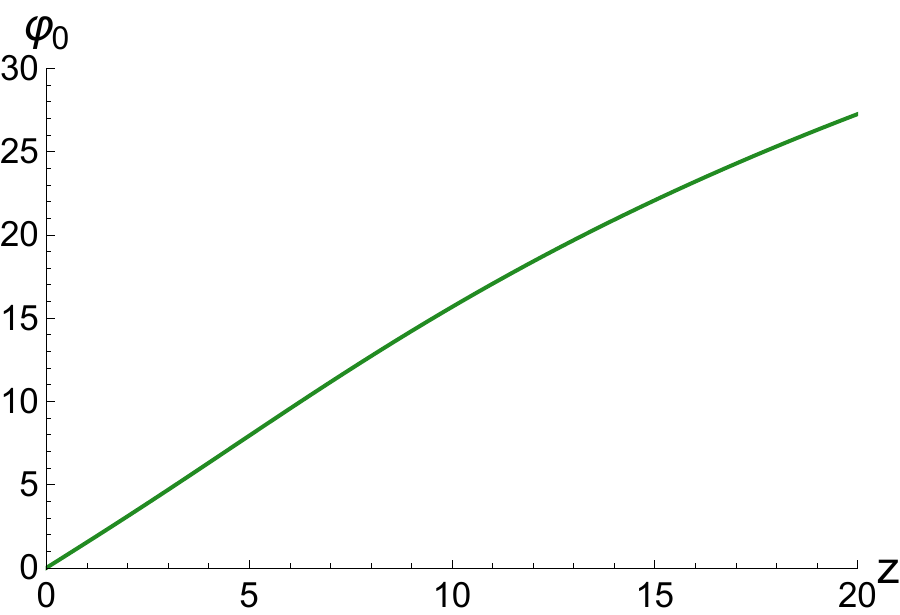}
    \caption{The logarithm of coupling constant $\log\alpha_0(z)$ for light quarks with the zero boundary condition $z_0=0$  ($\varphi_0(0)=0$); $[z]^{-1} =$ GeV.
    }
    \label{fig:phi0-0-LQ}
\end{figure}

The logarithm of coupling constant $\log\alpha_0(z)$ for the light quarks model with the zero boundary condition $z_0=0$  ($\varphi_0(0)=0$) is presented in Fig.\,\ref{fig:phi0-0-LQ}. Although $\log\alpha_0(z)$ increases as the energy scale $z$ increases (or, equivalently, the energy scale $E$ decreases), the thermodynamic properties of the model cannot be studied using this boundary condition.

\subsubsection{Running coupling for light quarks model with boundary condition  $z_0=z_h$}

In this subsection we consider the boundary condition \eqref{bch}. As mentioned above, this boundary condition produces the dependence of the coupling constant on $T$ and $\mu$. Indeed, we  have 
\bea\label{alphazh}
\alpha_{z_h}(z)&=&\alpha_0(z)\,\fG(z_h),\quad\mbox{where}
\quad \fG (z_h)=e^{-\varphi_{0}(z_h)},
\eea 
or we can rewrite \eqref{alphazh} as  \be \label{alphazh-m}
\alpha_{z_h}(z)=\alpha_{z_h}(z;T,\mu)=e^{\varphi_{z_h}(z;\mu,T)}=\alpha_0(z)\,\fG(z_h(T,\mu)).
\ee 
Here  $z_{h}=z_{h}(T,\mu)$ is the value of the horizon for the stable black hole corresponding to given chemical potential $\mu$ and temperature $T$ that can be obtained via $T(z_h)$ plot described in SubSect.\,\ref{phaseLQ}.

\begin{figure}[h!]
  \centering
\begin{minipage}{9.5cm}
    \includegraphics[scale=0.5]{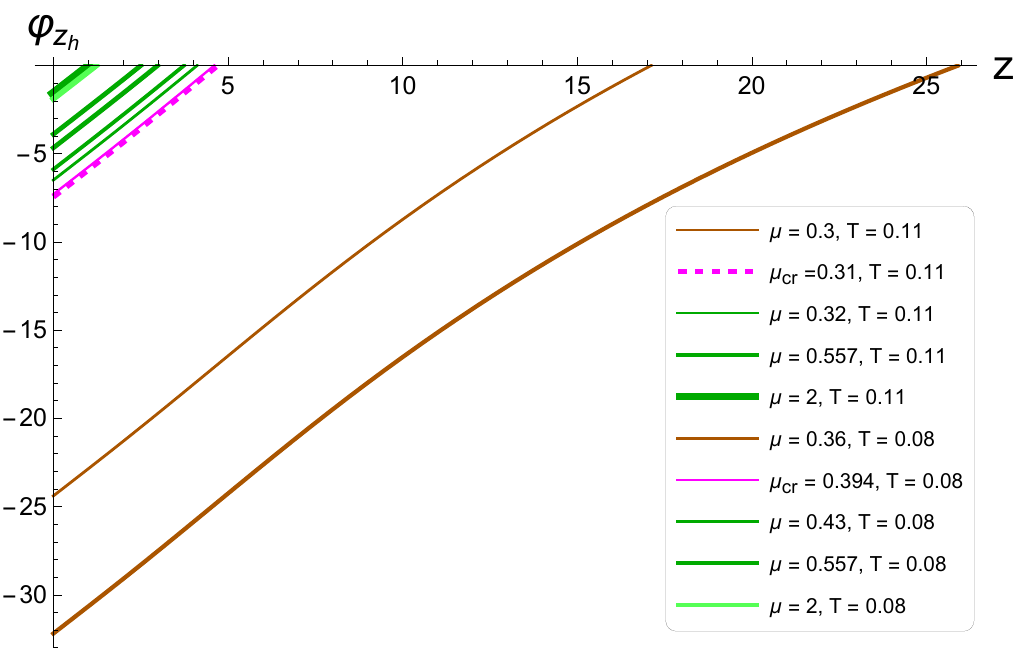} 
\end{minipage}
\begin{minipage}{5cm}
    \includegraphics[scale=0.56]{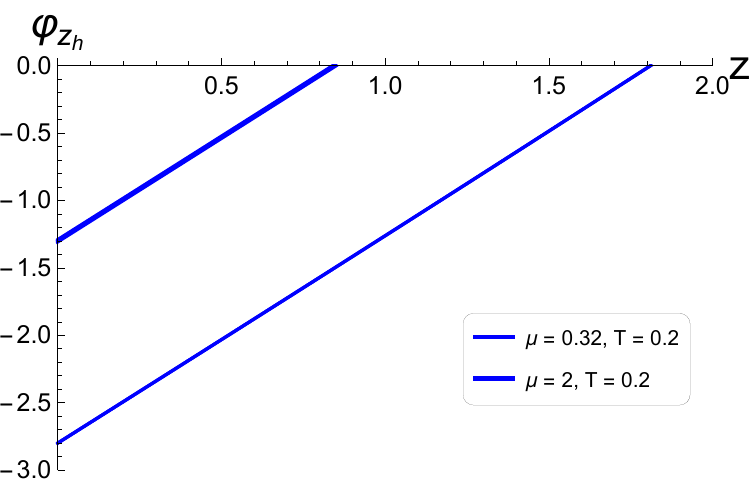}
\end{minipage}\\
\hspace{80pt}A\hspace{210pt}B
 \caption{The logarithm of coupling constant $\log\alpha_{z_h}(z;\mu,T)$ for light quarks model considering the first boundary condition $z_0=z_h$ at $T=0.08, 0.11$ (A) and  $T=0.2$ (B) for different $\mu$; $[\mu]=[T]=[z]^{-1} =$ GeV.
 }
  \label{Fig:LQphi-phi0-zh-mu-T}
\end{figure}

The logarithm of coupling constant $\varphi_{z_h}(z;\mu,T)$ for the light quarks model with the first boundary condition $z_0=z_h$ is depicted in Fig.\,\ref{Fig:LQphi-phi0-zh-mu-T}A for $T=0.08, 0.11$ (GeV) and Fig.\,\ref{Fig:LQphi-phi0-zh-mu-T}B for $T=0.2$ GeV with different values of the chemical potential corresponding to the points of different regions of light quarks phase diagram in Fig \ref{Fig:T-zh-mu-LQ-A}. Although $\log\alpha_{z_h}(z)$ increases as the energy scale $z$ increases (or, equivalently, the energy scale $E$ decreases), we will see that the resulted running coupling in terms of thermodynamic parameters $T$ and $\mu$ are not compatible with lattice calculations.

\begin{figure}[h!]
  \centering
  \includegraphics[scale=0.3]{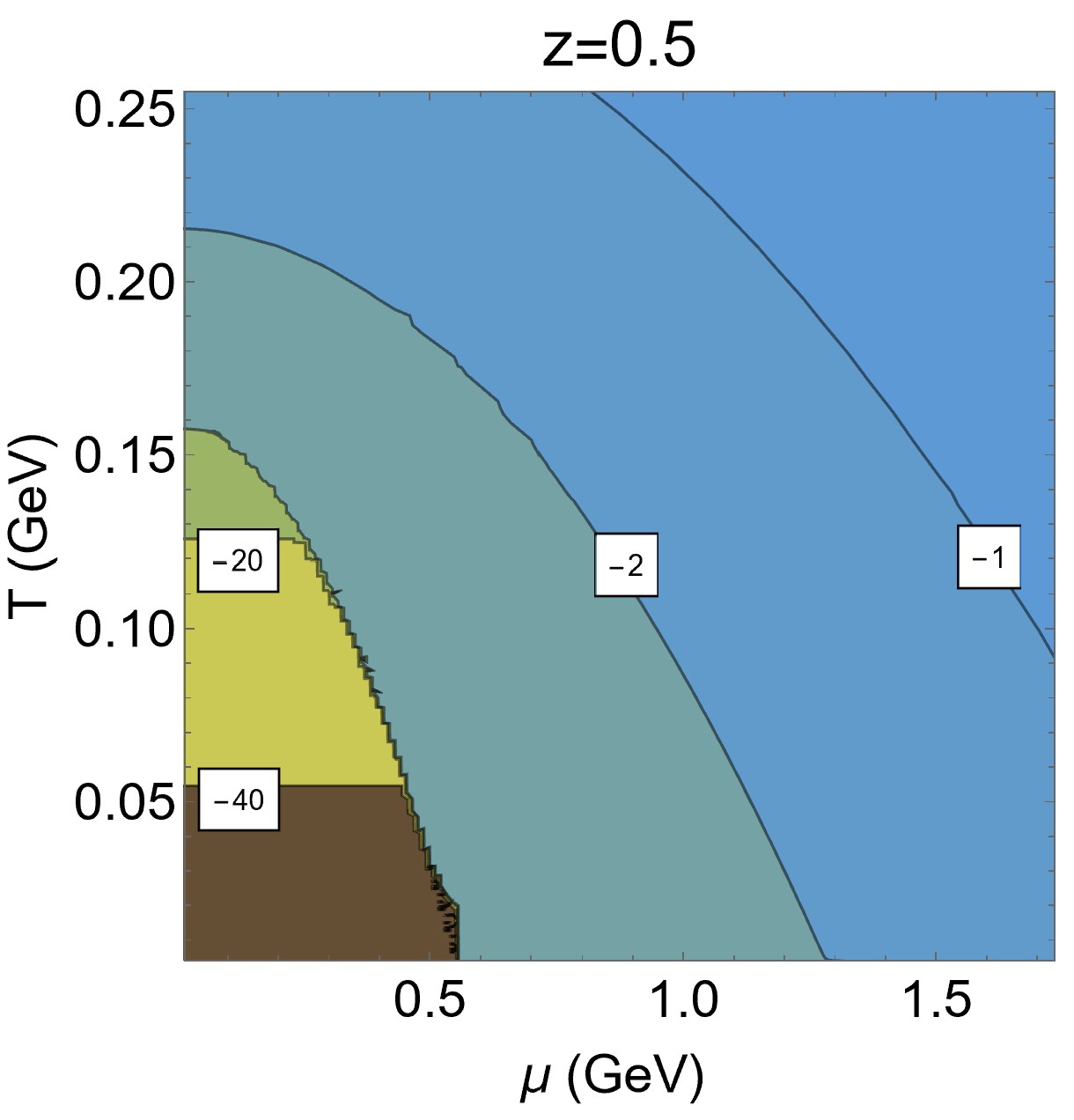}
\includegraphics[scale=0.3]{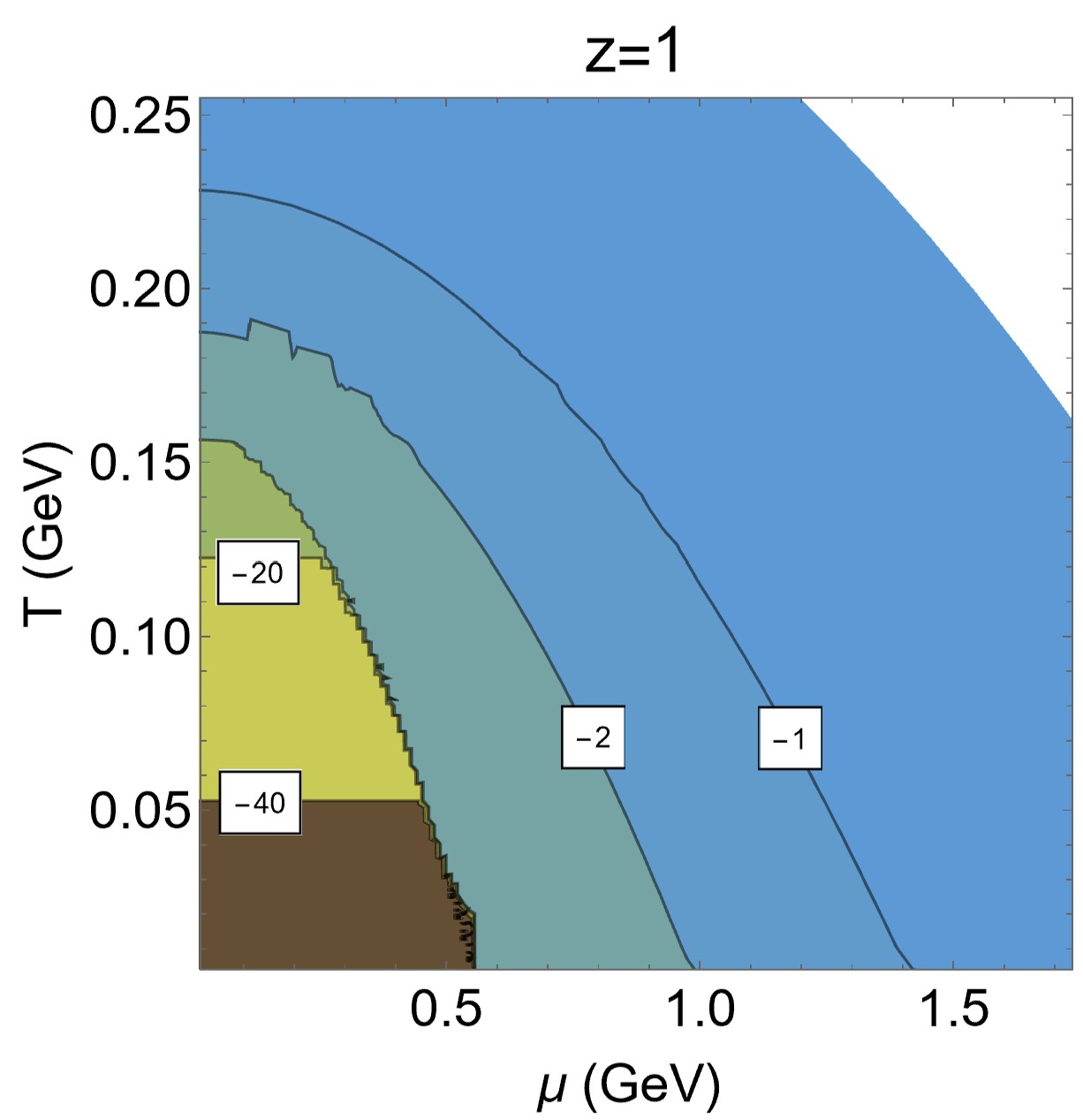}
 \includegraphics[scale=0.3]{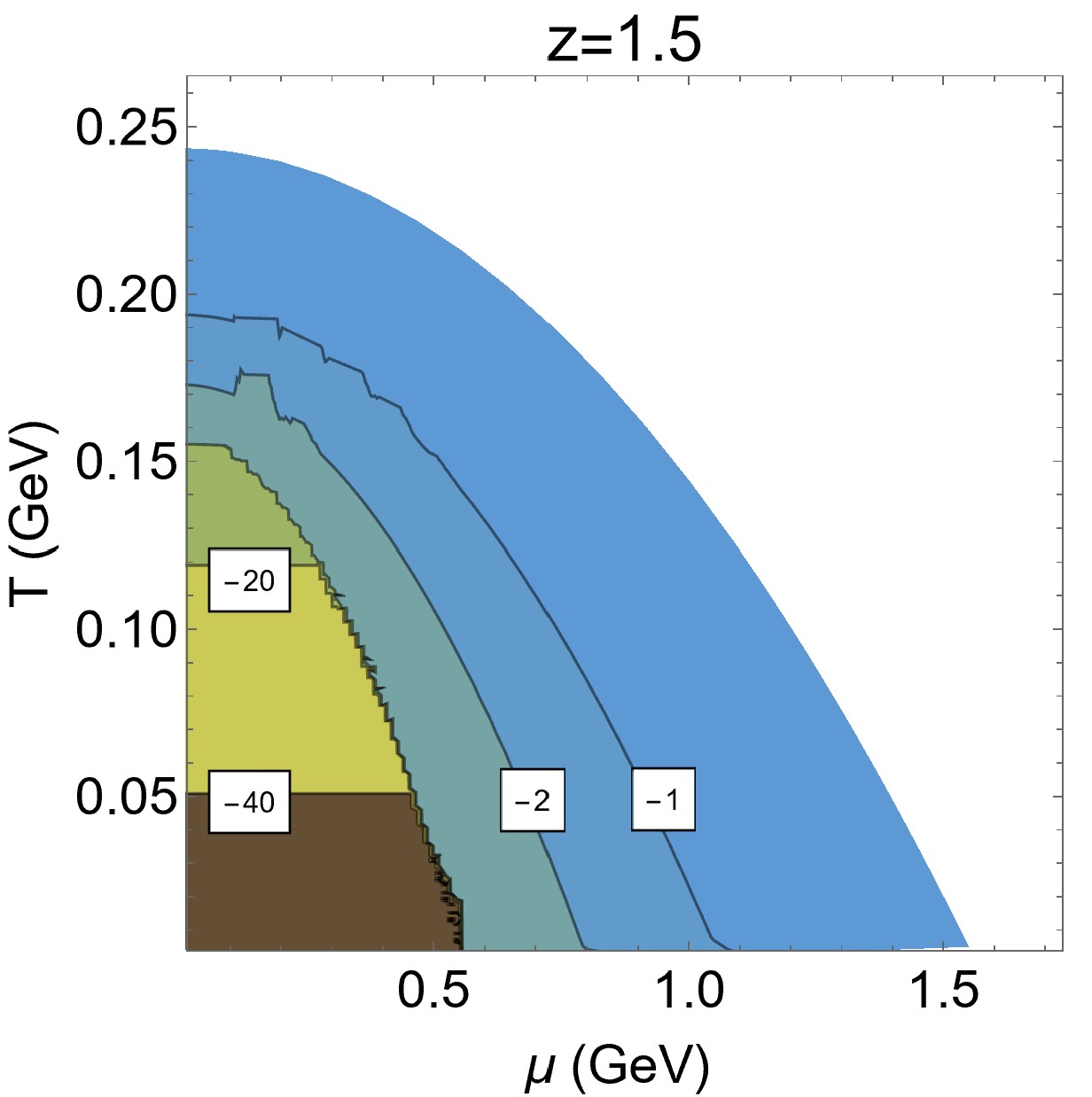}\\
 \includegraphics[scale=0.29]{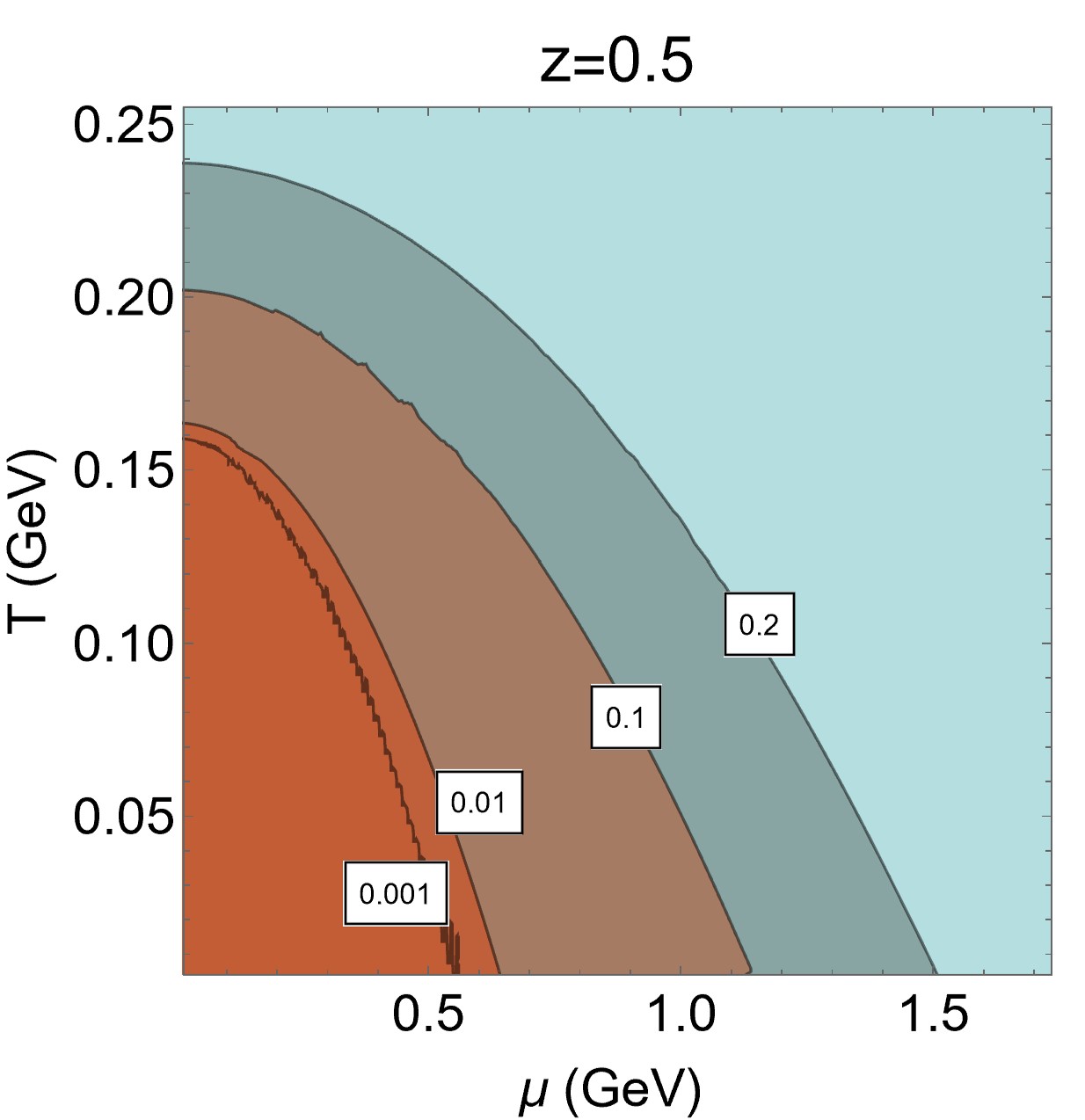}
 \includegraphics[scale=0.29]{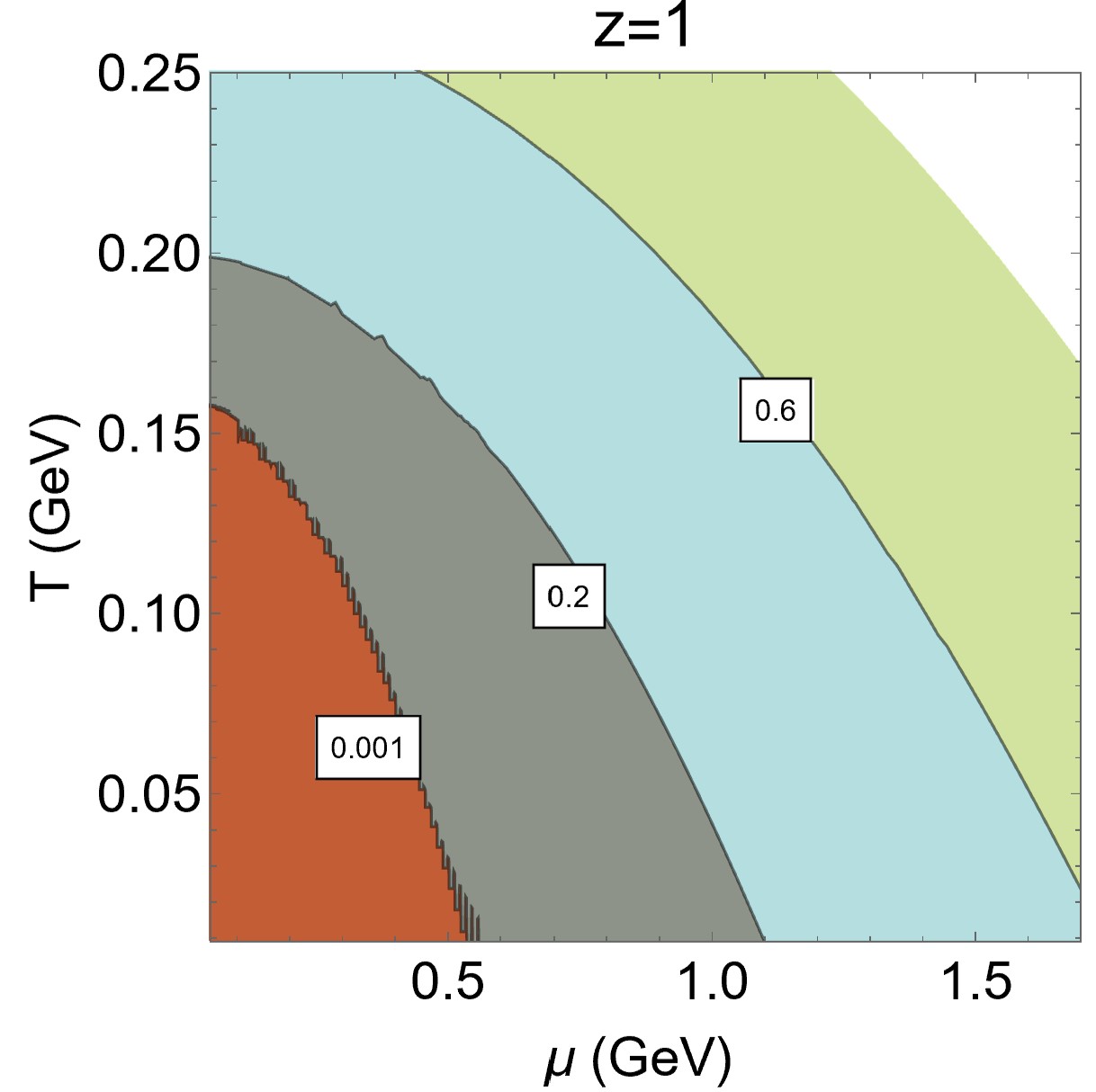}
 \includegraphics[scale=0.29]{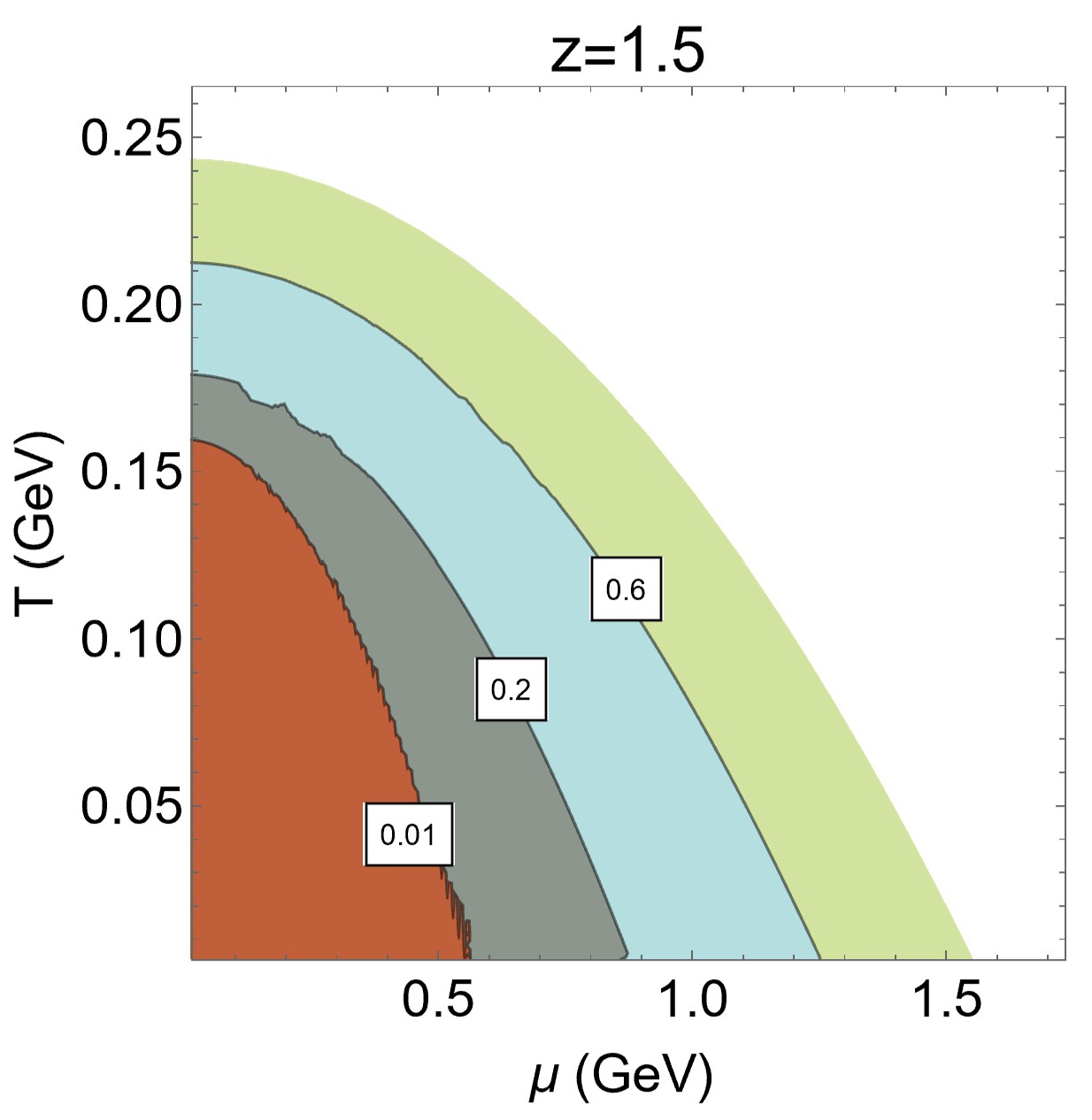}
 \\
\caption{The density plots with contours for $\log\alpha_{z_h}(z;\mu,T)$ (the upper line) and $\alpha_{z_h}(z;\mu,T)$ (the bottom line) for light quarks model at different energy scales $z=0.5,1,1.5$; $[z]^{-1} =$ GeV.\\
}
 \label{Fig:Cz0-z1-z15}
\end{figure}

\begin{figure}[h!]
  \centering
\includegraphics[scale=0.30]{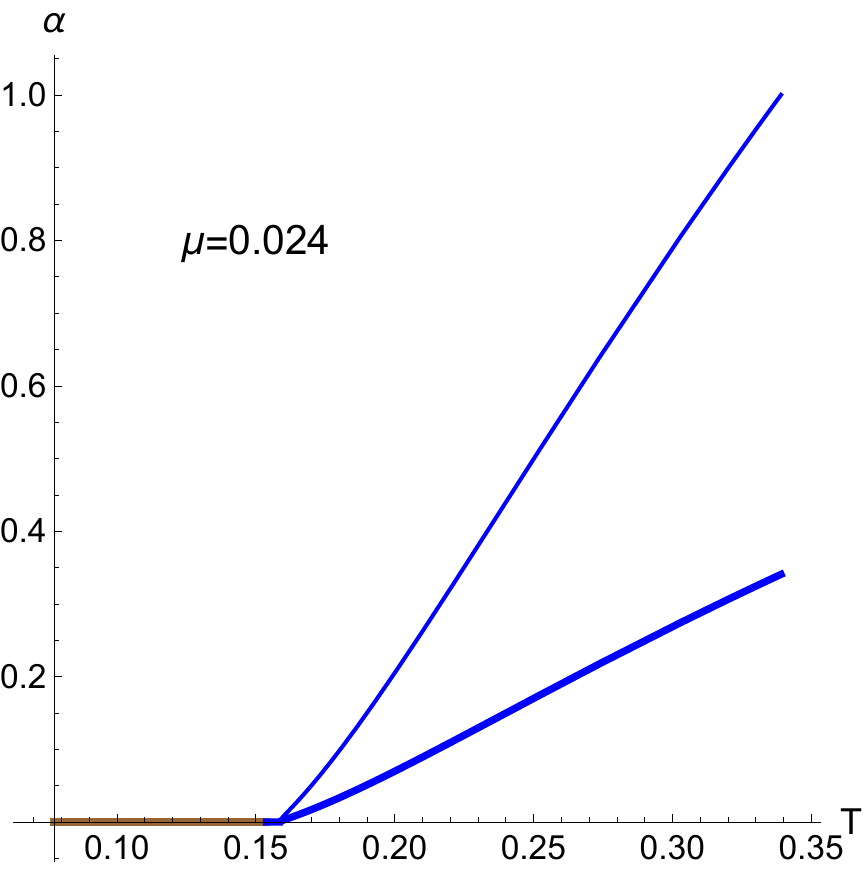} \qquad \qquad \qquad
\includegraphics[scale=0.30]{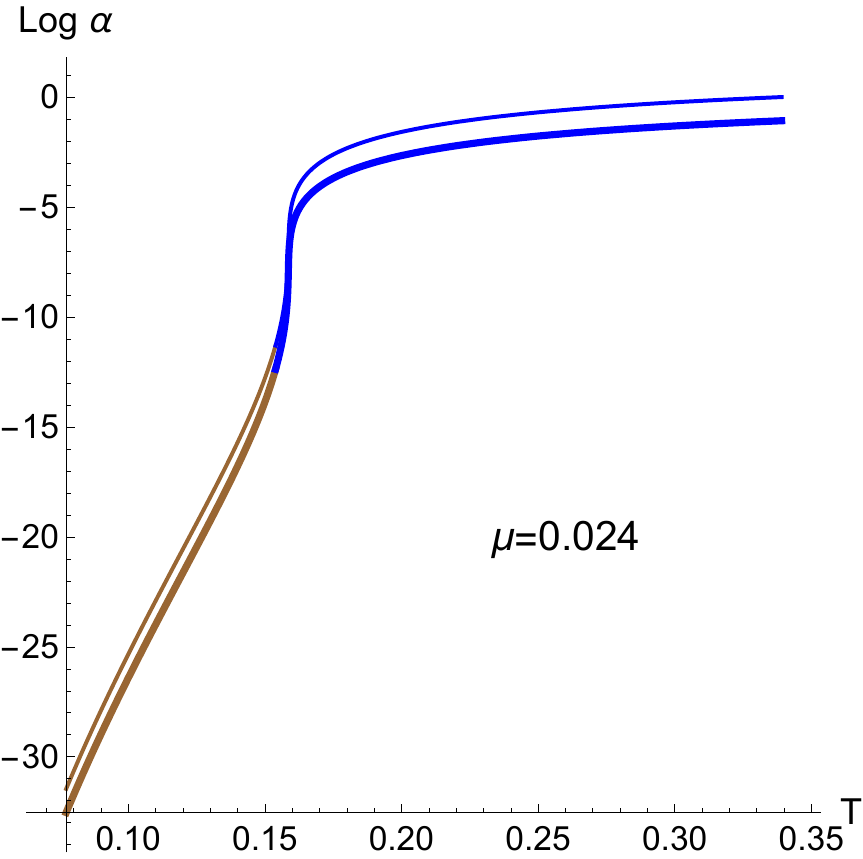}\\
 A\hspace{200pt}B \\\,\\
\includegraphics[scale=0.30]{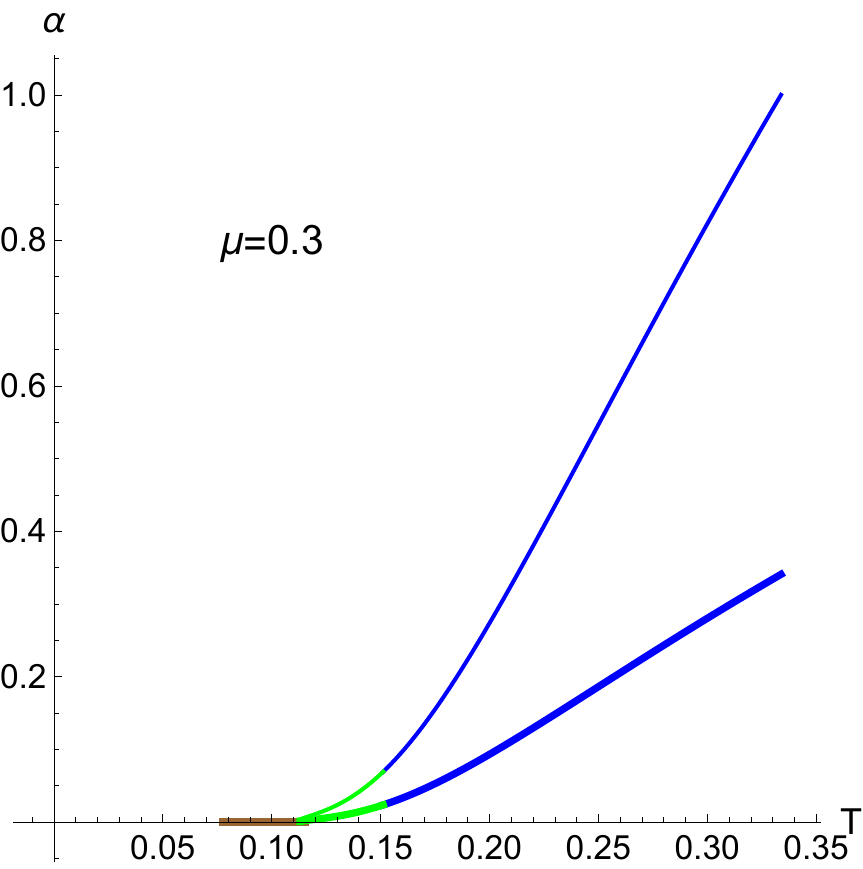} \qquad \qquad \qquad
\includegraphics[scale=0.30]{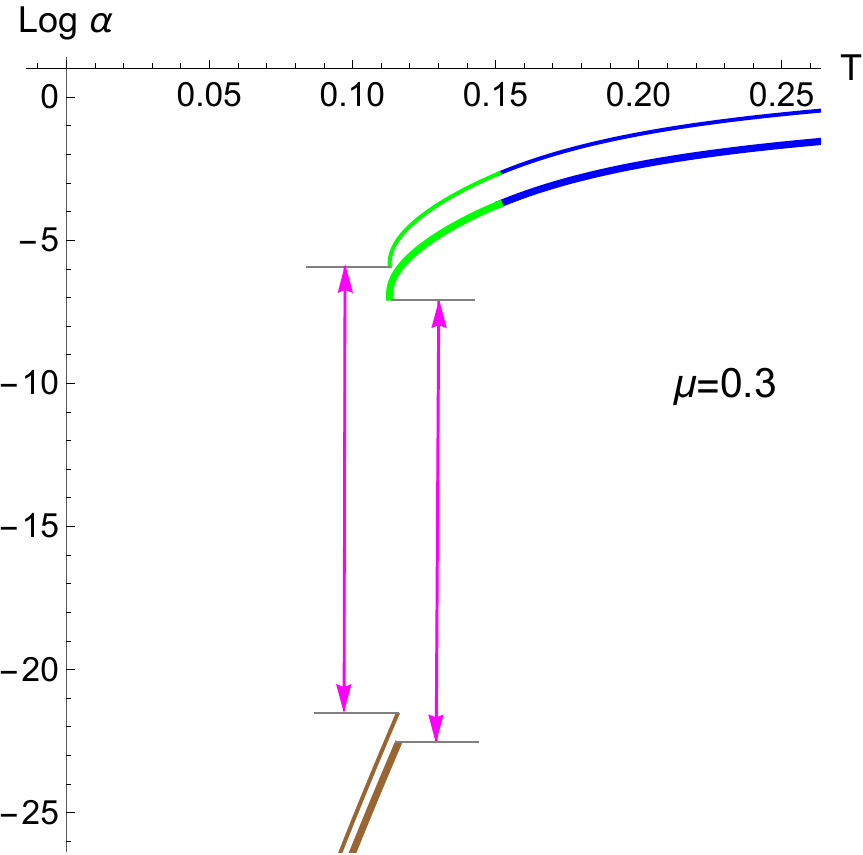}\\
 C\hspace{200pt}D \\\,\\
\includegraphics[scale=0.30]{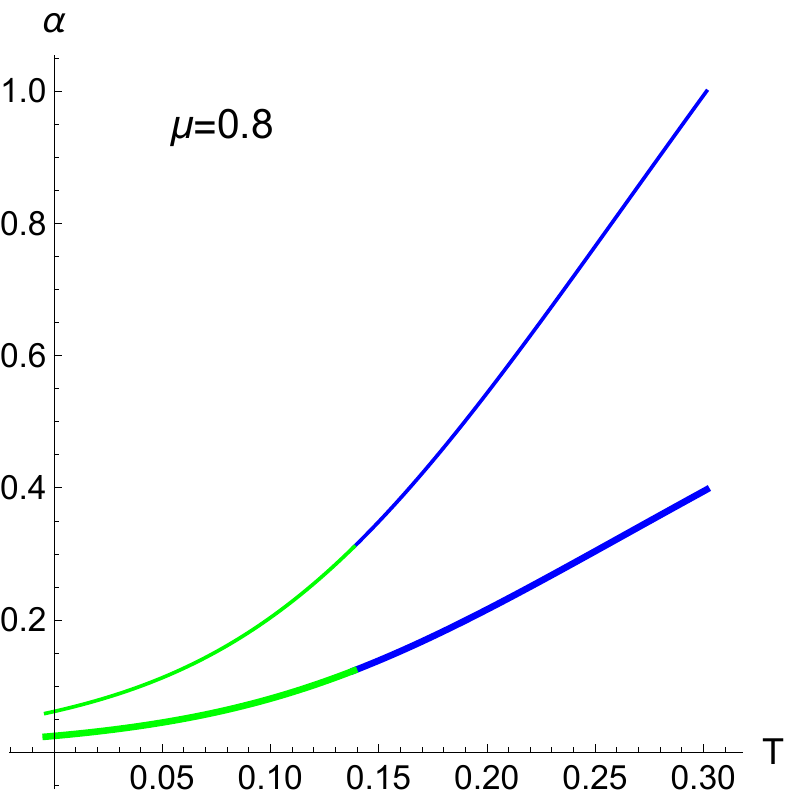} \qquad \qquad  \qquad
\includegraphics[scale=0.30]{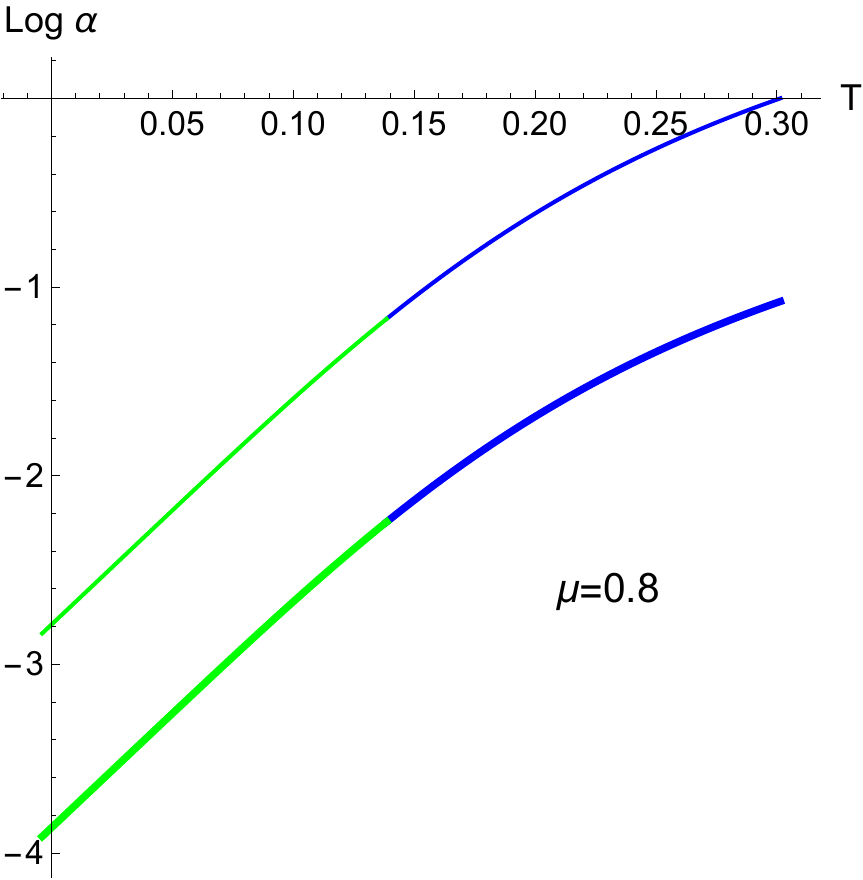}\\
 E\hspace{200pt}F \\
 \,
\caption{Coupling constant $\alpha=\alpha_{z_h}(z;\mu,T)$  and  $ \log \, \alpha_{z_h}(z;\mu,T) $ for light quarks at fixed $\mu$ at different energy scales  $z=1$ (thin lines) and $z=0.3$ (thick lines).  Hadronic, QGP and quarkyonic phases are denoted by brown, blue and green lines, respectively. Magenta arrows in D show the jumps at the 1st order phase transition. $[\mu]=[T]=[z]^{-1} =$ GeV.
}
\label{Fig:LQ-alpha-2D-zh}
\end{figure}

In Fig.\,\ref{Fig:Cz0-z1-z15}, the density plots with contours of the logarithm of coupling constant $\log\alpha_{z_h}(z;\mu,T)$ (the upper line) and coupling constant $\alpha_{z_h}(z;\mu,T)$ (the bottom line) for light
 quarks at different energy scales $z=0.5,1,1.5$ (GeV${}^{-1}$) are shown. In hadronic phase, $\log\alpha_{z_h}(z;\mu,T)$ has much dependence on the $T$ in comparison to dependence on the $\mu$. This result also is confirmed by result that obtained in Fig.\,\ref{Fig:LQ-alpha-2D-zh}.
 For coupling constant $\alpha_{z_h}(z;\mu,T)$ the dependence on $\mu$ and $T$ are  approximately the same for different phases at different scales of energy.

\begin{figure}[h!]
  \centering
\includegraphics[scale=0.39]{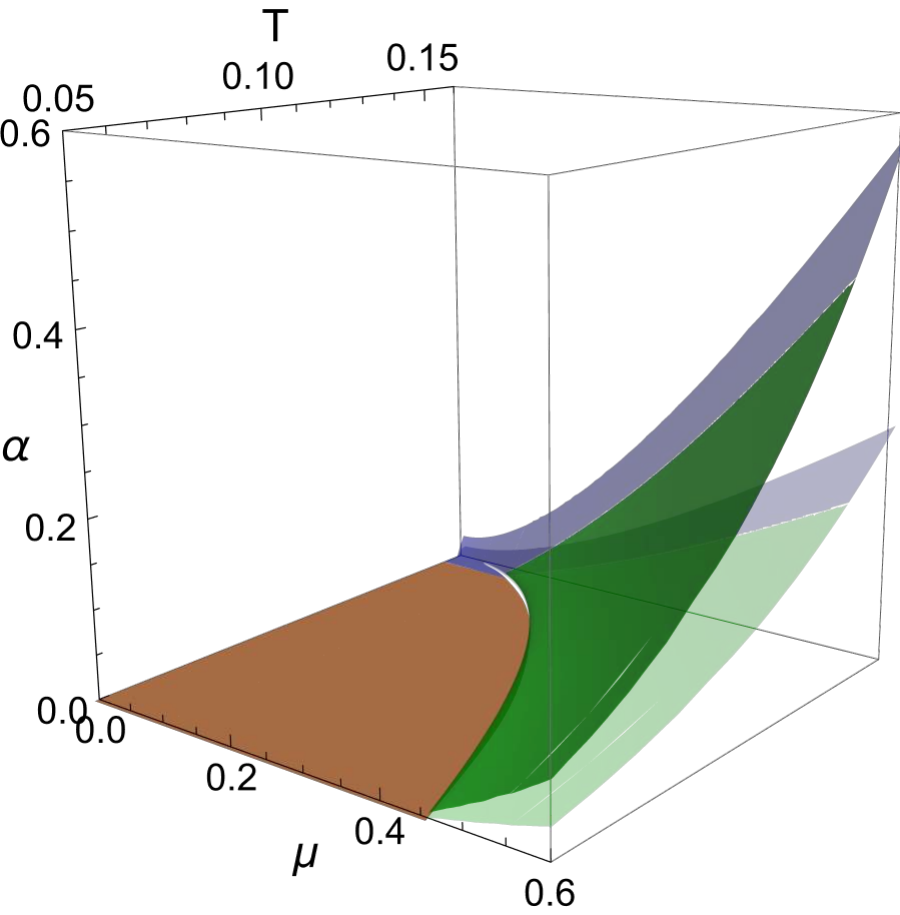}\qquad\qquad
\includegraphics[scale=0.4]{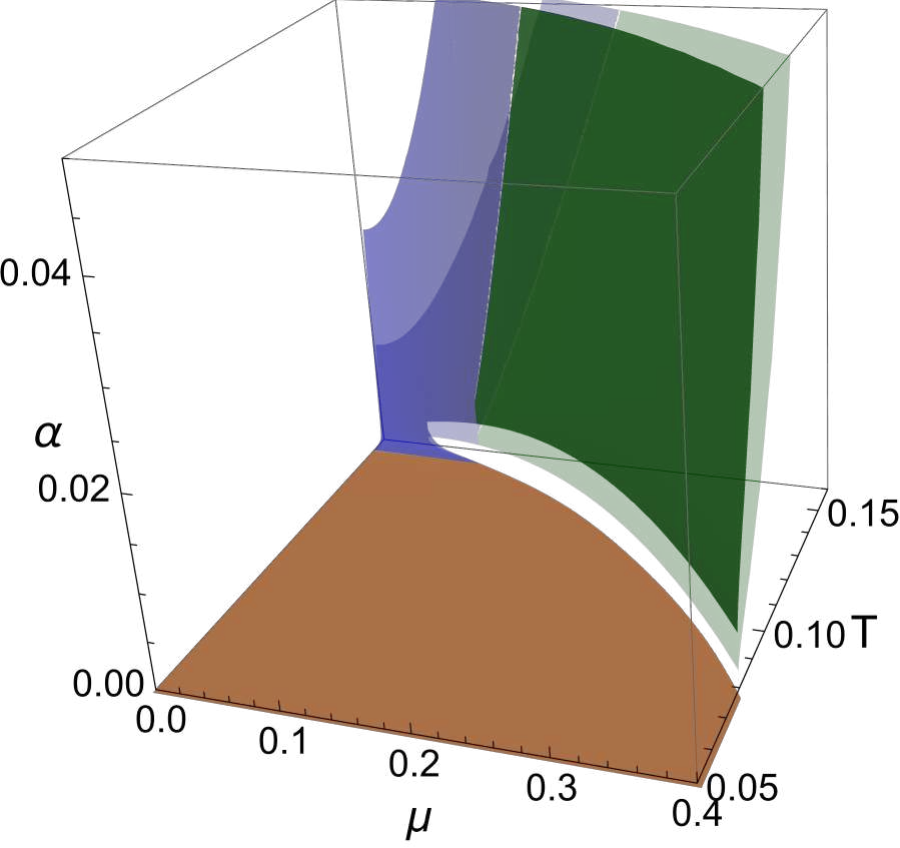}\\
A\hspace{200pt}B\\$\,$\\
\includegraphics[scale=0.4]{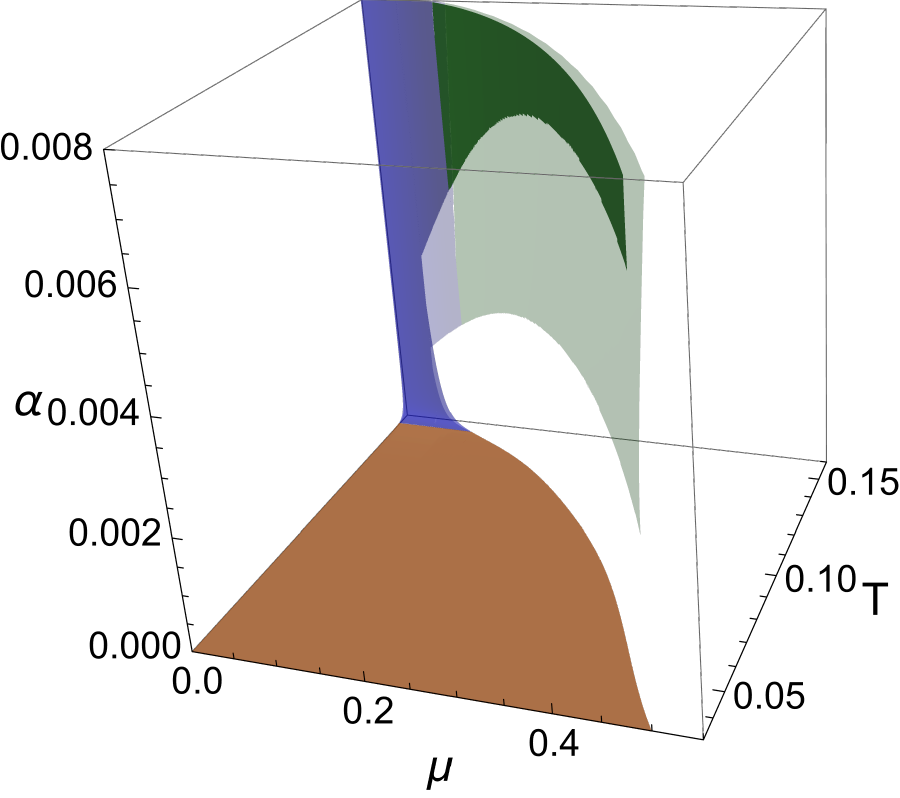}\qquad \qquad
\includegraphics[scale=0.4]{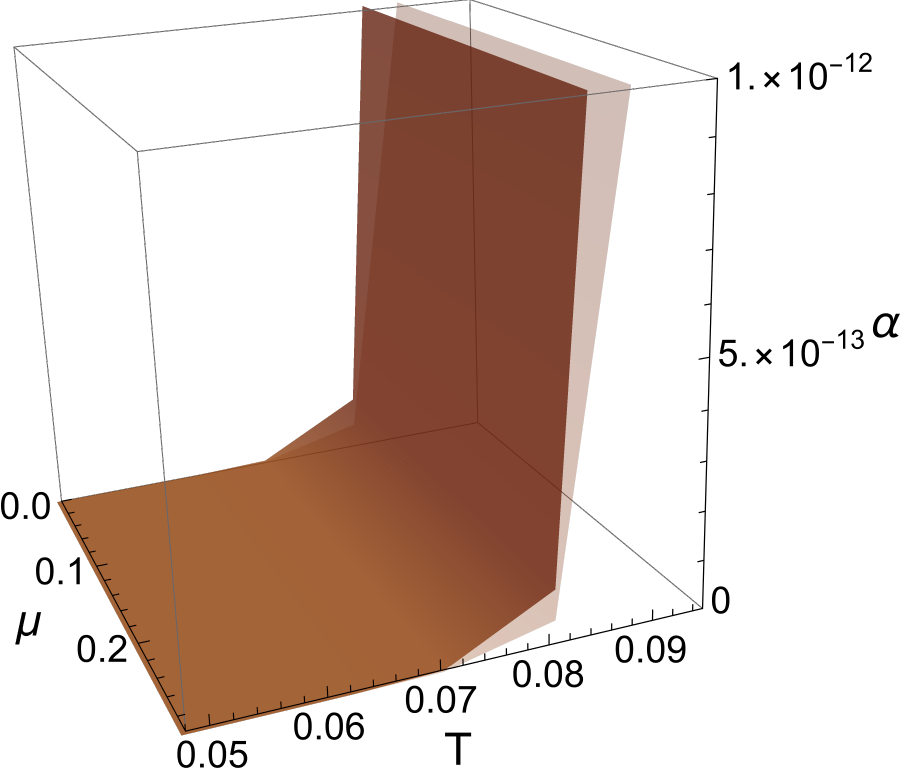}\\
C\hspace{200pt}D
\caption{Comparison of coupling constant $\alpha=\alpha_{z_h}(z;\mu,T)$ for light quarks in two energy scales, $z=1$ (light surfaces) and $z=1.5$ (dark surfaces). Hadronic, QGP and quarkyonic phases are denoted by brown, blue and green, respectively; $[\mu]=[T]=[z]^{-1} =$ GeV.
}
 \label{Fig:Cz1-z15-2}
\end{figure}

Coupling constant $\alpha(z)=\alpha_{z_h}(z;\mu,T)$ for light quarks model is depicted in Fig.\,\ref{Fig:Cz1-z15-2} for two scales of energy $z=1$ GeV${}^{-1}$ (light surfaces) and $z=1.5$ GeV${}^{-1}$ (dark surfaces). Hadronic, QGP and quarkyonic phases are denoted by brown, blue and green, respectively. At both scales, coupling constant in hadronic phase is less than in QGP and quarkyonic phases.
For very small values of $\mu$, i.e. crossover region $\alpha$ increases continuously, while for $\mu>0.048$ there is an increment with a jump that shows 1st order phase transition. 
The jump of $\alpha$ between hadronic and quarkyonic phases is larger for lower energy (larger $z$), see panel (C).

 Coupling constant $\alpha_{z_h}(z;\mu,T)$  and  $ \log \, \alpha_{z_h}(z;\mu,T)$ as a function of $T$ for light quarks at fixed $\mu=0.024$ GeV (A-B), $\mu=0.3$ GeV (C-D), $\mu=0.8$ GeV (E-F) at different energy scales  $z=1$ (thin lines) and $z=0.3$ GeV${}^{-1}$ (thick lines) are shown in Fig.\,\ref{Fig:LQ-alpha-2D-zh}.  Hadronic, QGP and quarkyonic phases are denoted by brown, blue and green lines, respectively. Magenta arrows in (D) at $\mu=0.3$ GeV show the jumps at the 1st order phase transition between hadronic and quarkyonic phases, while there is a phase transition without any jump between hadronic and QGP at $\mu=0.024$ GeV (A,B), and between quarkyonic and QGP at $\mu=0.8$ GeV (E,F). For different fixed $\mu$ the coupling increases when we increase the temperature and physical system experiences different phase transitions from hadronic to quarkyonic and then from quarkyonic to QGP phases.

\newpage
$$\,$$
\newpage
$$\,$$
\newpage

\subsubsection{Running coupling for light quarks model with boundary 
condition $z_0=\fz_{LQ}(z_h)$} \label{gbc10}

In this subsection we consider the second boundary condition 
\eqref{phi-z0-gen}, \eqref{phi-fz-LQ}. This boundary condition for light quarks leads to the physical results in agreement with lattice calculations. As has been mentioned above, this 
boundary condition produces the dependence of the coupling constant 
on $(T,\mu)$. Indeed,
\bea
\alpha_{\fz_{_{LQ}}}(z)=\alpha_{\fz_{_{LQ}}}(z;\mu,T)&=&e^{\varphi_{\fz_{_{LQ}}}(z;T,\mu))}=\alpha_0(z)\fG(\fz_{_{LQ}}(T,\mu)),
\\ \fG(\fz_{_{LQ}}(T,\mu))&=&
e^{-\varphi_0 (\fz_{_{LQ}}(z_h))}=
e^{-\varphi_0 (\fz_{_{LQ}}(z_h(T,\mu)))}
\eea
  Here  $\fz_{\,_{LQ}}(z_h)$ is given by \eqref{phi-fz-LQ} and  $z_{h}=z_{h}(T,\mu)$, as in previous consideration,   is the value of the horizon for the stable black hole corresponding to given chemical potential $\mu$ and temperature $T$.

\begin{figure}[h!]
\centering
\includegraphics[scale=0.43]{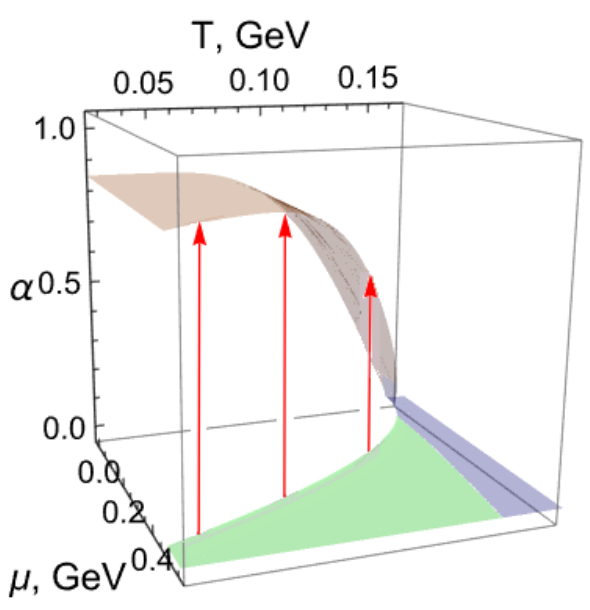}\\
 A\\
 \includegraphics[scale=0.29]{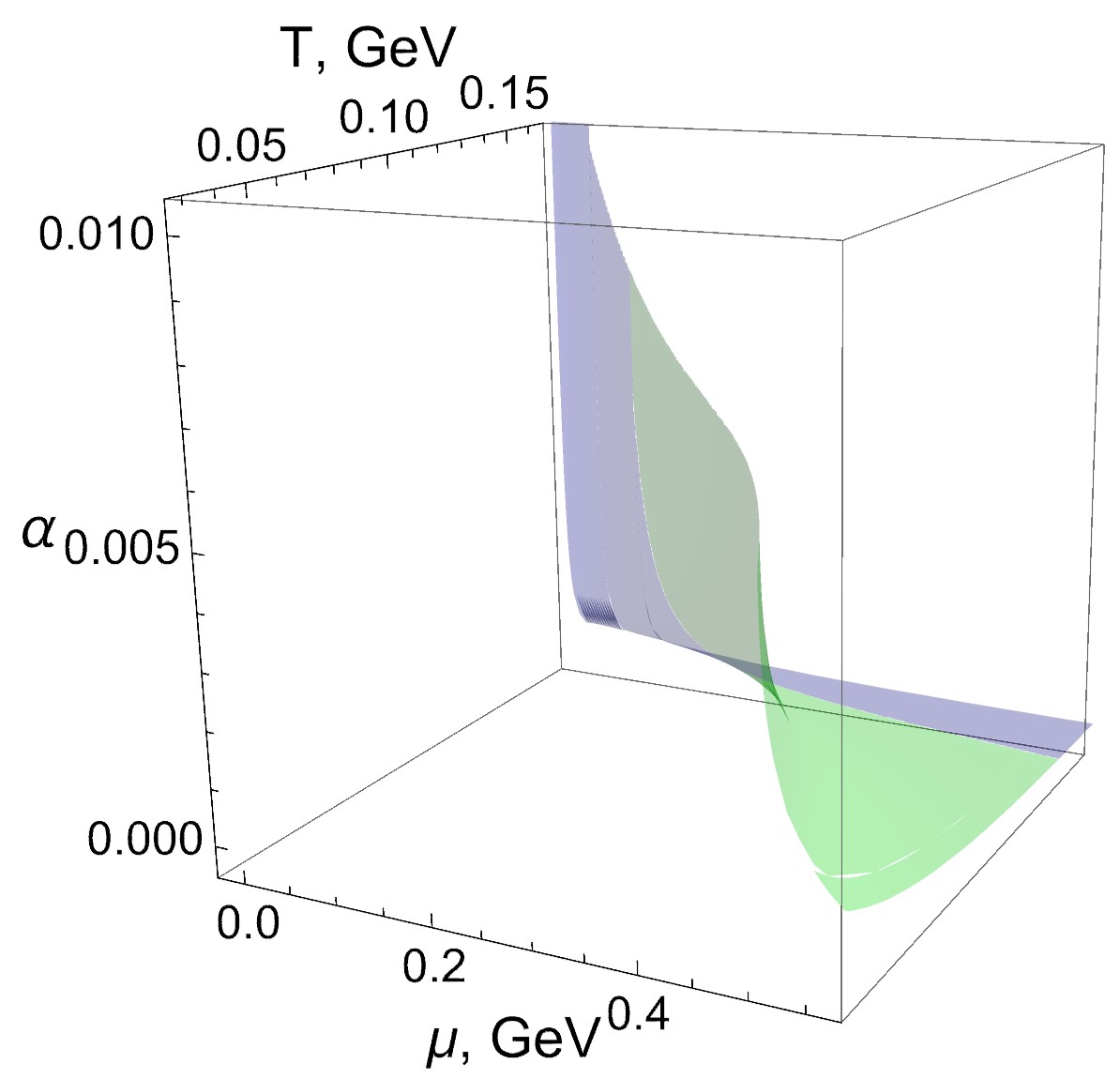} \qquad\qquad
\includegraphics[scale=0.39]{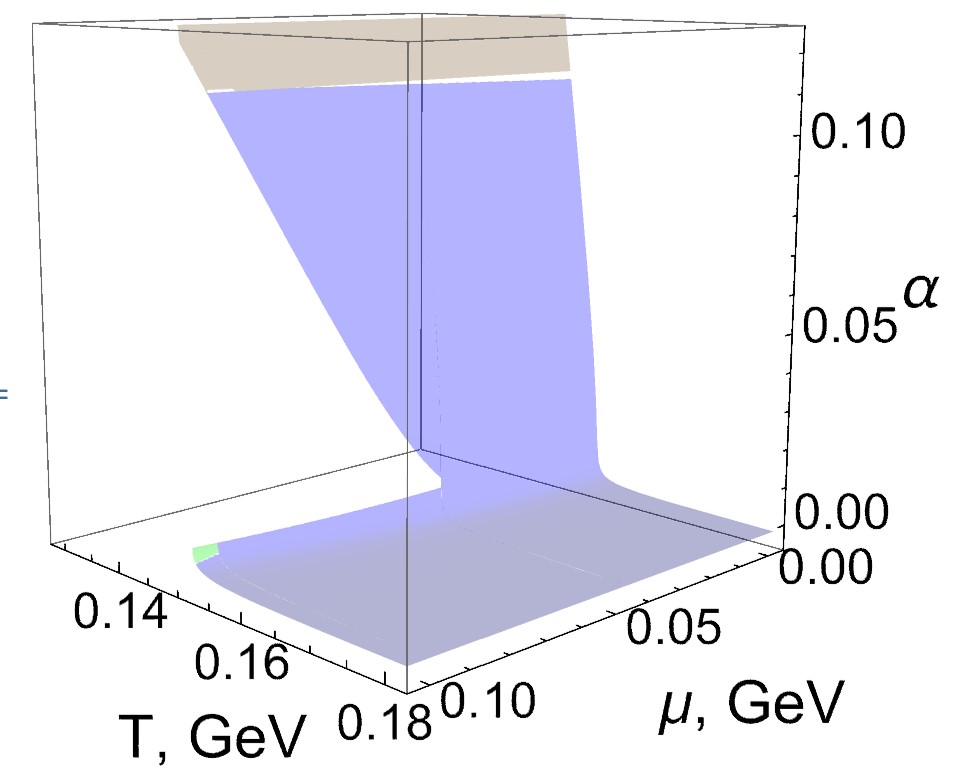}\\
B\hspace{150pt}C
\caption{Coupling constant $\alpha=\alpha_{\fz_{_{LQ}}}(z;\mu,T)$ for light quark model with the second boundary condition \eqref{phi-fz-LQ} at the energy scale $z=0$ (A) and its zooms (B,C). Red arrows show the jump for running coupling constant between hadronic and quarkyonic phases; $[z]^{-1} =$ GeV.
}
 \label{Fig:alphamuT-0}
\end{figure}

The coupling constant  $\alpha=\alpha_{\fz_{_{LQ}}}(z;\mu,T)$ for light quark model with the second boundary condition \eqref{phi-fz-LQ} at the energy scale $z=0$ (A) and its zooms (B,C) is depicted in Fig.\,\ref{Fig:alphamuT-0}. At the 1st order phase transition line the coupling constant feels a jump and magnitude of the jump increases by increasing $\mu$. The coupling in hadronic phase decreases faster than QGP phase by increasing $T$. Although, by changing $\mu$ in hadronic and QGP phase, the coupling constant does not change significantly and decreases slowly.  

\begin{figure}[t!]
 \centering
\includegraphics[scale=0.21]{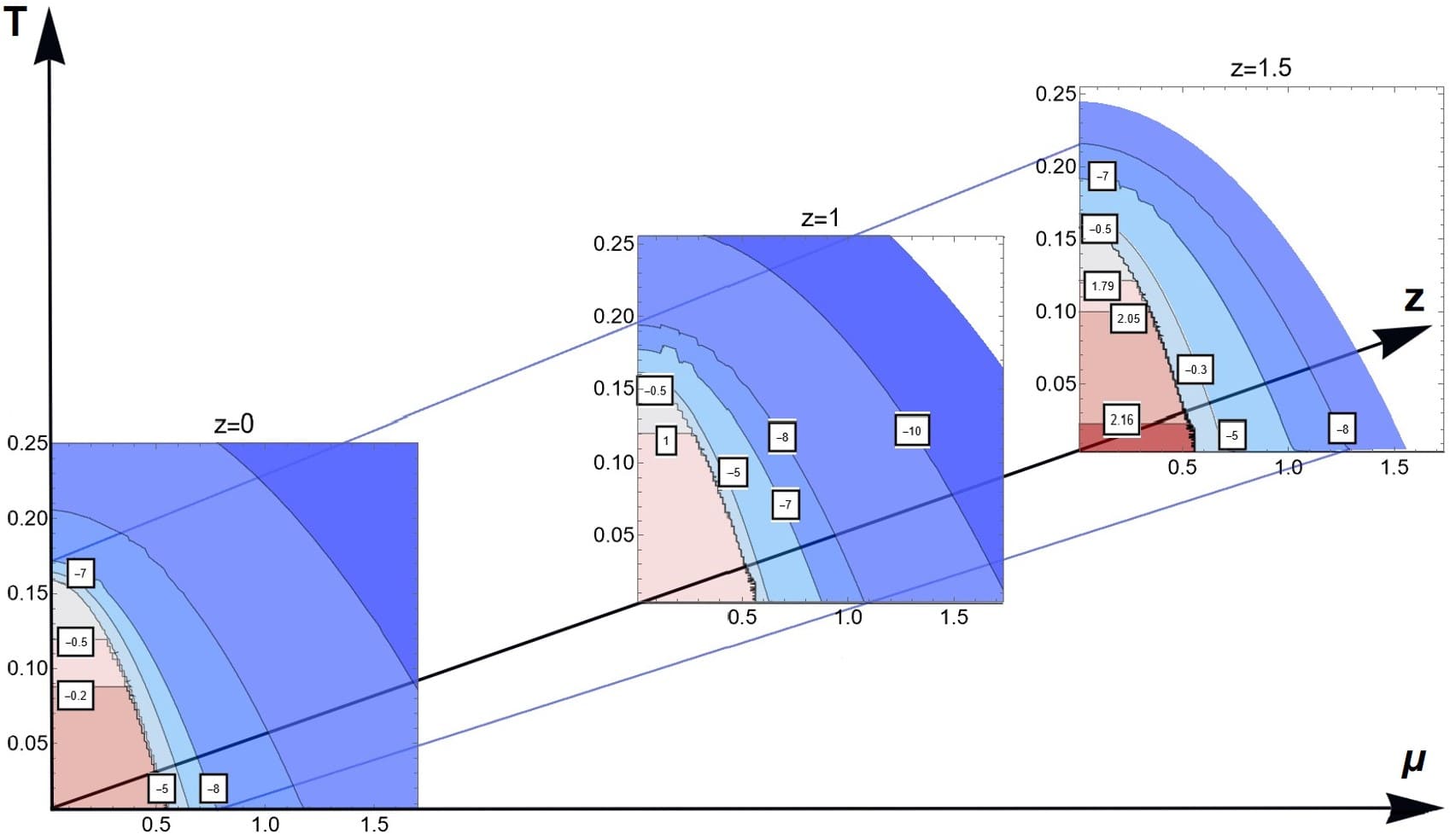}  
\caption{Density plots with contours for  $\log\alpha_{\fz_{_{LQ}}}(z;\mu,T)$ for light quarks at different energy scales $z=0, 1, 1.5$; $[\mu]=[T]=[z]^{-1} =$ GeV.
}
    \label{LQ-alpha-z0115-comp}
\end{figure}

In Fig.\,\ref{LQ-alpha-z0115-comp} the density plots with contours for  $\log\alpha_{\fz_{_{LQ}}}(z;\mu,T)$ for light quarks at different energy scales $z=0, 1$ and $1.5$ (GeV${}^{-1}$). At each (fixed) scale of energy $z$ running coupling decreases by increasing $T$ and by increasing $\mu$ in hadronic phase there is not significant change in coupling and while in QGP phase coupling decreases slowly. Fig.\,\ref{LQ-alpha-z0115-comp} shows that at fixed coupling, increasing the energy scale $z$ corresponds to larger values in contour, i.e. larger in $T$ and $\mu$ in the phase diagram.

\begin{figure}[h!]
    \centering
 \includegraphics[scale=0.44]{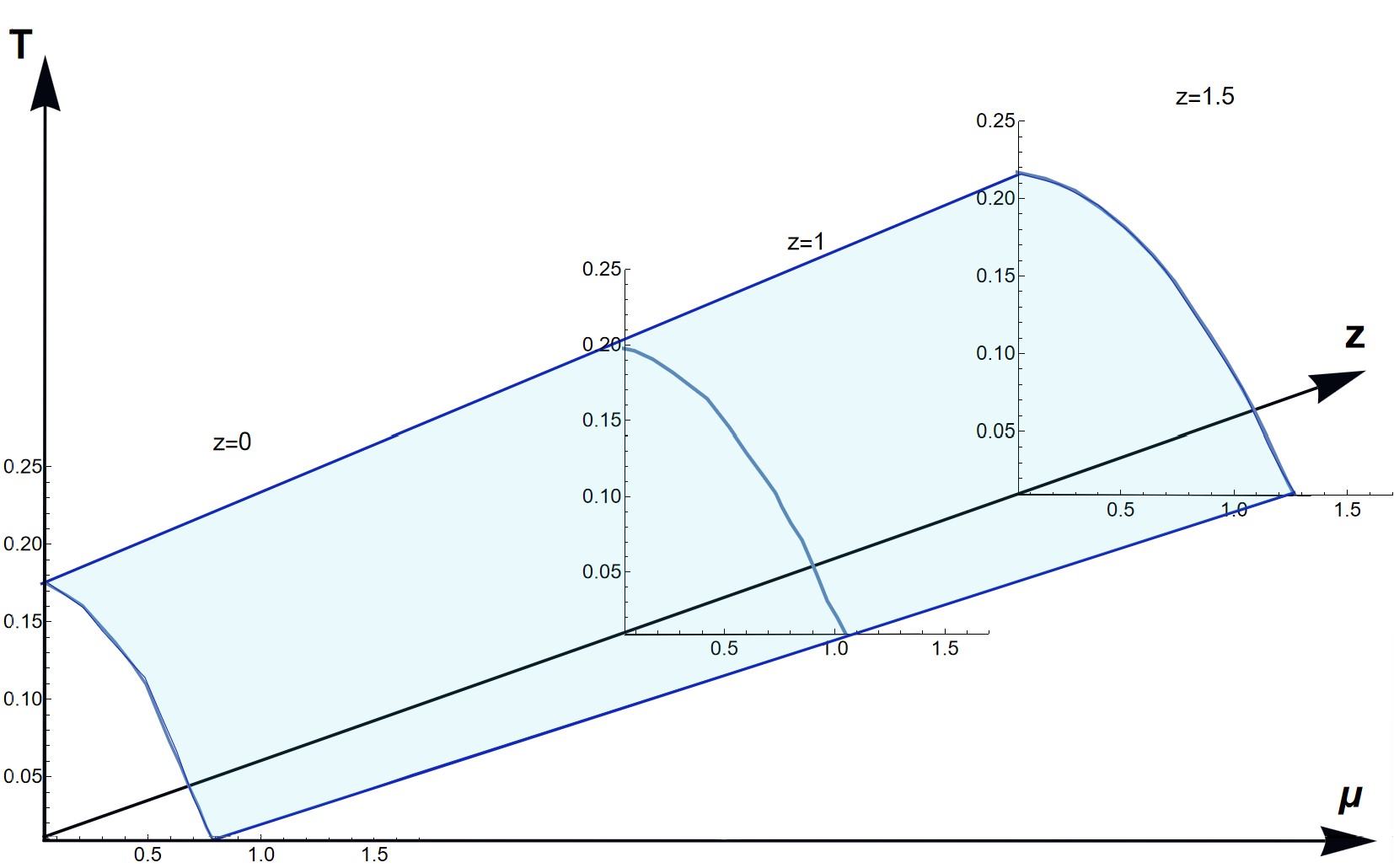}    \caption{The 2-dim surface with fixed value of the $\log\alpha_{\fz_{_{LQ}}}(z;\mu,T)$ for light quarks at different energy scales $z=0, 1, 1.5$; $[\mu]=[T]=[z]^{-1} =$ GeV.
}
    \label{LQ-alpha-z1-z15-2d}
\end{figure}

To have a complementary description of physics in Fig.\,\ref{LQ-alpha-z0115-comp}, the 2-dimensional surface with fixed value of the running coupling is depicted in Fig.\,\ref{LQ-alpha-z1-z15-2d}. This figure shows that at fixed value of coupling, increasing the energy scale increases the values of temperature and chemical potential.

\begin{figure}[t!]
    \centering
   \includegraphics[scale=0.2]{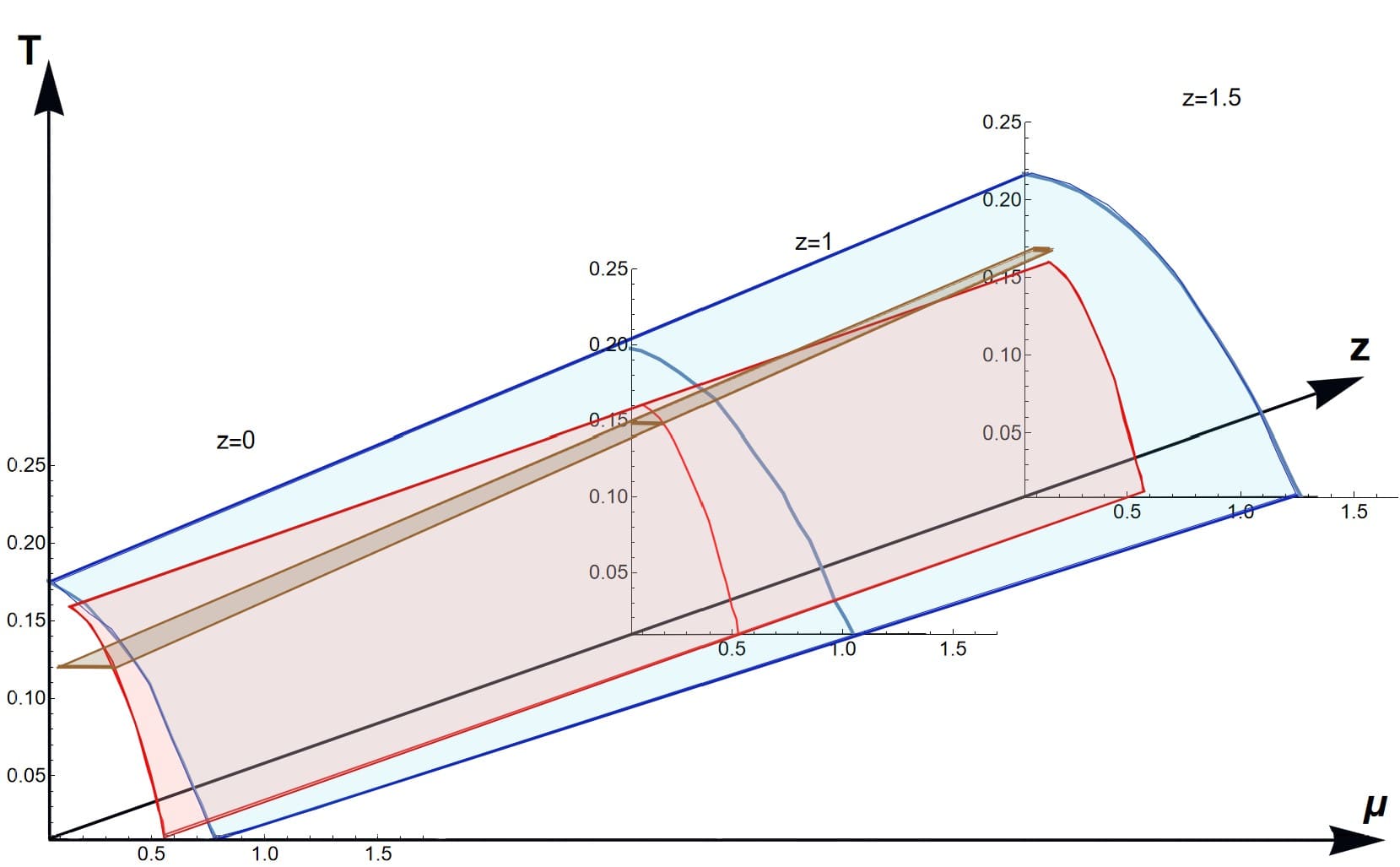} 
   \caption{2-dim surfaces in 3-dim space $(\mu,T,z)$ with fixed values of the $\log\alpha_{\fz_{_{LQ}}}(z;\mu,T)$ for light quarks at different energy scales $z=0, 1, 1.5$. Cyan and brown correspond with QGP and hadronic phases, respectively. The red surface shows the location of the 1st order phase transition line; $[\mu]=[T]=[z]^{-1} =$ GeV. 
   } 
    \label{LQ-alpha-z1-z15-2d3d}
\end{figure}

In Fig.\,\ref{LQ-alpha-z1-z15-2d3d}, 2-dim surfaces in 3-dim space $(\mu,T,z)$ with fixed values of the logarithm of running coupling  $\log\alpha_{\fz_{_{LQ}}}(z;\mu,T)$ for light quarks at different energy scales $z=0, 1, 1.5$ (GeV${}^{-1}$) are shown. Cyan and brown correspond with QGP and hadronic phases, respectively. The red surface shows the location of the 1st order phase transition line that does not depend on the energy scale parameter $z$ and just depends on the thermodynamics of the model. Also, the brown surface in the hadronic region shows that the coupling constant depends on the temperature and not chemical potential.

\begin{figure}[h!]
    \centering
 \includegraphics[scale=0.24]{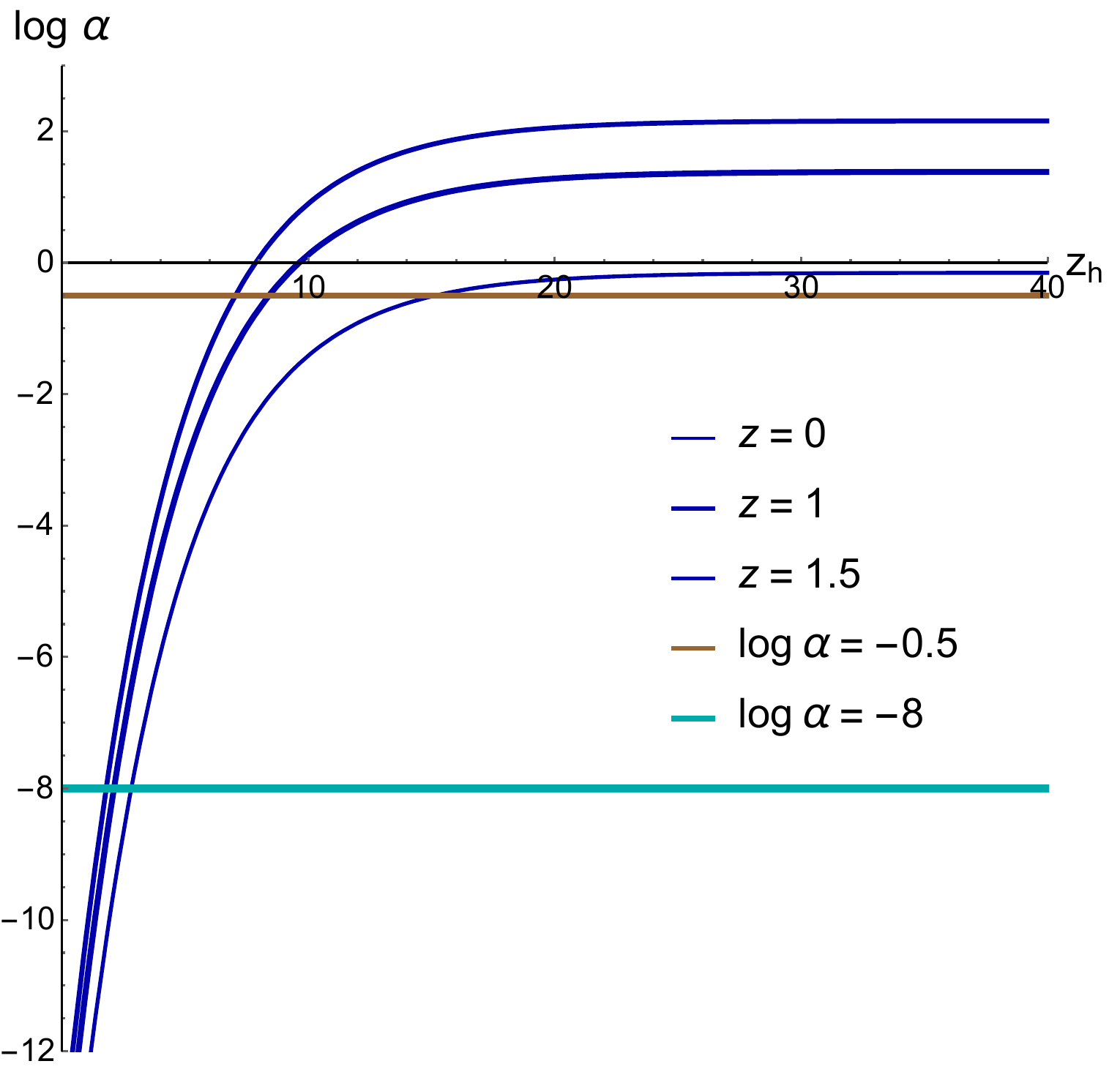} \qquad 
 \includegraphics[scale=0.31]{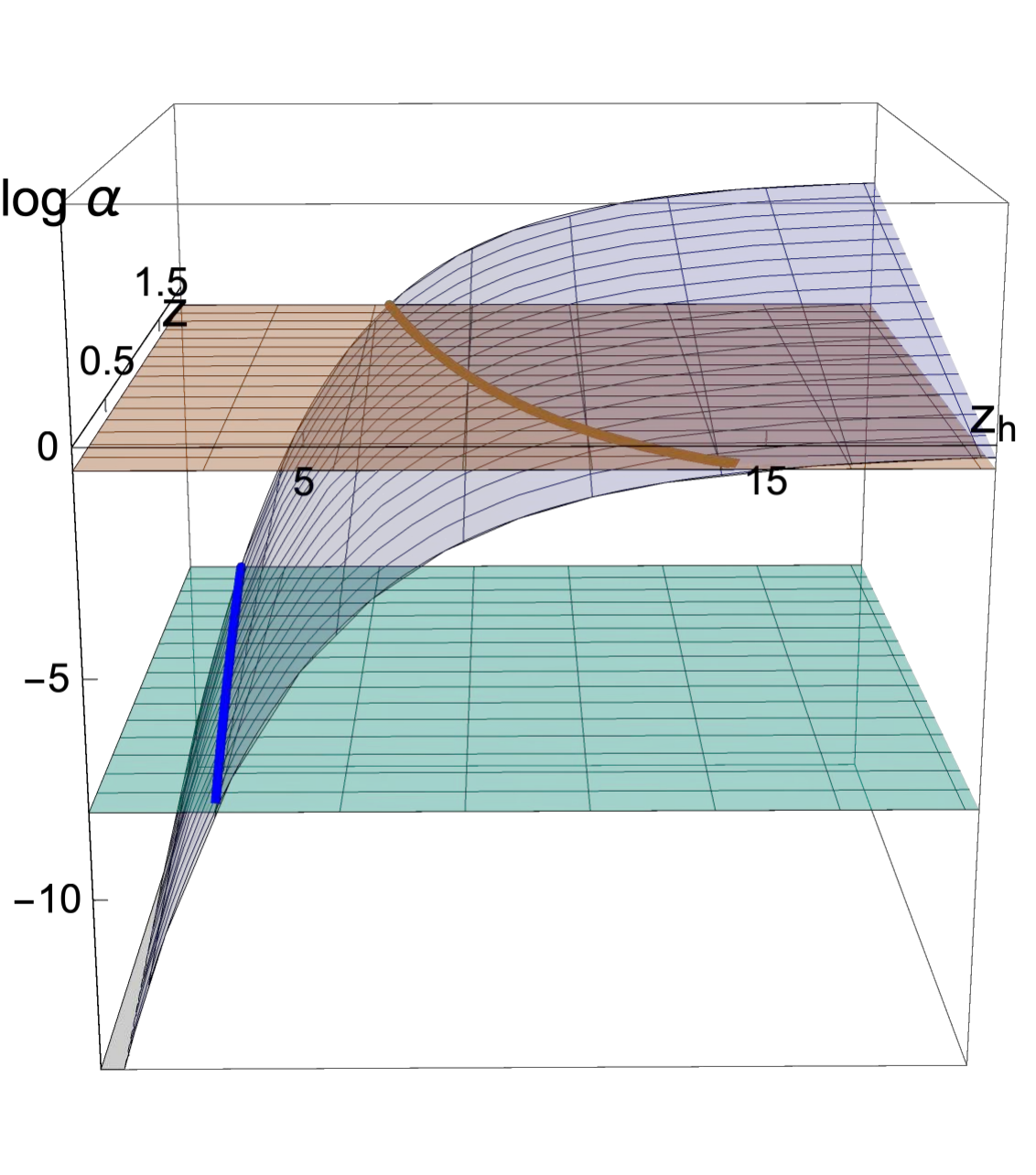}\\
 A\hspace{200pt}B
   \caption{2-dim presentation of the logarithm of coupling constant $\log\alpha_{\fz_{_{LQ}}}(z)$ (A) and its 3-dim presentation (B) as a function of $z_h$ for different values of the holographic coordinate (energy scale) $z$. Cyan and brown colors correspond with QGP and hadronic phases, respectively; $[z_h]^{-1} =$ GeV. }
    \label{LQ-alpha-z1-z15-zh}
\end{figure}

The 2-dim presentation of the logarithm of coupling constant $\log\alpha_{\fz_{_{LQ}}}(z)$ as a function of $z_h$ for different values of the holographic coordinate (energy scale) $z$ is depicted in Fig.\,\ref{LQ-alpha-z1-z15-zh}A 
 and its 3-dim presentation in Fig.\,\ref{LQ-alpha-z1-z15-zh}B. Cyan and brown colors correspond with QGP and hadronic phases, respectively. This figure is complementary description for Fig.\,\ref{LQ-alpha-z1-z15-2d3d}.

\begin{figure}[h!]
  \centering
\includegraphics[scale=0.30]{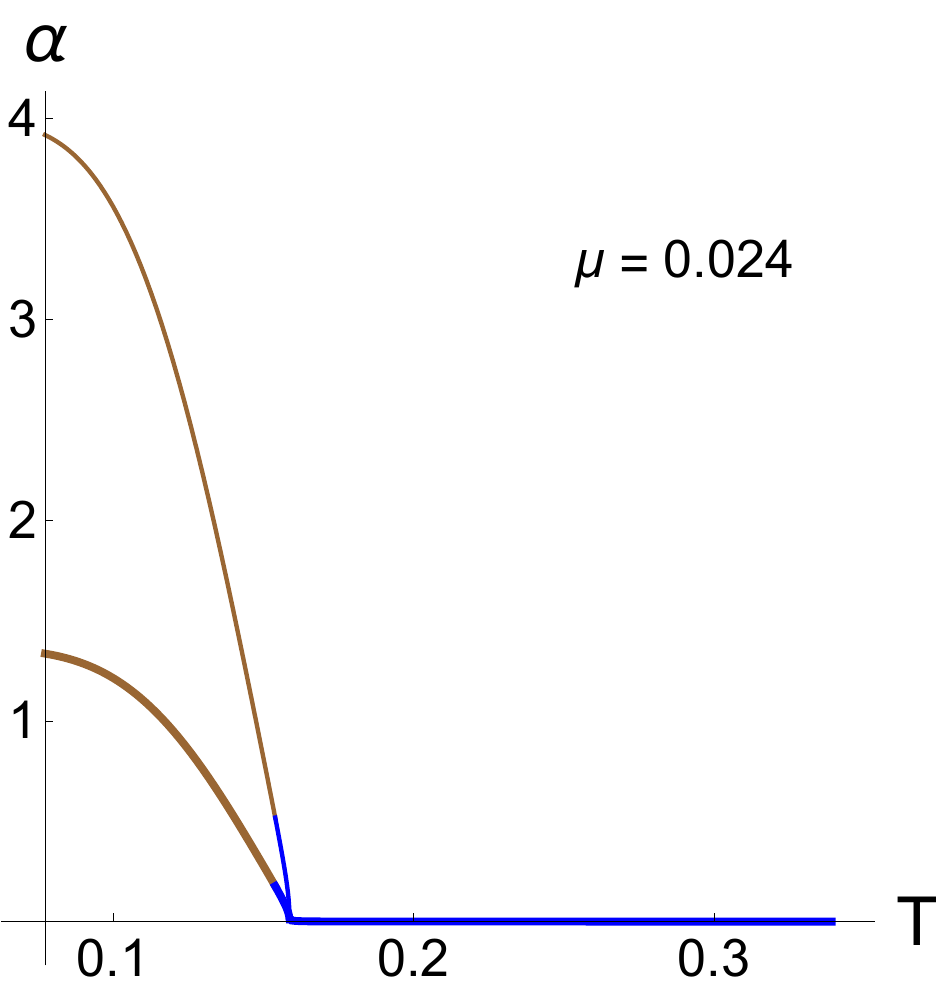} \qquad \qquad \qquad
\includegraphics[scale=0.30]{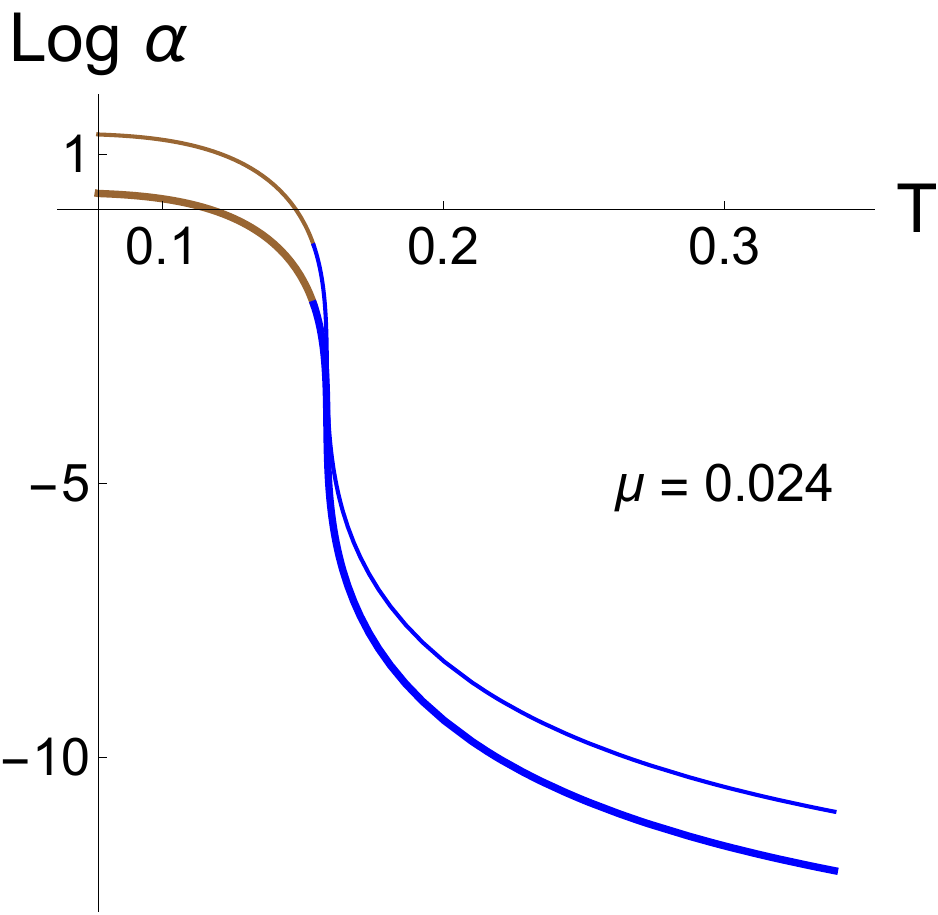}\\
 A\hspace{200pt}B \\
\includegraphics[scale=0.30]{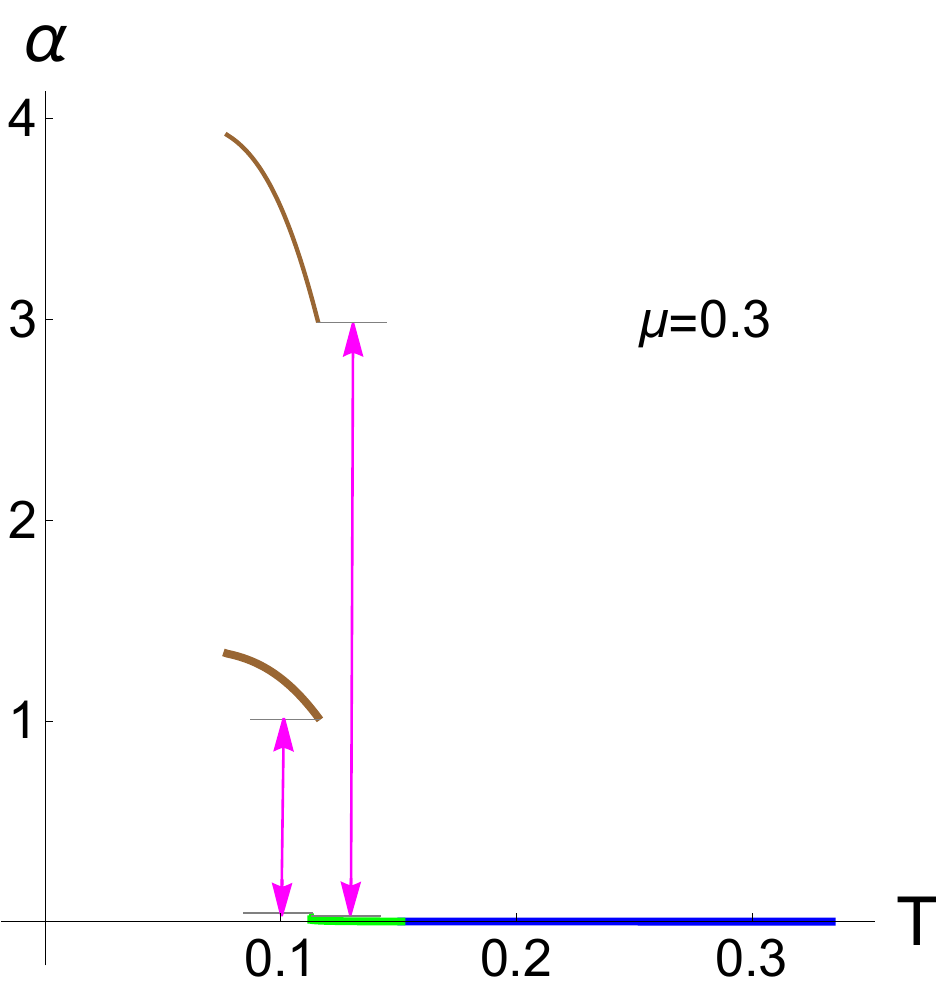} \qquad \qquad \qquad
\includegraphics[scale=0.30]{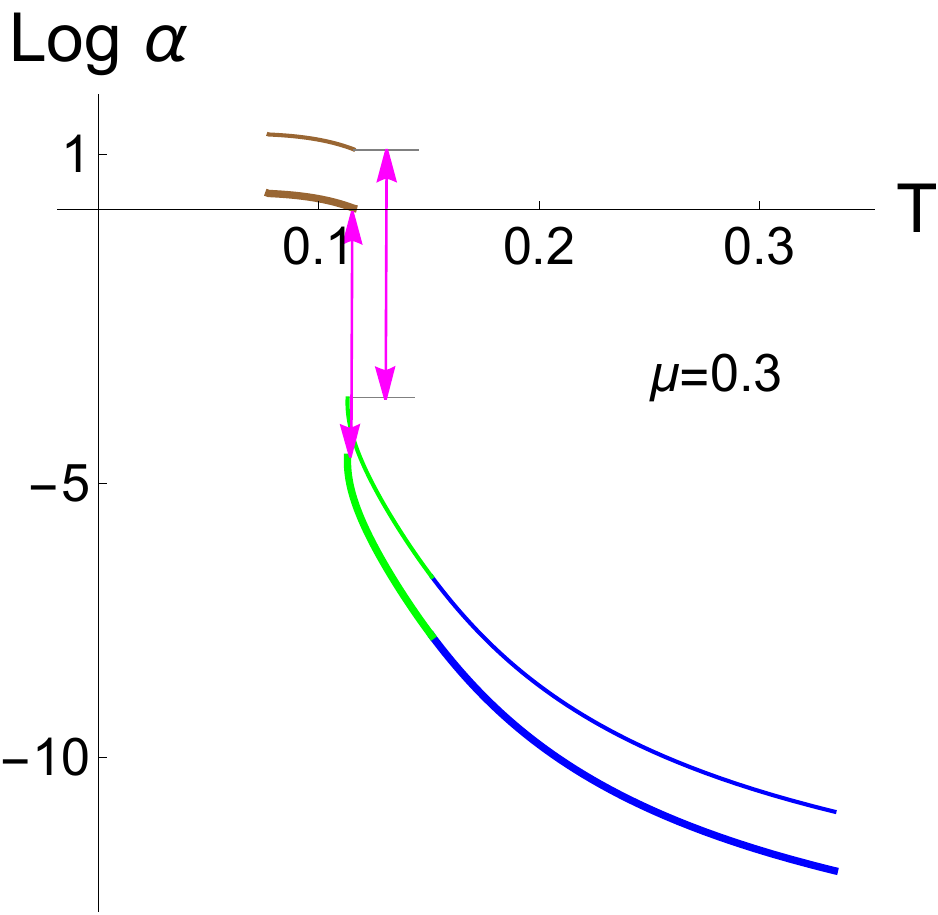}\\
 C\hspace{200pt}D \\
\includegraphics[scale=0.30]{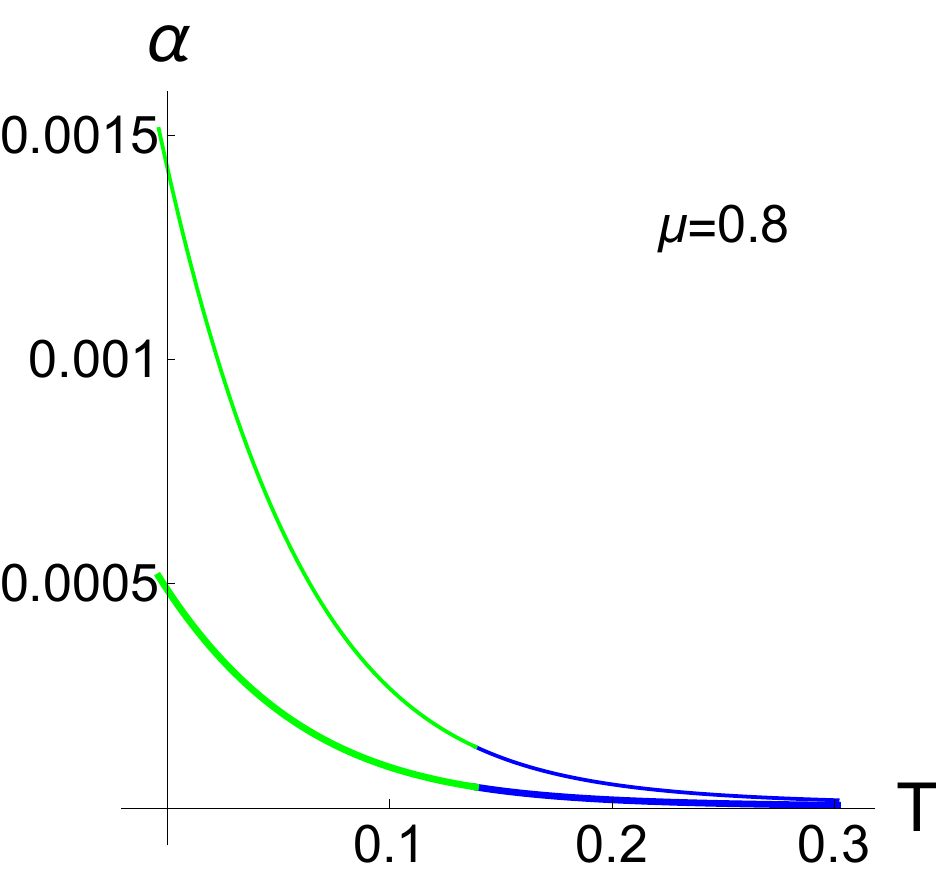} \qquad \qquad  \qquad
\includegraphics[scale=0.30]{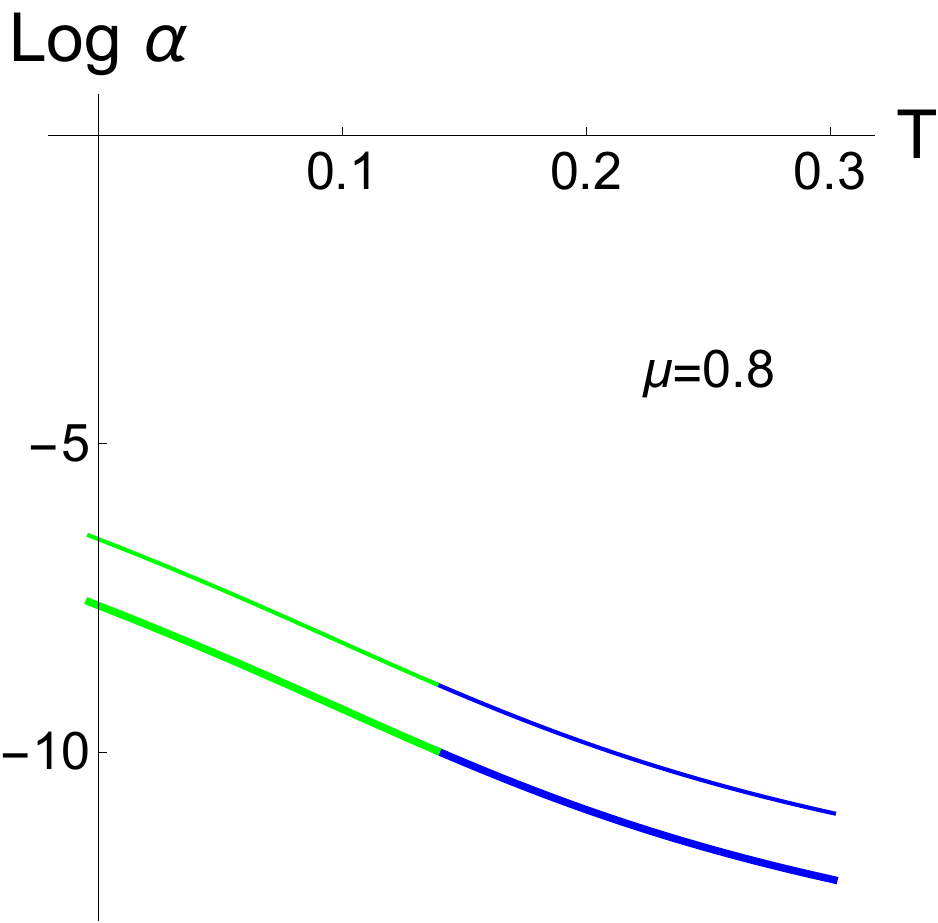}\\
 E\hspace{200pt}F
\caption{Coupling constant $\alpha=\alpha_{\fz_{_{LQ}}}(z;\mu,T)$  and  $ \log \, \alpha=\log \, \alpha_{\fz_{_{LQ}}}(z;\mu,T)$ for light quarks at fixed $\mu$ at different energy scales  $z=1$ (thin lines) and $z=0.3$ (thick lines).  Hadronic, QGP and quarkyonic phases are denoted by brown, blue and green lines, respectively. Magenta arrows in (C), (D) show the jumps at the 1st order phase transition. $[\mu]=[T]=[z]^{-1} =$ GeV. 
}
\label{Fig:LQ-alpha-2Dnbc}
\end{figure}

 Coupling constant $\alpha=\alpha_{\fz_{_{LQ}}}(z;\mu,T)$  and  $ \log \, \alpha=\log \, \alpha_{\fz_{_{LQ}}}(z;\mu,T)$  as a function of $T$ for light quarks at fixed $\mu=0.024$ GeV (A-B), $\mu=0.3$ GeV (C-D), $\mu=0.8$ GeV (E-F) at different energy scales  $z=1$ GeV${}^{-1}$ (thin lines) and $z=0.3$ GeV${}^{-1}$ (thick lines) are shown in Fig.\,\ref{Fig:LQ-alpha-2Dnbc}.  Hadronic, QGP and quarkyonic phases are denoted by brown, blue and green lines, respectively. Magenta arrows in (D) at $\mu=0.3$ GeV show the jumps at the 1st order phase transition between hadronic and quarkyonic phases, while there is a phase transition without any jump between hadronic and QGP at $\mu=0.024$ GeV (A,B), and between quarkyonic and QGP at $\mu=0.8$ GeV (E,F). For different fixed $\mu$ the coupling decreases when we increase the temperature and physical system experiences different phase transitions from hadronic to quarkyonic and then from quarkyonic to QGP phases. The magnitude of the jump increases for larger energy scale $z$ (or equivalently, for lower  energy scale $E$).

\begin{figure}[h!]
  \centering
\includegraphics[scale=0.38]{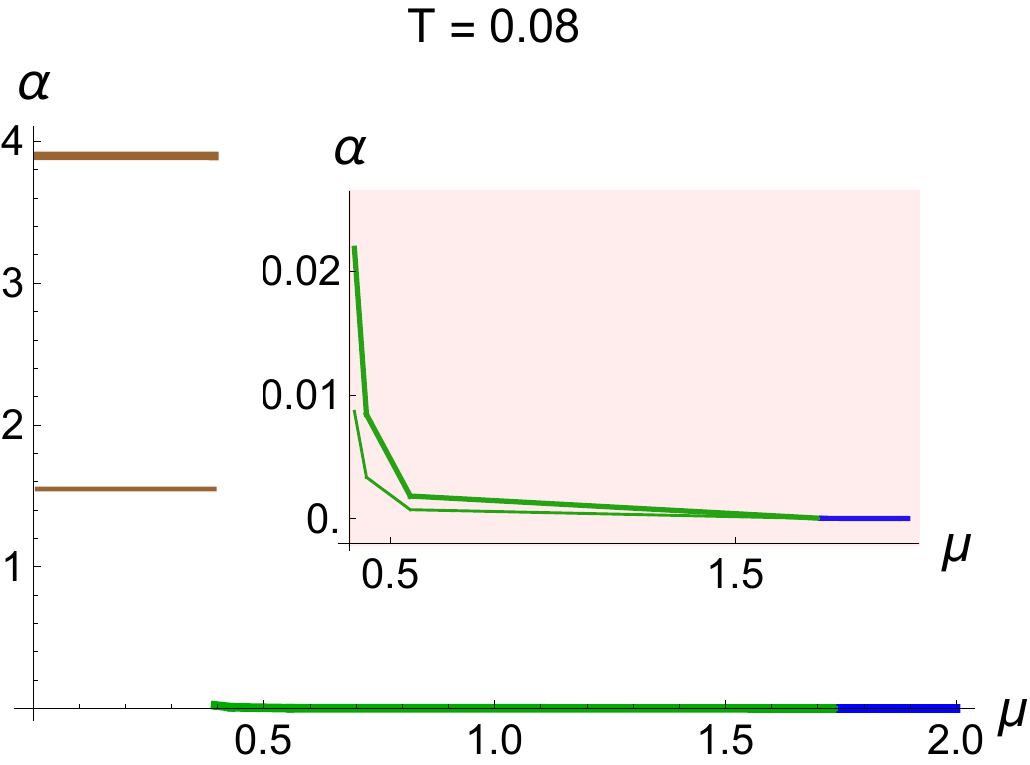}
\qquad\includegraphics[scale=0.39]{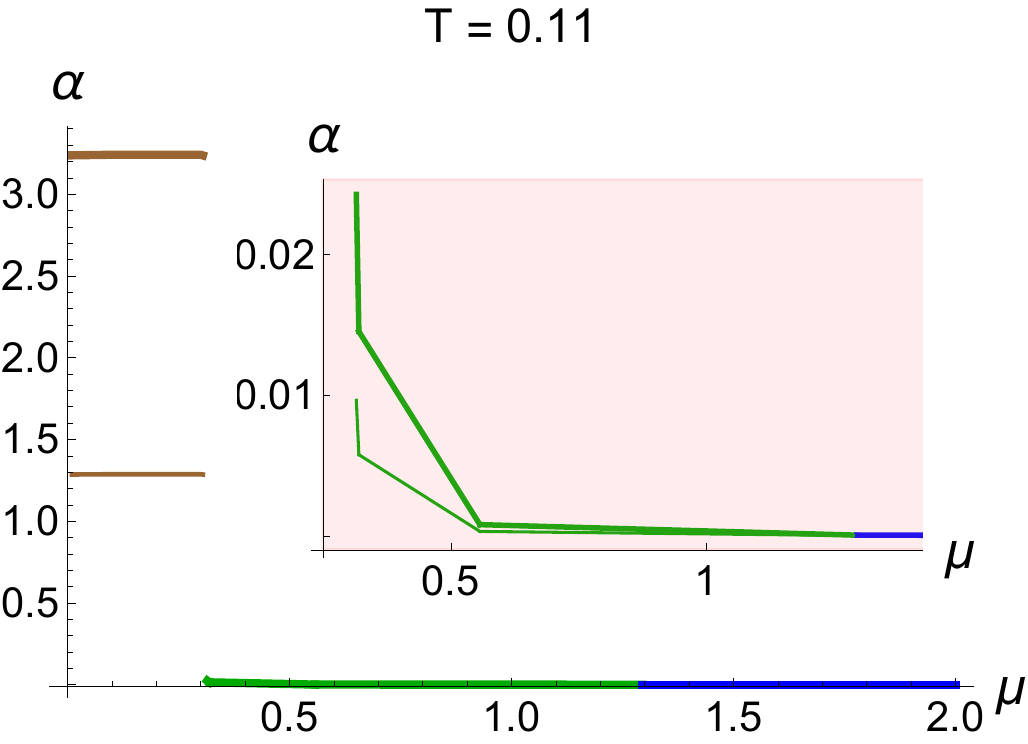}
\\A\hspace{200pt}B\\$\,$\\
\includegraphics[scale=0.42]{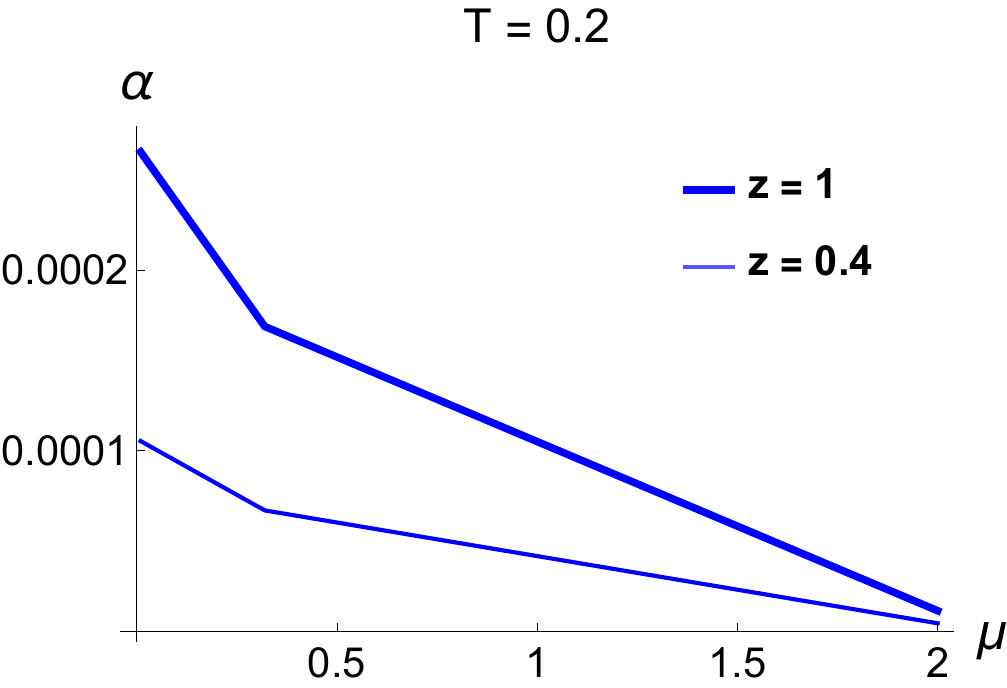}\\
  C
\caption{Coupling constant $\alpha=\alpha_{\fz_{_{LQ}}}(z;\mu,T)$ as a function of $\mu$ for light quarks at fixed $T=0.08$ (A), $T=0.11$ (B) and $T=0.2$ (C) for different scales of energy $z=0.4$ (thin lines) and $z=1$ (thick lines). Hadronic and QGP phases are denoted by brown and blue lines, respectively. $[\mu]=[T]=[z]^{-1} =$ GeV. 
}
\label{Fig:LQ-alpha-2D-17}
\end{figure}

The dependence of coupling constant $\alpha=\alpha_{\fz_{_{LQ}}}(z;\mu,T)$ on $\mu$ for light quarks at fixed $T=0.08$ GeV (A), $T=0.11$ GeV (B) and $T=0.2$ GeV (C) for different scales of energy $z=0.4$ GeV${}^{-1}$ (thin lines) and $z=1$ GeV${}^{-1}$ (thick lines) is depicted in Fig.\,\ref{Fig:LQ-alpha-2D-17}. Hadronic, QGP and quarkyonic phases are denoted by brown, blue and green lines, respectively. In all regions, coupling constant does not change significantly and decreases very slowly at fixed $T$ and energy scale $z$.

\begin{figure}[h!]
\centering
\includegraphics[scale=0.19]{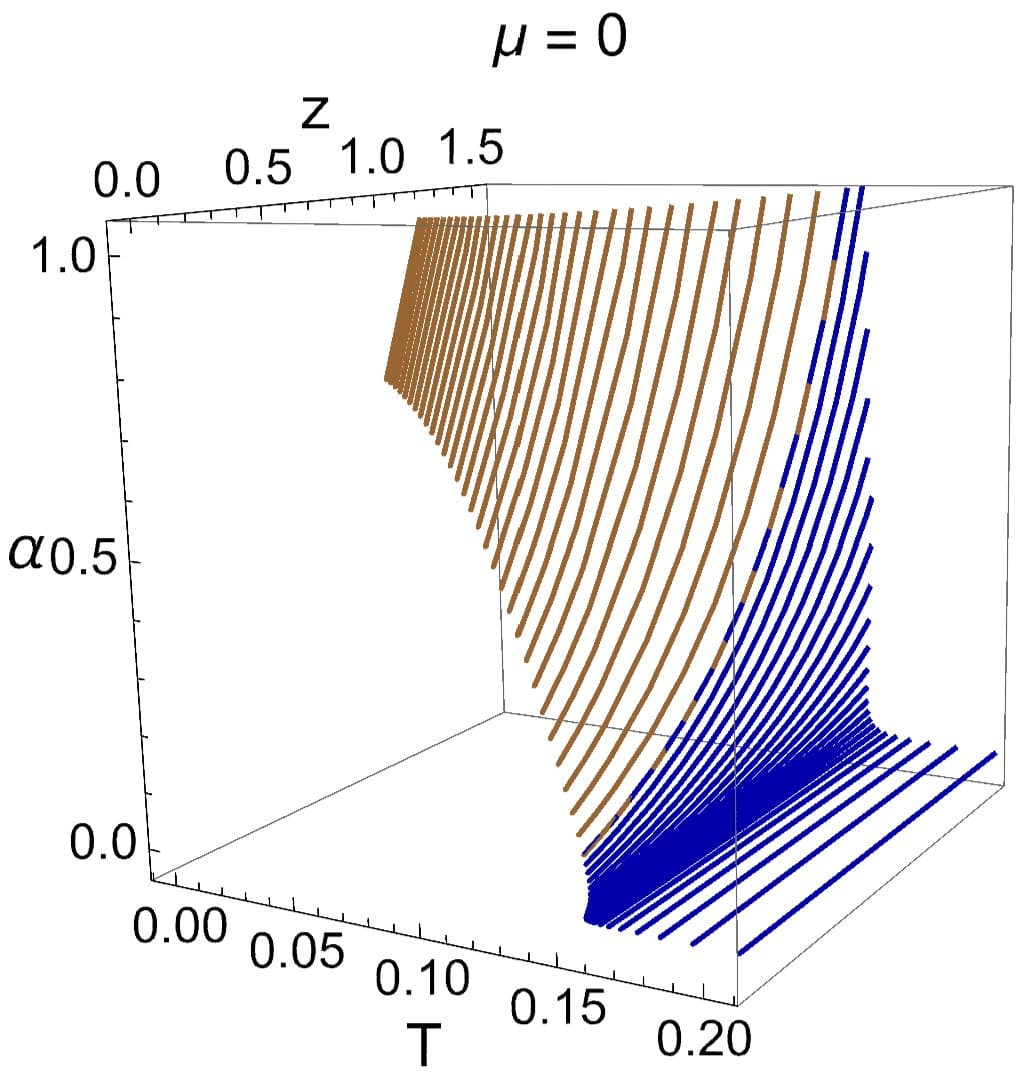}\\
A\\
\includegraphics[scale=0.16]{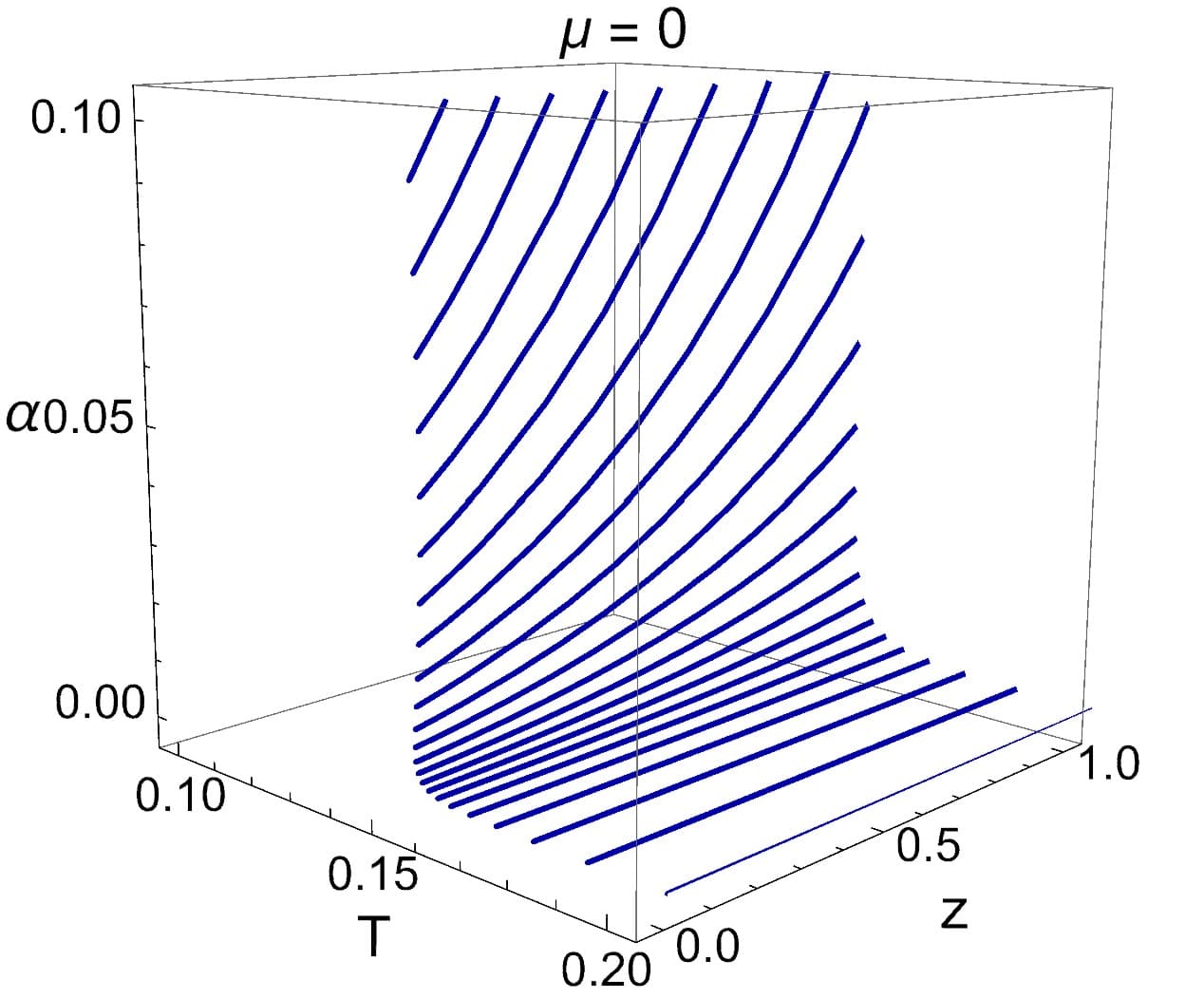} \quad
 \includegraphics[scale=0.45]{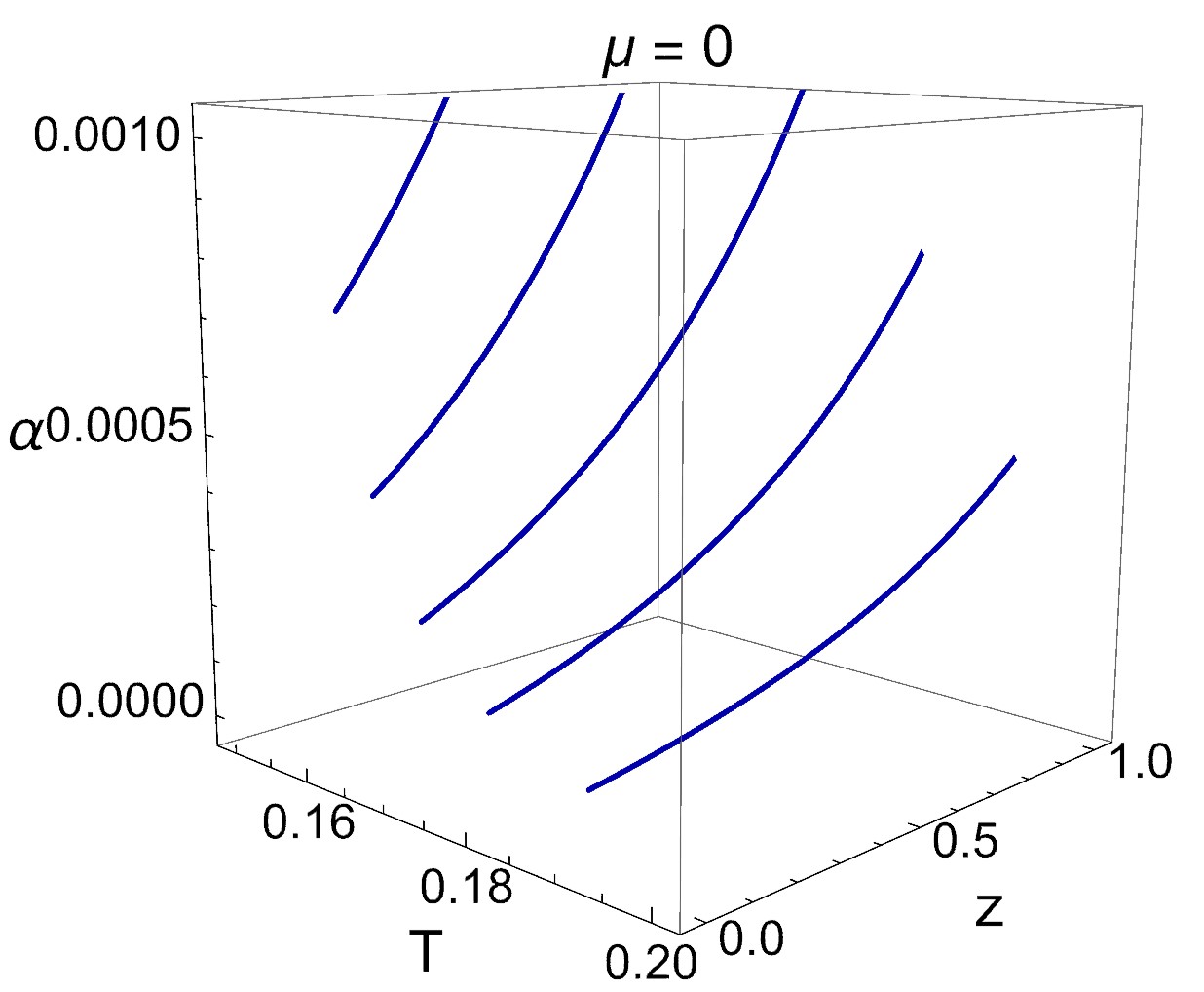} \\
 B\hspace{170pt}C
\caption{3D-plot of coupling constant $\alpha=\alpha_{\fz_{_{LQ}}}(z;T,\mu)$ for light quarks at fixed chemical potential $\mu=0$ (A) and its zooms (B,C).  Hadronic and QGP phases are denoted by brown and blue lines, respectively; $[\mu]=[T]=[z]^{-1} =$ GeV.\\
}
 \label{Fig:alphazT-mu0}
\end{figure}

In Fig.\,\ref{Fig:alphazT-mu0}, 3D-plot of coupling constant $\alpha=\alpha_{\fz_{_{LQ}}}(z;T,\mu)$ for light quarks at fixed chemical potential $\mu=0$ (A) and its zooms (B,C) is plotted.  Hadronic, QGP and quarkyonic phases are denoted by brown, blue and green lines, respectively. The coupling constant has no jump at $\mu=0$ between hadronic and QGP phases because of crossover region.

\begin{figure}[h!]
\centering
 \includegraphics[scale=0.15]{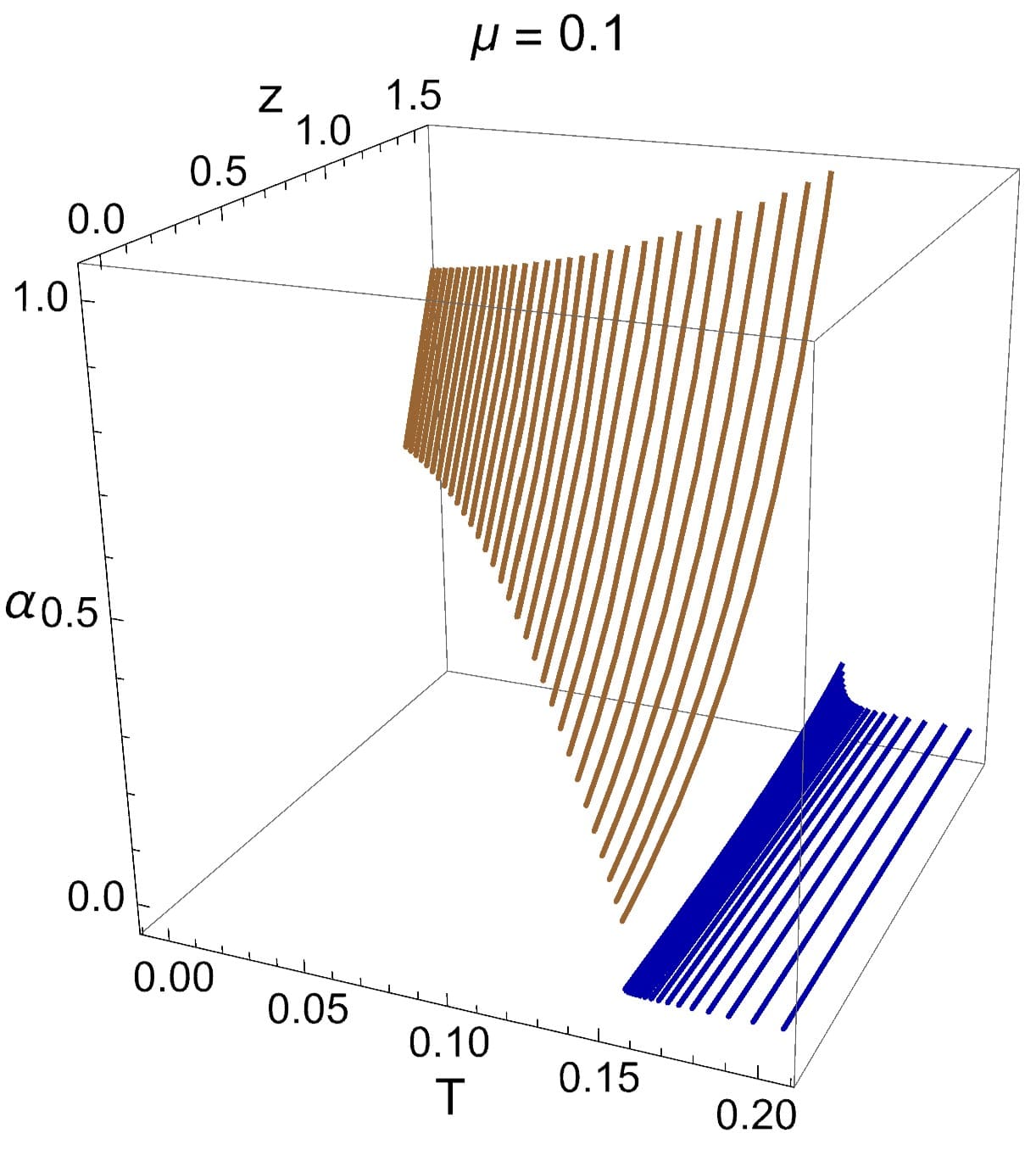}\quad\quad
 \includegraphics[scale=0.17]{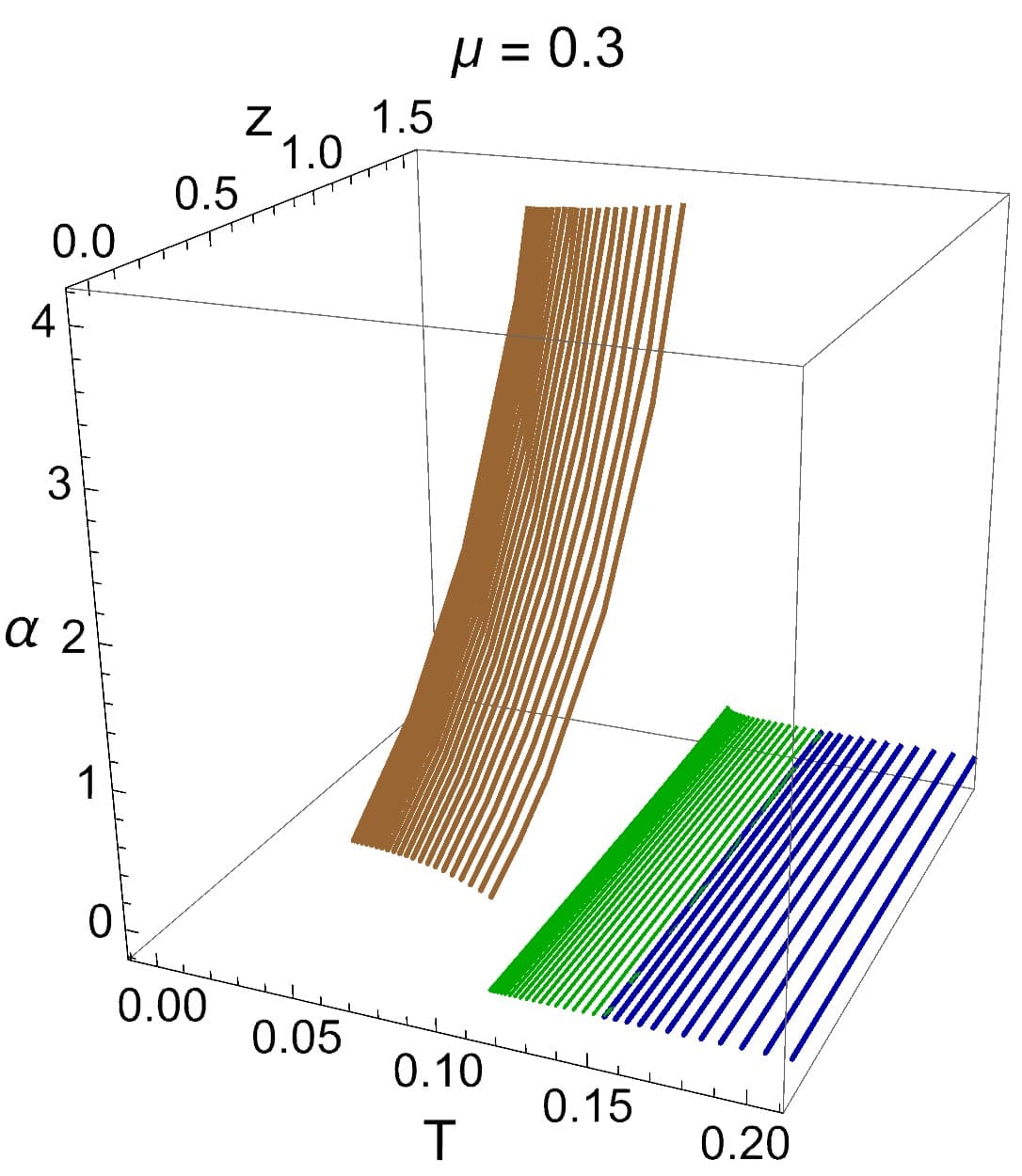}\\
 A\hspace{150pt}B\\
 \includegraphics[scale=0.17]{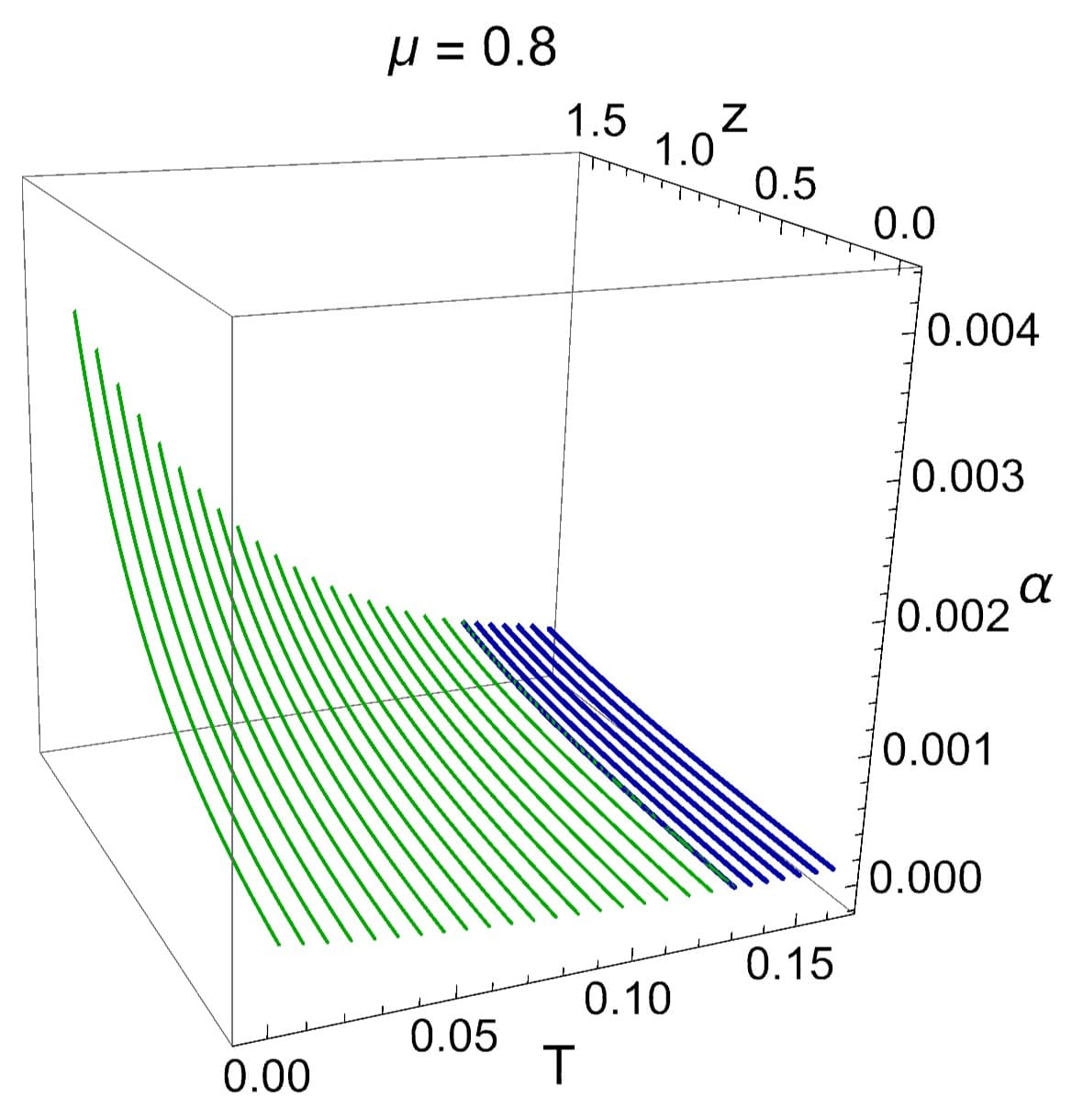}  \\C
\caption{3D-plots of coupling constant $\alpha=\alpha_{\fz_{_{LQ}}}(z;T,\mu)$ for light quarks at fixed $\mu=0.1$ (A), $\mu=0.3$ (B) and $\mu=0.8$ (C). Hadronic, QGP and quarkyonic phases are denoted by brown, blue and green lines, respectively; $[\mu]=[T]=[z]^{-1} =$ GeV.\\
}
 \label{Fig:alphazT-mu01-03-08}
\end{figure}

The 3D-plot of coupling constant $\alpha=\alpha_{\fz_{_{LQ}}}(z;T,\mu)$ for light quarks at fixed chemical potentials $\mu=0.1$ GeV (A), $\mu=0.3$ GeV (B) and $\mu=0.8$ GeV (C) is depicted in Fig.\,\ref{Fig:alphazT-mu01-03-08}. Hadronic, QGP and quarkyonic phases are denoted by brown, blue and green lines, respectively. Because of 1st order phase transition there are jumps between hadronic and QGP phases at $\mu=0.1$ GeV (A) and between hadronic and Quarkyonic phases at $\mu=0.3$ GeV (B). At larger $\mu=0.8$ GeV (C) there is phase transition without any jump in coupling constant.

\newpage
$$\,$$
\newpage

\subsubsection{Running coupling versus energy scale $E$ for light quarks model} \label{gbcELQ}

Considering the formula \eqref{BBL} for light quarks model one can cover UV limit of QFT for small values of $z$ and IR limit for larger values of $z$. Therefore, the bulk theory can cover all possible energy scales of the boundary field theory.
The results of running coupling for light quarks are obtained at fixed energy scale $E$ in different $T$ and $\mu$. For this reason, the running coupling  can vary in wide ranges by changing $T$ and $\mu$.
In other words, if we want to discuss UV or IR domain of the theory we need to express $\alpha$ as a function of the energy scale $E$ at fixed $T$ and $\mu$.

\begin{figure}[h!]
\centering
\includegraphics[scale=0.52]{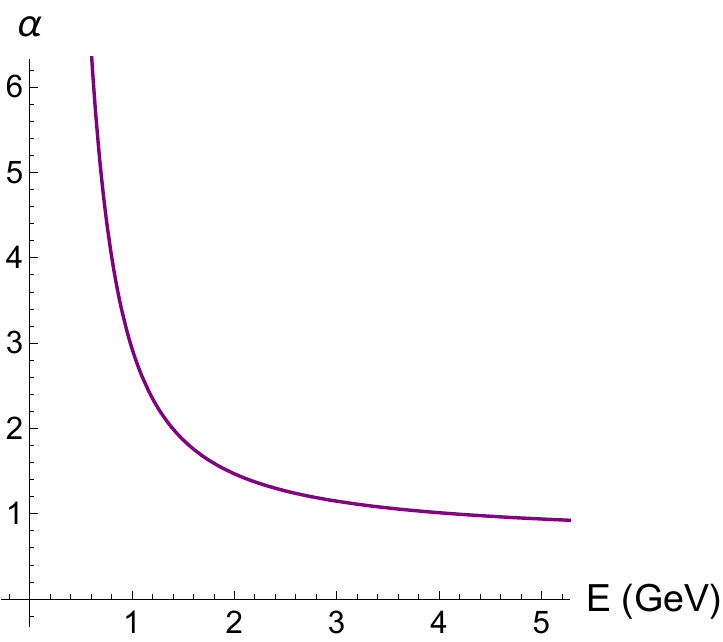}\quad\includegraphics[scale=0.53]{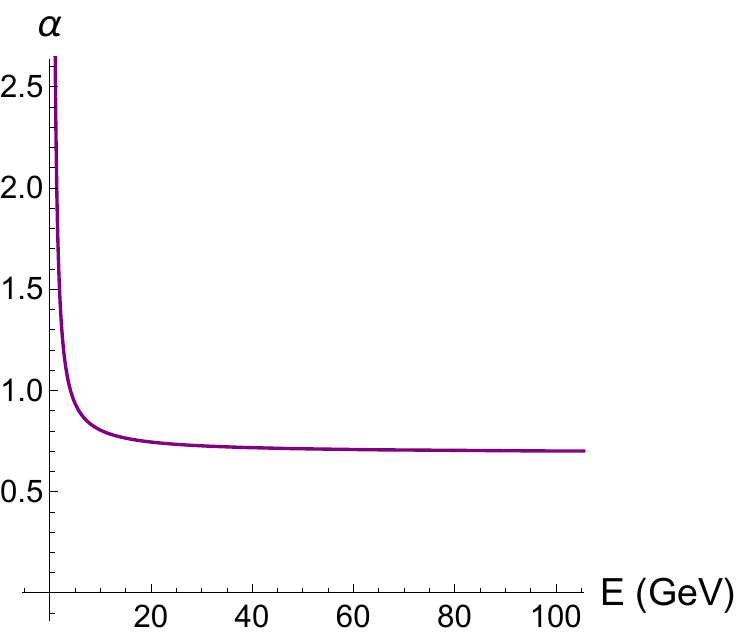}\\
 A\hspace{200pt}B
\caption{Running coupling $\alpha=\alpha_{_{LQ}}(E;\mu,T)$ for light quarks model as a function of the energy scale $E$ in QFT at fixed $T = 0.11$ and $\mu = 0.04$. Panel\,B presents panel\,A in different plot range; $[\mu]=[T]=$ GeV.}
\label{alphavsELQ}
\end{figure}

Running coupling $\alpha=\alpha_{_{LQ}}(E;\mu,T)$ as a function of the energy scale $E$ in QFT for light quarks at fixed $T = 0.11$ GeV and $\mu = 0.04$ GeV is depicted in Fig.\,\ref{alphavsELQ}. The Fig.\,\ref{alphavsELQ}B is the rescaling of the Fig.\,\ref{alphavsELQ}A.
The Fig.\,\ref{alphavsELQ}A shows the monotonic behavior of running coupling $\alpha(E)$ as a decreasing function that qualitatively is compatible with the perturbation results \cite{ParticleDataGroup:2022pth,Deur:2016tte}.
Interestingly, the Fig.\,\ref{alphavsELQ}B shows that although we can reach to the higher values of the energy scale $E$, but we cannot cover the very small values of the running coupling $\alpha$ in our holographic model. In other words, our model automatically covers the strongly coupled regime of the QFT (IR regime) and a small part of UV. In fact, our model can not describe ultra-UV regime for light quarks model.

\subsection{Running coupling for heavy quarks model} \label{HQ-running10}

In this section we investigate the running coupling constant by considering zero, first and second boundary conditions correspond to \eqref{bc0}, \eqref{bch}  and \eqref{bce}, respectively. let us note that just choosing special form of the function for second boundary condition, i.e. \eqref{bceHQ}
leads to the proper physical results for heavy quarks.

\subsubsection{Running coupling for heavy quarks model with boundary condition $z_0=0$}

\begin{figure}[h!]
  \centering
  \includegraphics[scale=0.5]{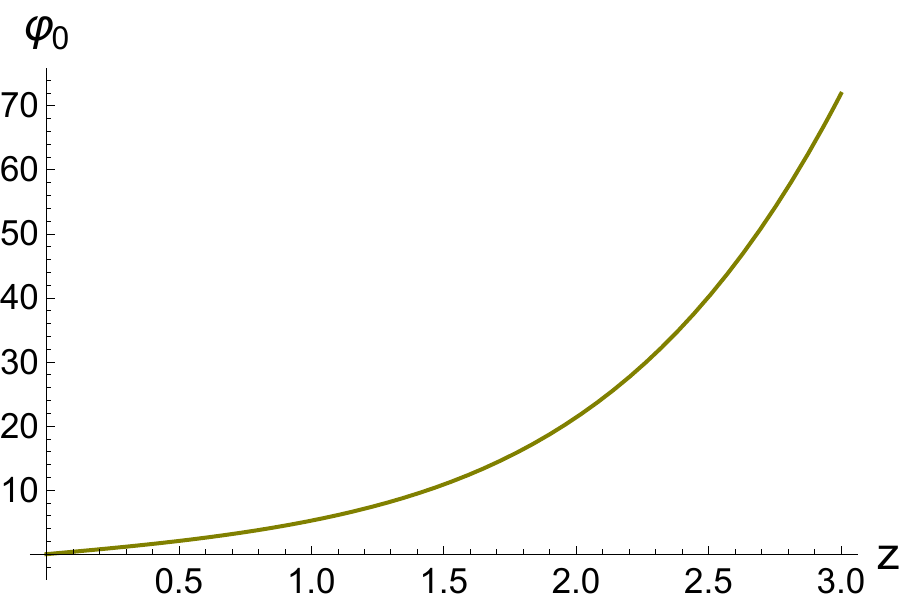}
  \caption{The logarithm of coupling constant $\log\alpha_0(z)$ for the heavy quarks with the boundary condition $z_0=0$ or $\varphi_0(0)=0$; $[z]^{-1} =$ GeV.
  }
    \label{fig:phi0-0-HQ}
\end{figure}

The logarithm of coupling constant $\log\alpha_0(z)$ for heavy quarks model with the zero boundary condition $z_0=0$  ($\varphi_0(0)=0$) is presented in Fig.\,\ref{fig:phi0-0-HQ}. Although $\log\alpha_0(z)$ increases as the energy scale $z$ increases (or, equivalently, the energy scale $E$ decreases), the thermodynamic properties of the model cannot be studied using this boundary condition.

\subsubsection{Running coupling for heavy quarks model with boundary condition $z_0=z_h$}

In this subsection we consider the boundary condition \eqref{bch}. As mentioned above, this boundary condition produces the dependence of the coupling constant $T$ and $\mu$. In fact, we  have 
\bea\label{alphazhHQ}
\alpha_{z_h}(z)&=&\alpha_0(z)\,\fG(z_h),\quad\mbox{where}
\quad \fG (z_h)=e^{-\varphi_{0}(z_h)},
\eea 
or we can rewrite \eqref{alphazhHQ} as  \be \label{alphazh-mHQ}
\alpha_{z_h}(z)=\alpha_{z_h}(z;T,\mu)=e^{\varphi_{z_h}(z;\mu,T)}=\alpha_0(z)\,\fG(z_h(T,\mu)).
\ee
Here  $z_{h}=z_{h}(T,\mu)$ is the value of the horizon for the stable black hole corresponding to given chemical potential $\mu$ and temperature $T$ that can be obtained via $T(z_h)$ plot described in SubSect.\,\ref{phaseHQ}.

\begin{figure}[t!]
  \centering
  \includegraphics[scale=0.48]{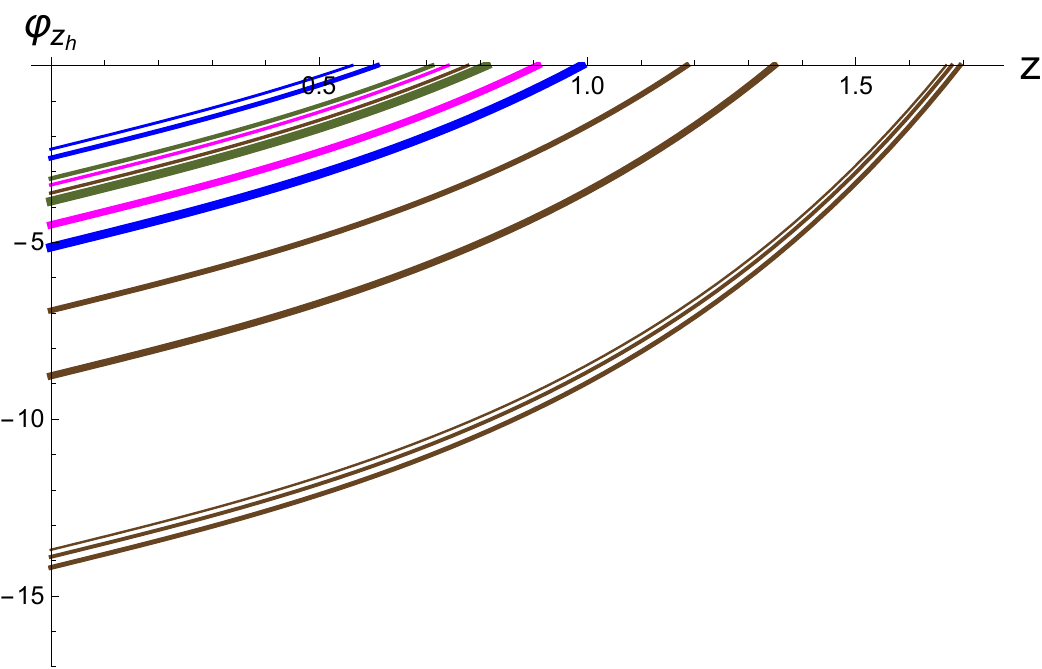} \quad  \includegraphics[scale=0.48]{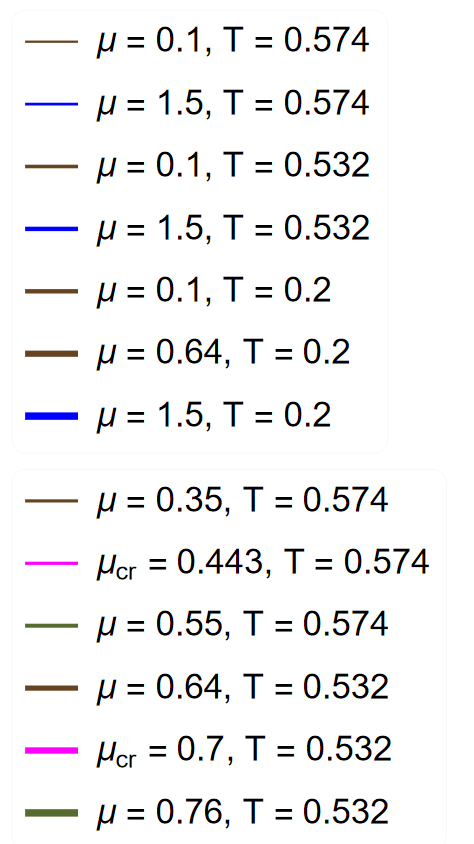} 
 \caption{The logarithm of coupling constant $\log\alpha_{z_h}(z;\mu,T)$ for heavy quarks model considering the first boundary condition $z_0=z_h$ at $T=0.2, 0.532$ and  $0.574$ for different $\mu$; $[\mu]=[T]=[z]^{-1} =$ GeV. 
 }
  \label{Fig:HQphi-phi0-zh-mu-T}
\end{figure}

In Fig.\,\ref{Fig:HQphi-phi0-zh-mu-T} the logarithm of coupling constant $\log\alpha_{z_h}(z;\mu,T)$ for heavy quarks model considering the first boundary condition $z_0=z_h$ at $T=0.2, 0.532$ and  $0.574$ (GeV) is depicted for different chemical potentials corresponding to the points of different regions of heavy quarks phase diagram in Fig.\,\ref{Fig:Ph-TmuHQ}. This figure shows that $\log\alpha_{z_h}(z)$ increases as the energy scale $z$ increases (or, equivalently, the energy scale $E$ decreases). But, we will see that the resulted running coupling in terms of thermodynamic parameters $T$ and $\mu$ are not compatible with lattice calculations.

\begin{figure}[h!]
  \centering
 \includegraphics[scale=0.35]{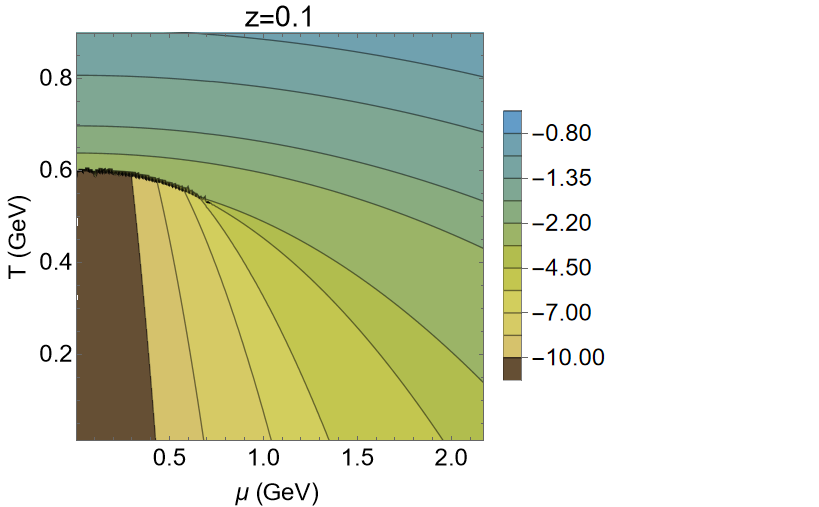}
\includegraphics[scale=0.35]{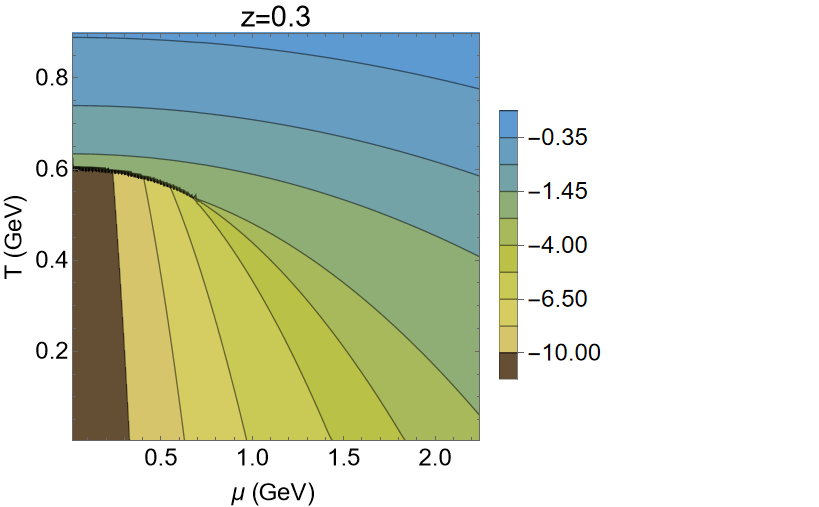}
\\ $\,$\\
   \includegraphics[scale=0.35]{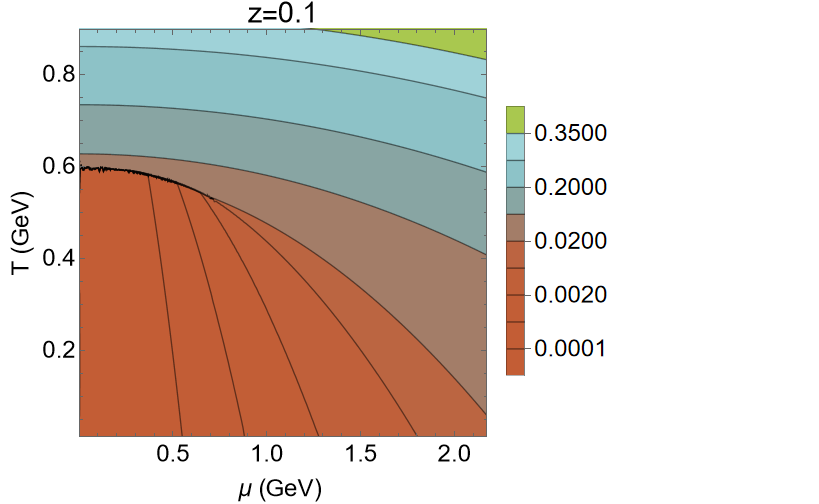} 
\includegraphics[scale=0.35]{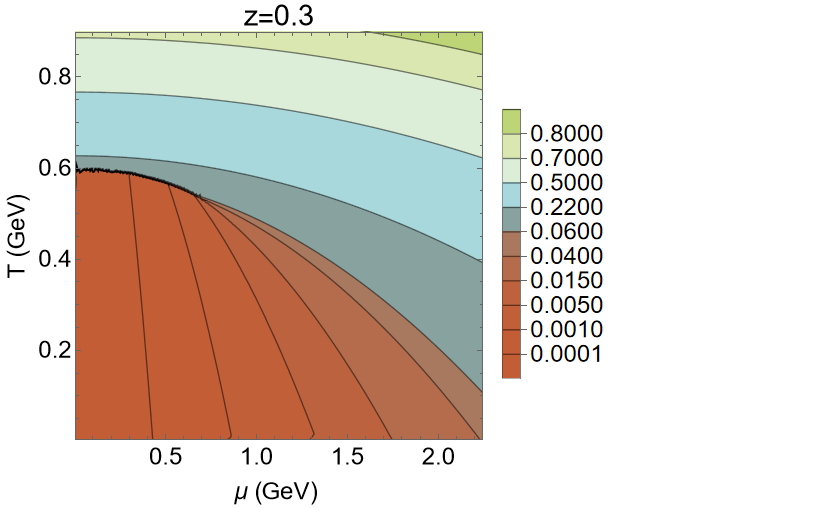}
\caption{ Density plots with contours 
 for logarithm of coupling constant $\log\alpha_{z_h}(z;\mu,T)$ (upper line) and coupling constant $\alpha=\alpha_{z_h}(z;\mu,T)$ (bottom line) for heavy
 quarks at different energy scales $z=0.1$ and $z=0.3$; $[z]^{-1} =$ GeV. \\
}
 \label{Fig:Cz01-z05-Intro-HQ}
\end{figure}

Density plots with contours for logarithm of coupling constant $\log\alpha_{z_h}(z;\mu,T)$ (upper line) and coupling constant $\alpha=\alpha_{z_h}(z;\mu,T)$ (bottom line) for heavy
 quarks at different energy scales $z=0.1$ and $0.3$ (GeV${}^{-1}$) is plotted in Fig.\,\ref{Fig:Cz01-z05-Intro-HQ}. It shows that as energy scale $z$ increases, the coupling constant increases slightly.
In particular, the running coupling in a hadronic region crucially depends on the chemical potential and not on the temperature. It is important to note that the behavior of the running coupling in the hadronic region for light quarks model is directly opposite. The reason of this opposite behavior is originated from different thermodynamics of heavy and light quarks models (see plots Fig.\,\ref{Fig:PhL3d} and Fig.\,\ref{Fig:PhH}).

\begin{figure}[h!]
  \centering
\includegraphics[scale=0.48]{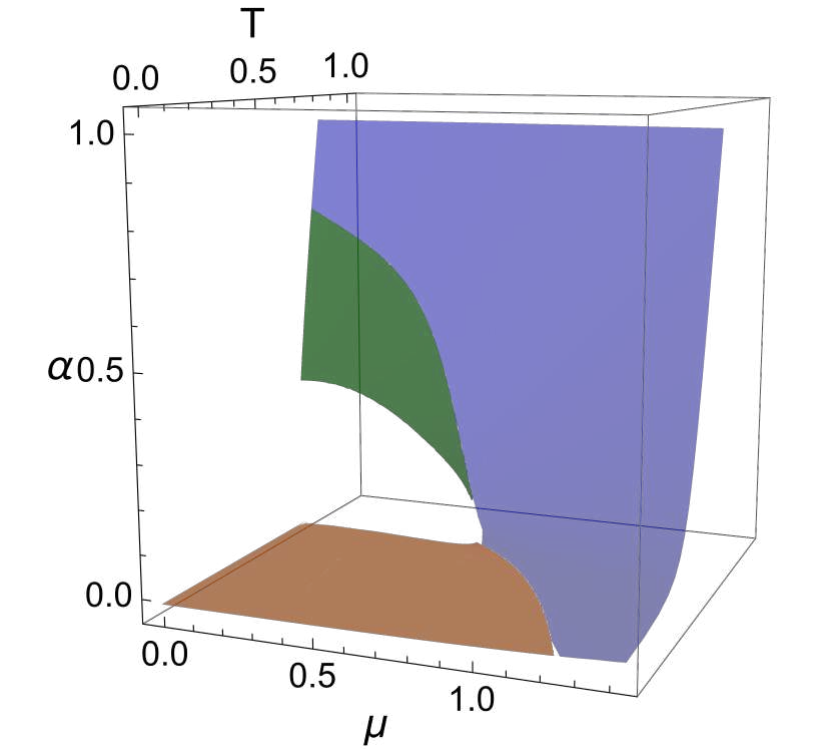}
\qquad
\includegraphics[scale=0.43]{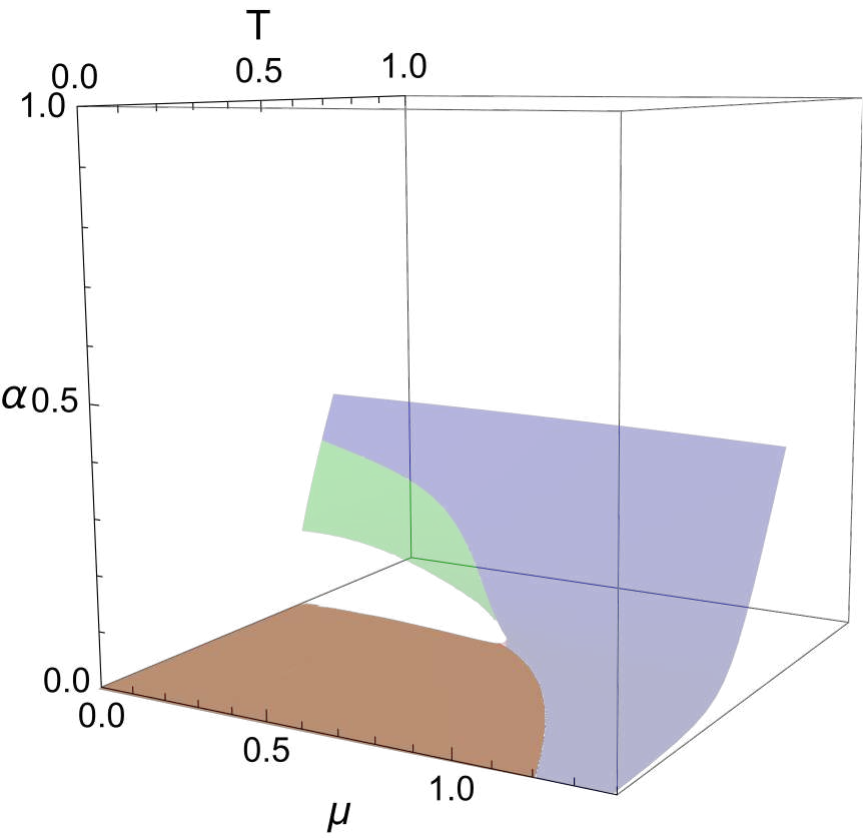}\\
A\hspace{150pt}B\\
\includegraphics[scale=0.47]{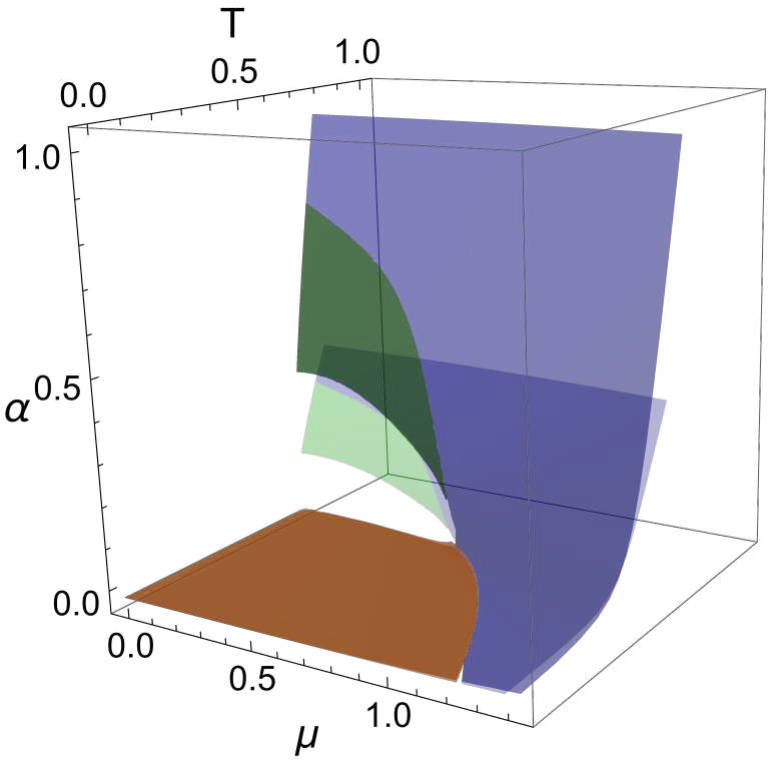}\qquad
\includegraphics[scale=0.45]{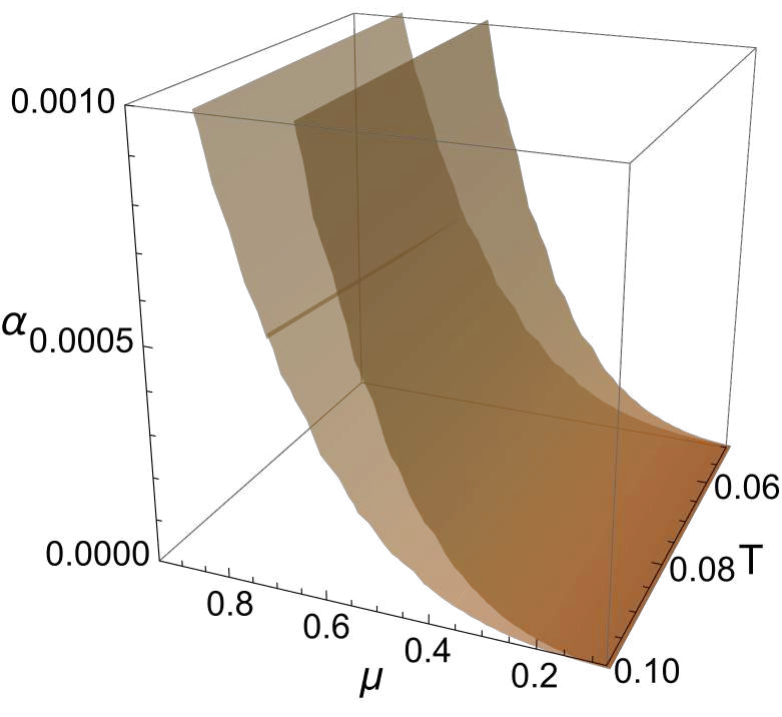}
\\
 C\hspace{150pt}D\\
\caption{Coupling constant $\alpha=
\alpha_{z_h}(z;\mu,T)$ for heavy quarks at two energy scales $z=0.5$ (dark surfaces) (A) and $z=0.3$ (bright surfaces) (B).  Hadronic, QGP and quarkyonic phases are denoted by brown, blue and green, respectively. Comparison of coupling constant for two scales of energy (C,D); $[\mu]=[T]=[z]^{-1} =$ GeV.
}
 \label{Fig:coupling-mu-T-HQ2}
\end{figure}

Coupling constant $\alpha(z;\mu,T)$ for heavy quarks model at two energy scales $z=0.5$ GeV${}^{-1}$ (dark images) (A), $z=0.3$ GeV${}^{-1}$ (bright images) (B) and their comparison (C,D) is presented in Fig.\,\ref{Fig:coupling-mu-T-HQ2}.  Hadronic, QGP and quarkyonic phases are denoted by brown, blue and green, respectively. For $0<\mu<0.7$ there is transition between hadronic and quarkyonic phases as a jump that shows 1st order phase transition. The chemical potential $\mu=0.7$ GeV denotes the intersection of confinement/deconfinement and 1st order phase transition lines. For $0.7<\mu<1.18$  the transition occurs continuously between hadronic and QGP phases. 
At fixed $\mu$ and $T$ (panels (C) and (D)) the value of $\alpha$ increases by decreasing the energy of system (increasing $z$) the same as light quarks model and it is compatible with the running coupling of QCD \cite{ModernTextBook}. The jump of $\alpha$ between hadronic and quarkyonic phases is larger for lower energy (larger $z$), see panel (C).

\begin{figure}[h!]
  \centering
  \includegraphics[scale=0.34]{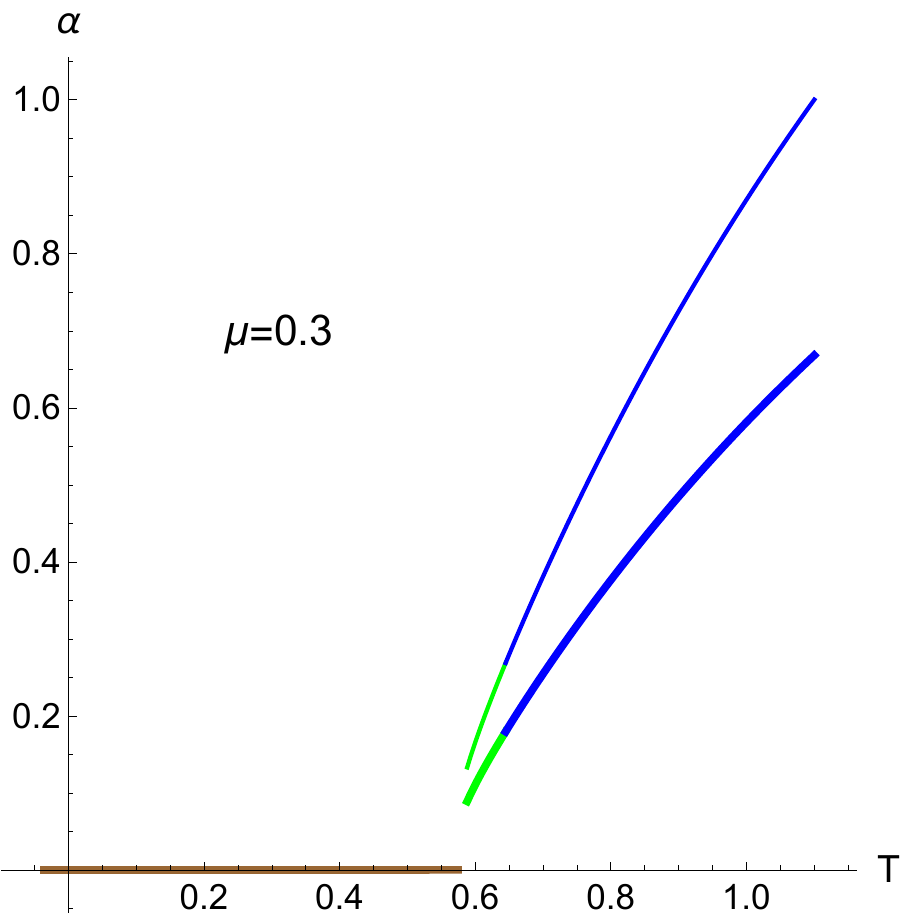} \qquad \qquad \qquad
\includegraphics[scale=0.34]{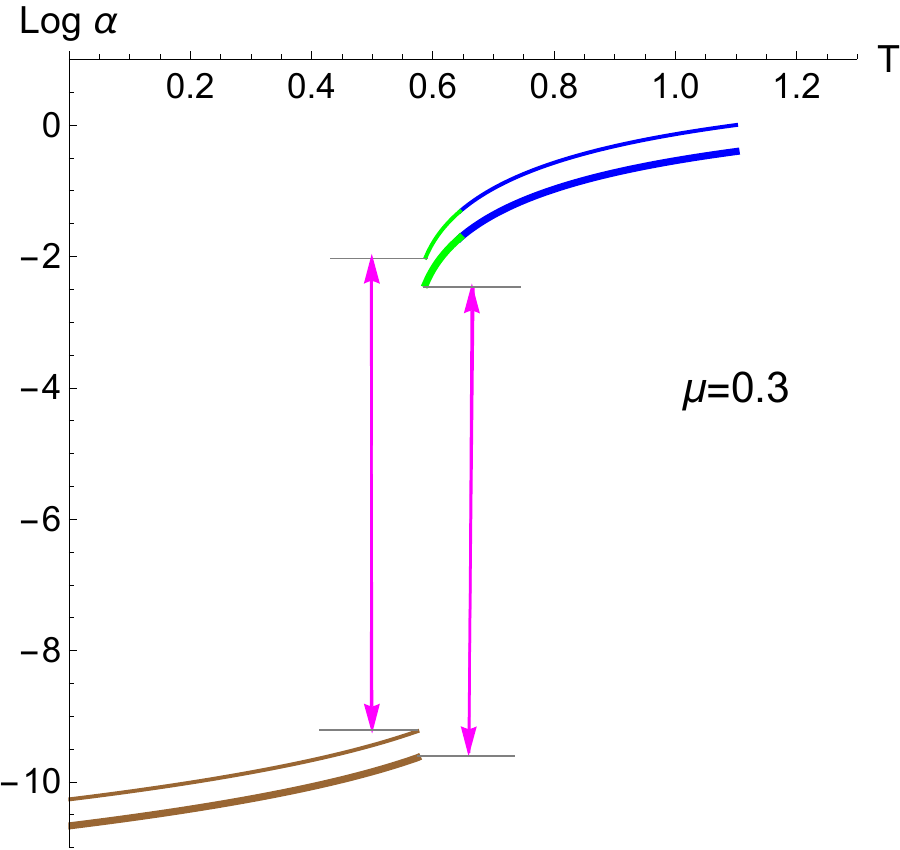}\\
 A\hspace{200pt}B \\\,\\
\includegraphics[scale=0.34]{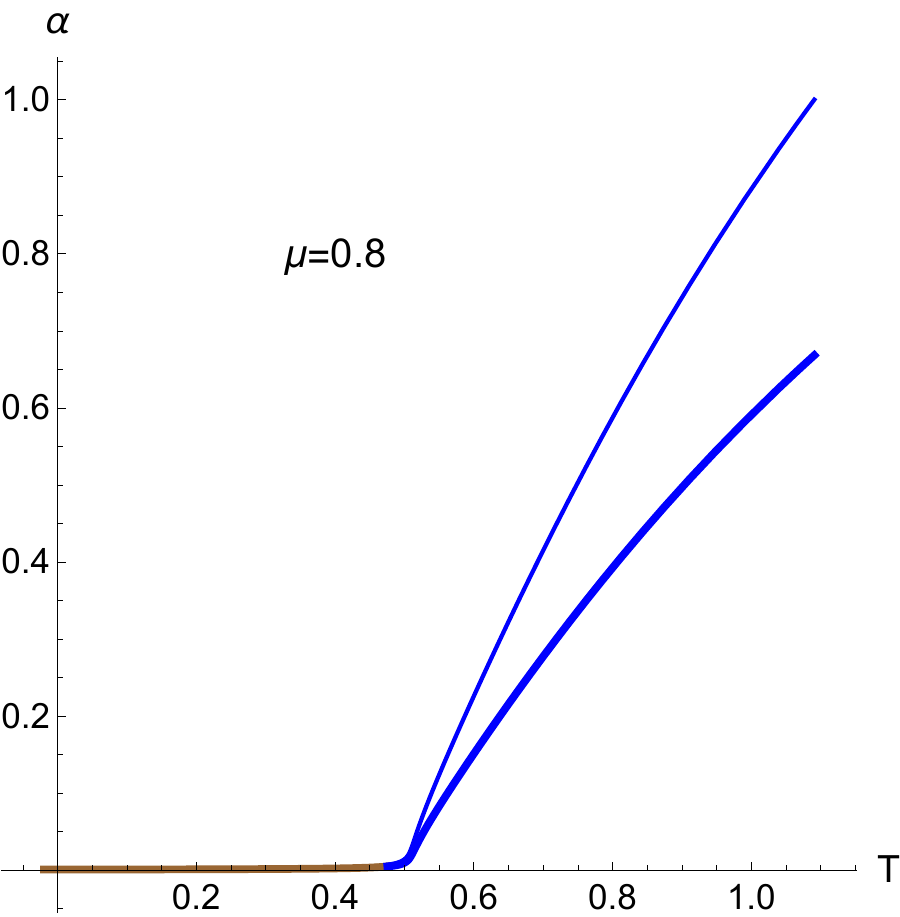} \qquad \qquad \qquad
\includegraphics[scale=0.34]{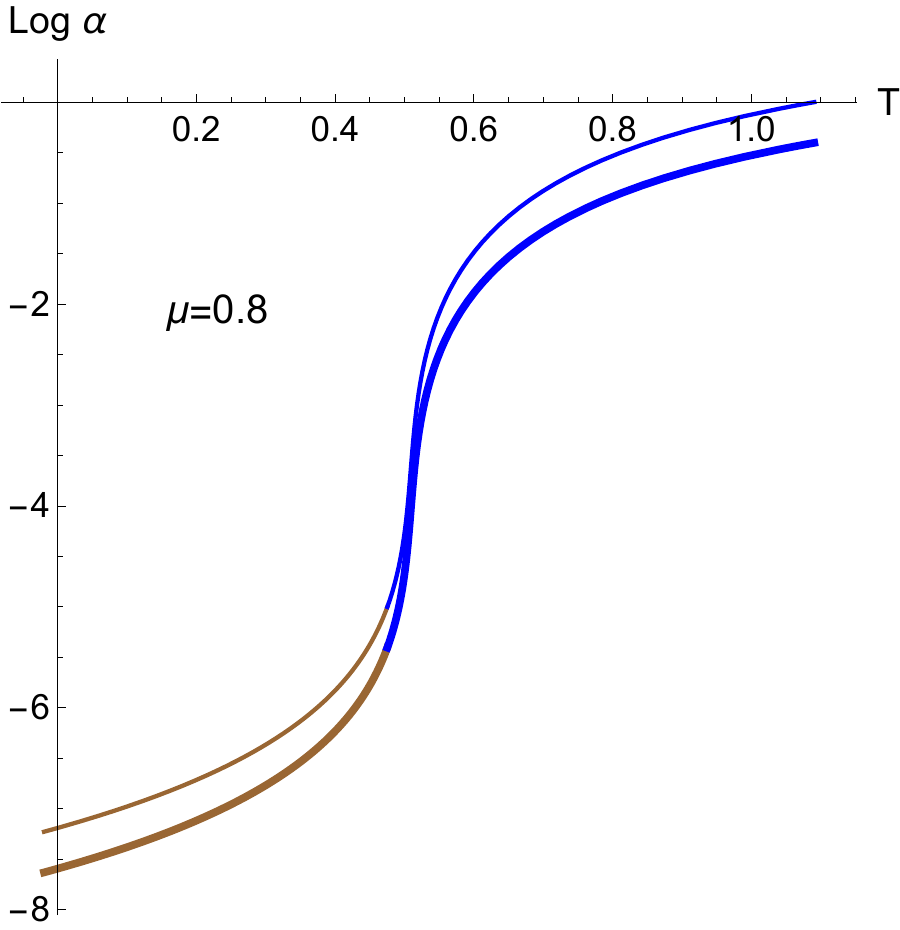}\\
 C\hspace{200pt}D \\
\caption{Coupling constant $\alpha=\alpha_{z_h}(z;\mu,T)$  and  $ \log\alpha=\log \, \alpha_{z_h}(z;\mu,T) $ for heavy quarks at fixed $\mu$ at different energy scales  $z=0.3$ (thin lines) and $z=0.2$ (thick lines).  Hadronic, QGP and quarkyonic phases are denoted by brown, blue and green lines, respectively. Magenta arrows in B show the jumps at the 1st order phase transition. $[\mu]=[T]=[z]^{-1} =$ GeV. 
}
\label{Fig:HQ-alpha-2D}
\end{figure}

Coupling constant $\alpha(z;\mu,T)$  and  $ \log \, \alpha(z;\mu,T) $ for heavy quarks at fixed $\mu=0.3$  and $0.8$ (GeV) at different energy scales $z=0.3$ GeV${}^{-1}$ (thin lines) and $z=0.2$ GeV${}^{-1}$ (thick lines) is depicted in Fig.\,\ref{Fig:HQ-alpha-2D}.  Hadronic, QGP and quarkyonic phases are denoted by brown, blue and green lines, respectively. Magenta arrows in (B) with $\mu=0.3$ GeV show the jumps at the 1st order phase transition between hadronic and quarkyonic phases while at $\mu=0.8$ GeV (C,D) there is a phase transition between hadronic and QGP phases without any jump. The coupling increases when the temperature increases and the physical system undergoes a transition from hadronic to quarkyonic and from quarkyonic to QGP phases.

\newpage

\subsubsection{Running coupling for heavy quarks model with boundary condition $z_0=\fz_{HQ}(z_h)$} \label{HQ-nbc11}

In this subsection we consider the second boundary condition for heavy quarks, i.e. \eqref{phi-z0-gen}, \eqref{bceHQ}. This boundary condition for heavy quarks leads to the physical results in agreement with lattice calculations. As has been mentioned above, this 
boundary condition produces the dependence of the coupling constant 
on $T$ and $\mu$. Indeed, we have
\bea
\alpha_{\fz_{_{HQ}}}(z)=\alpha_{\fz_{_{HQ}}}(z;\mu,T)&=&e^{\varphi_{\fz_{_{HQ}}}(z;T,\mu))}=\alpha_0(z)\fG(\fz_{_{HQ}}(T,\mu)),
\\ \fG(\fz_{_{HQ}}(T,\mu))&=&
e^{-\varphi_0 (\fz_{_{HQ}}(z_h))}=
e^{-\varphi_0 (\fz_{_{HQ}}(z_h(T,\mu)))}
\eea
  Here  $\fz_{\,_{HQ}}(z_h)$ is given by \eqref{bceHQ} and  $z_{h}=z_{h}(T,\mu)$, as in previous consideration,   is the value of the horizon for the stable black hole corresponding to given chemical potential $\mu$ and temperature $T$.

\begin{figure}[h!]
\centering
\includegraphics[scale=0.44]{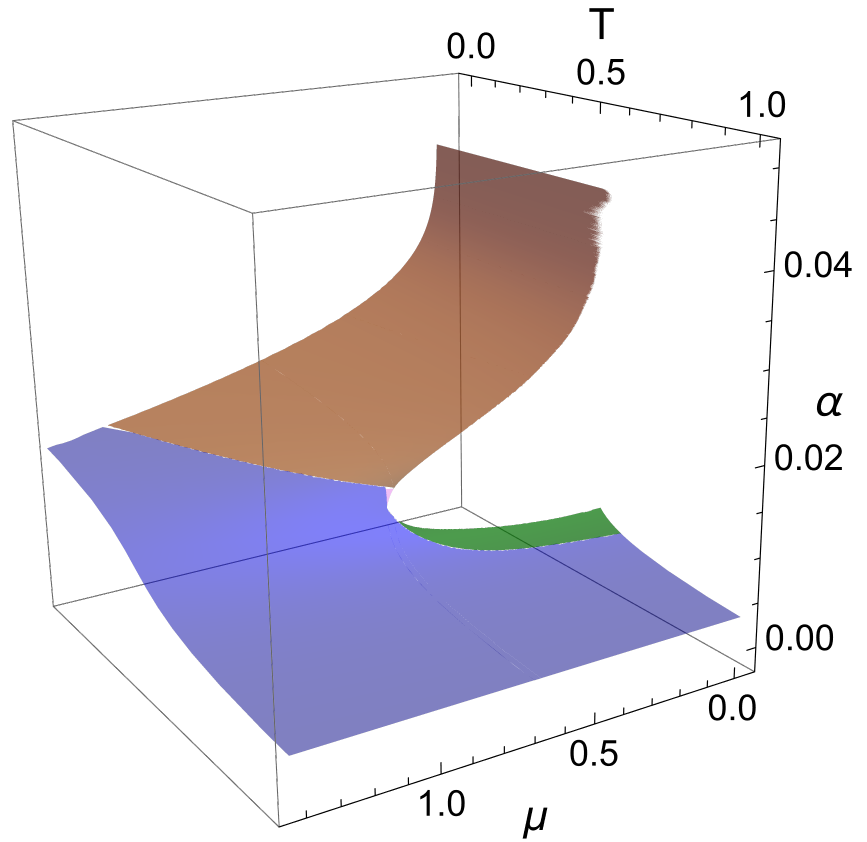}
\caption{The 3D-plot for coupling constant  $\alpha=\alpha_{\fz_{_{HQ}}}(z;\mu,T)$ for heavy quarks with the second boundary condition \eqref{bceHQ} at fixed energy scale $z=0$. Hadronic, QGP and quarkyonic phases are denoted by brown, blue and green lines, respectively; $[\mu]=[T]=[z]^{-1} =$ GeV. 
}
\label{Fig:HQ-alpha-3D2}
\end{figure}

The 3D-plot for coupling constant  $\alpha=\alpha_{\fz_{_{HQ}}}(z;\mu,T)$ for heavy quarks with the second boundary condition \eqref{bceHQ} at fixed energy scale $z=0$ is depicted in Fig.\,\ref{Fig:HQ-alpha-3D2}. At the 1st order phase transition line the coupling constant feels a jump and magnitude of the jump increases by decreasing $\mu$. The coupling in hadronic phase decreases faster than QGP phase by increasing $\mu$. Although, by changing $T$ in hadronic and QGP phase, the coupling constant does not change significantly and decreases slowly.

\begin{figure}[h!]
  \centering
 \includegraphics[scale=0.37]{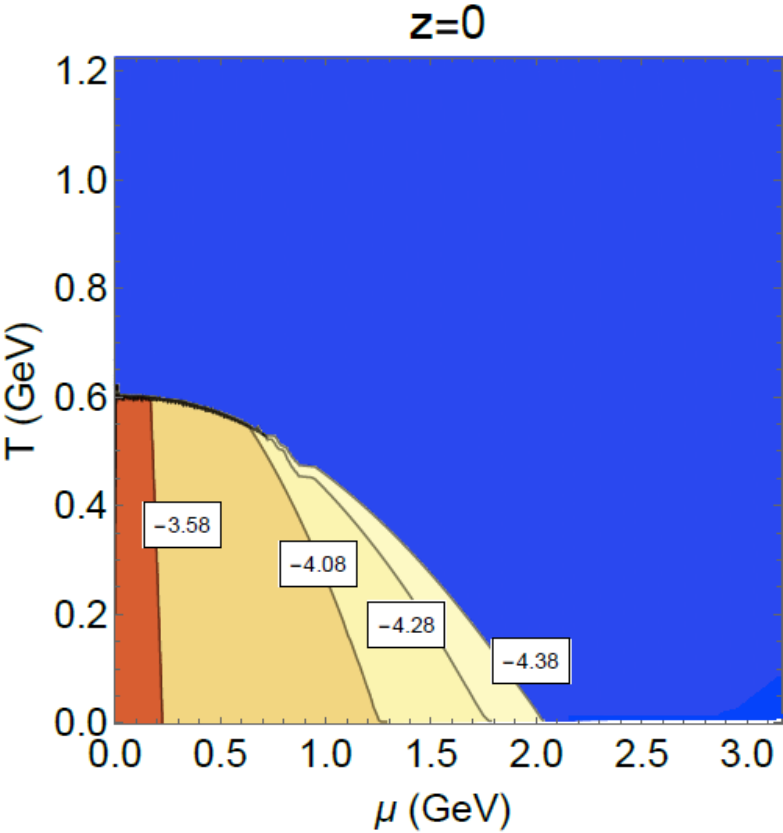} 
\includegraphics[scale=0.31]{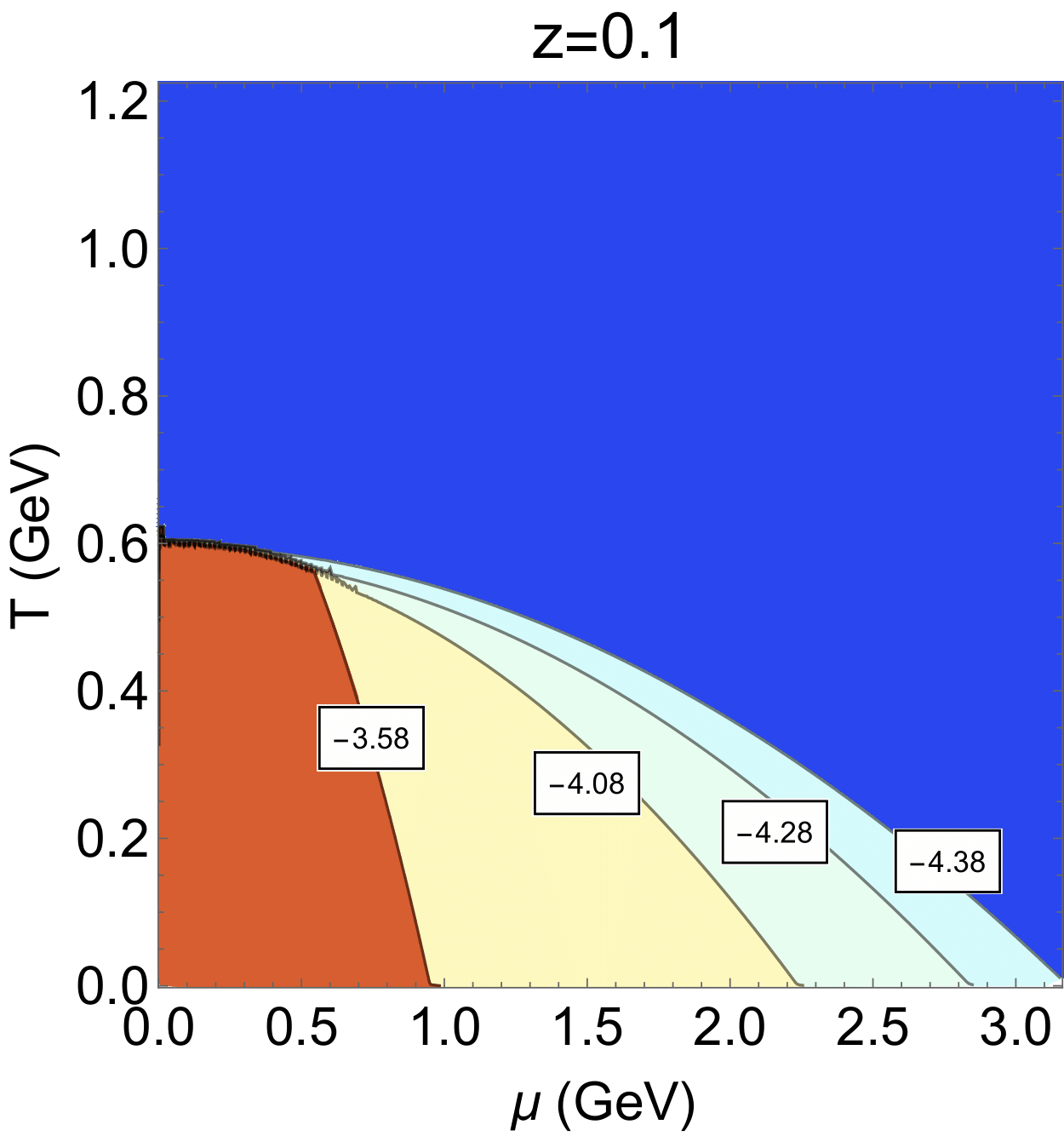}
 \includegraphics[scale=0.31]{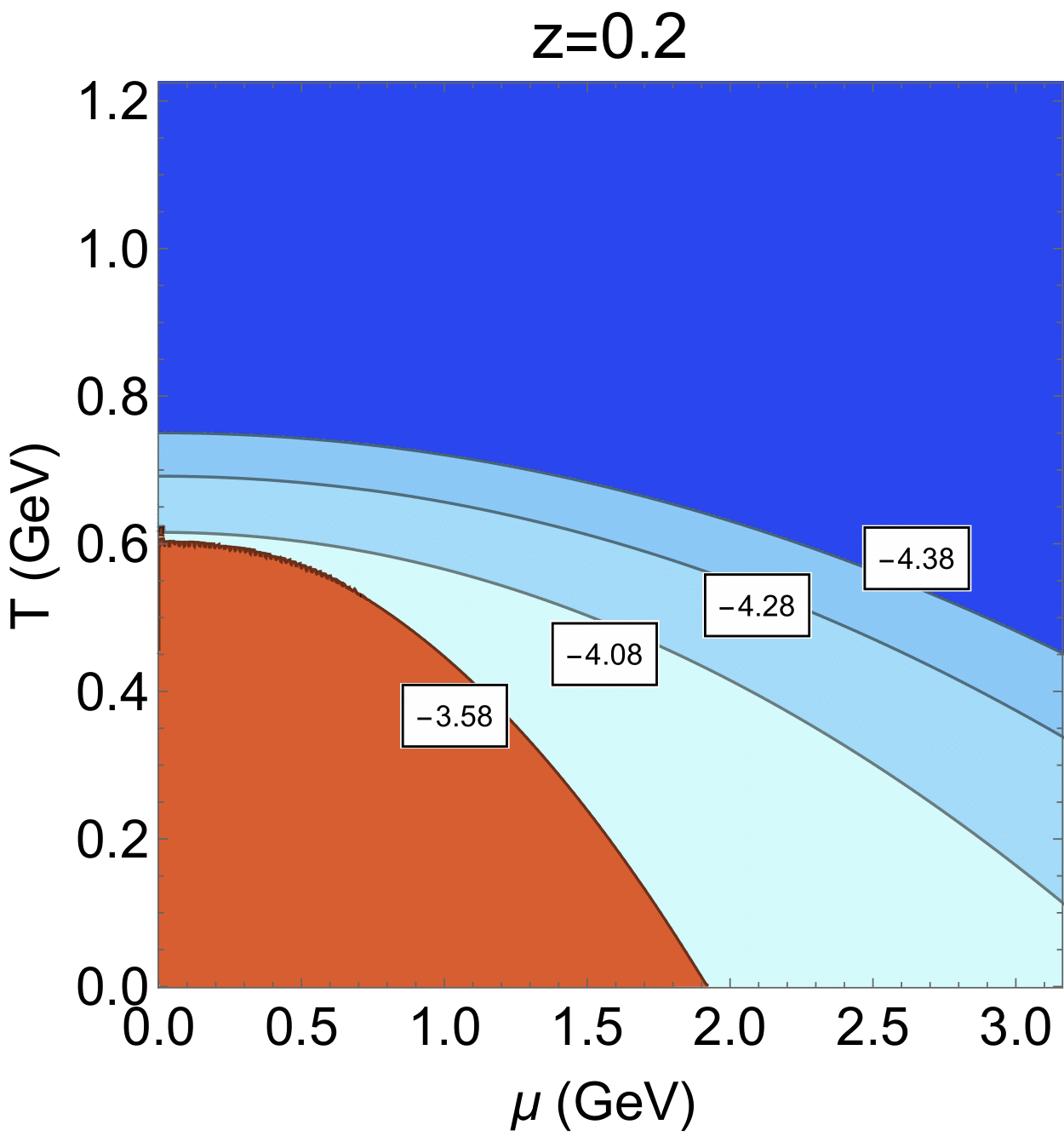}
 \includegraphics[scale=0.37]{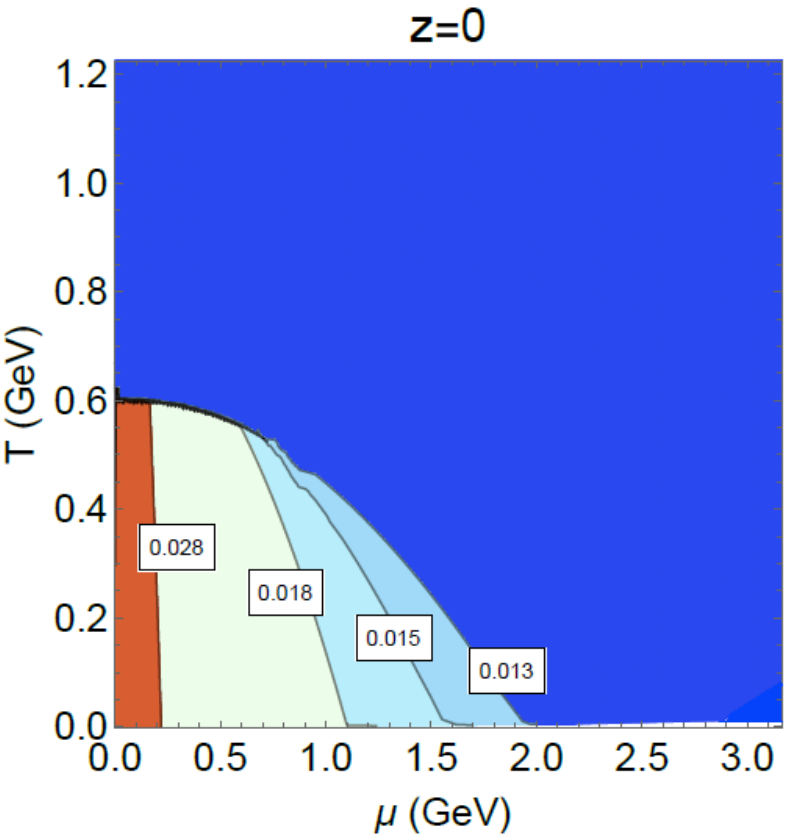} 
\includegraphics[scale=0.31]{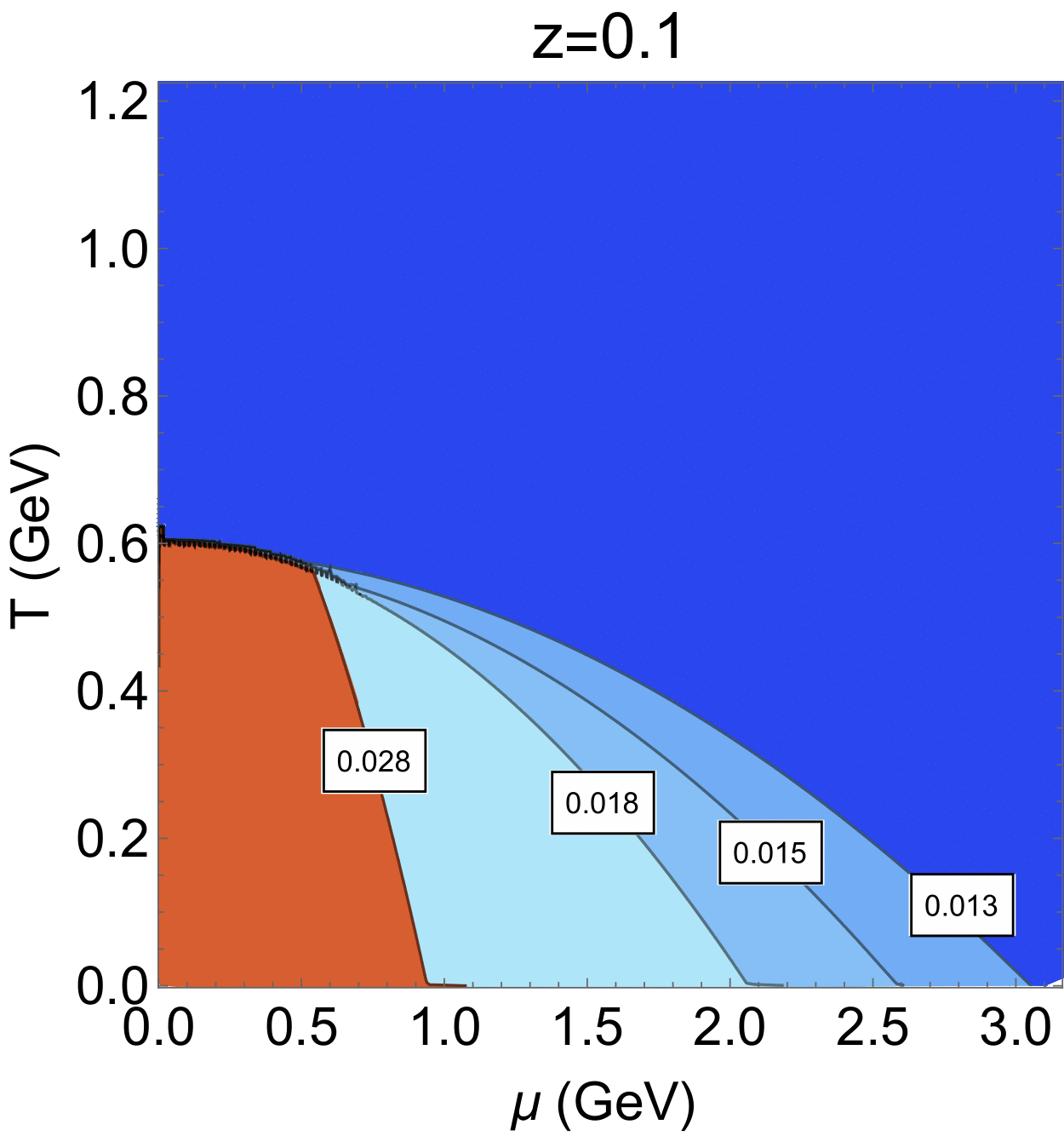}
 \includegraphics[scale=0.31]{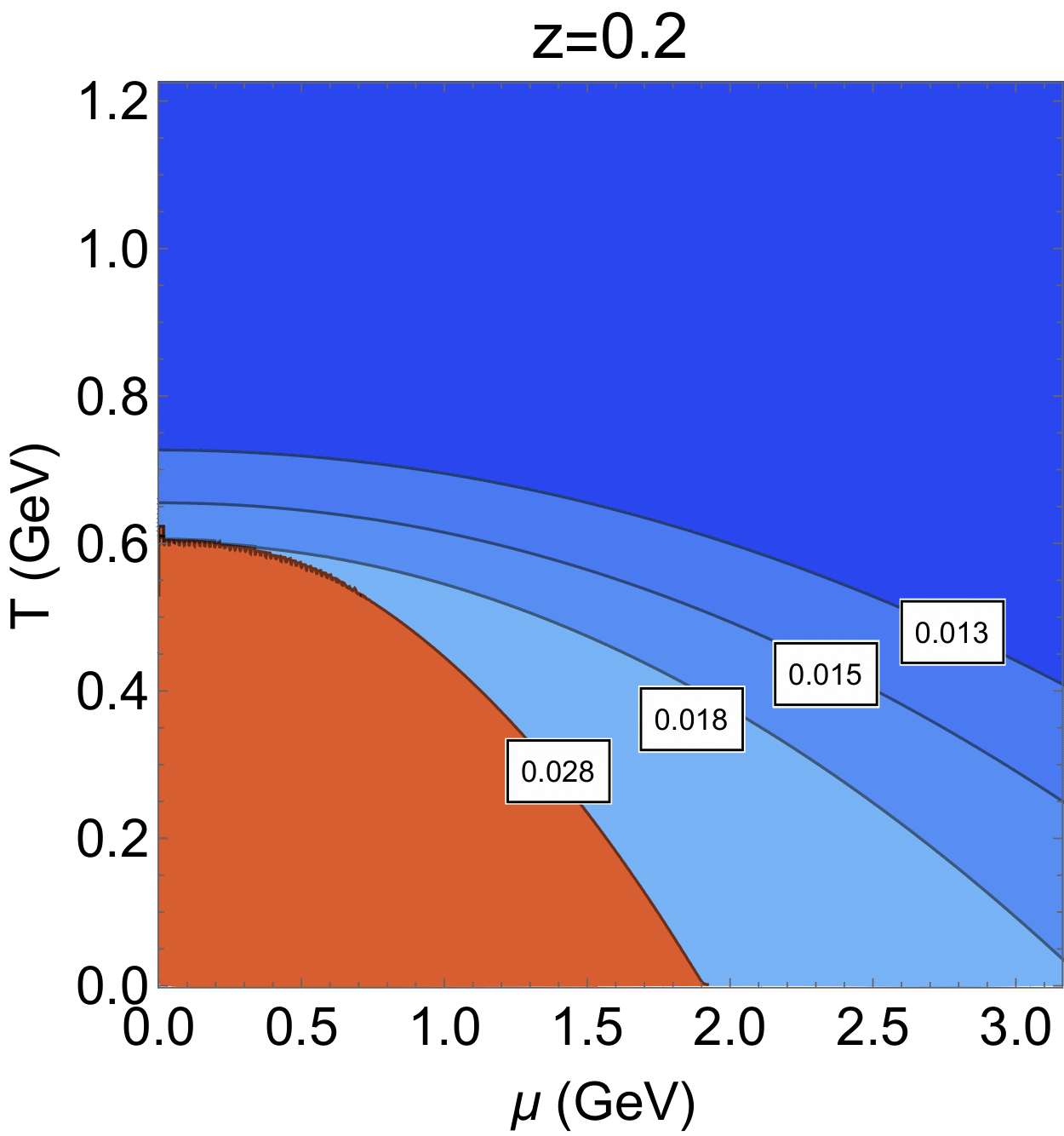}
 \caption{Density plots for heavy quarks with contours of the
 logarithm of coupling constant 
$\log\alpha_{\fz_{_{HQ}}}(z;\mu,T)$ 
 (upper line) and coupling constant $\alpha_{\fz_{_{HQ}}}(z;\mu,T)$ (bottom line) using the boundary condition \eqref{bceHQ} at different energy scales $z=0, 0.1, 0.2$; $[z]^{-1} =$ GeV.\\
}
 \label{Fig:Cz01-z05-Intro-HQ2}
\end{figure}

In Fig.\,\ref{Fig:Cz01-z05-Intro-HQ2} density plots for heavy quarks with contours of the logarithm of coupling constant 
$\log\alpha_{\fz_{_{HQ}}}(z;\mu,T)$ 
 (upper line) and coupling constant $\alpha_{\fz_{_{HQ}}}(z;\mu,T)$ (bottom line) using the boundary condition \eqref{bceHQ} at different energy scales $z=0, 0.1, 0.2$ (GeV${}^{-1}$) is depicted.
 At each (fixed) scale of energy $z$ running coupling decreases by increasing $T$ and by increasing $\mu$ in hadronic phase there is not significant change in coupling and while in QGP phase coupling decreases slowly. Fig.\,\ref{Fig:Cz01-z05-Intro-HQ2} shows that at fixed coupling,  increasing the energy scale $z$ corresponds to larger values in contour, i.e. larger in $T$ and $\mu$ in the phase diagram.

\begin{figure}[h!]
\centering
\includegraphics[scale=0.3]{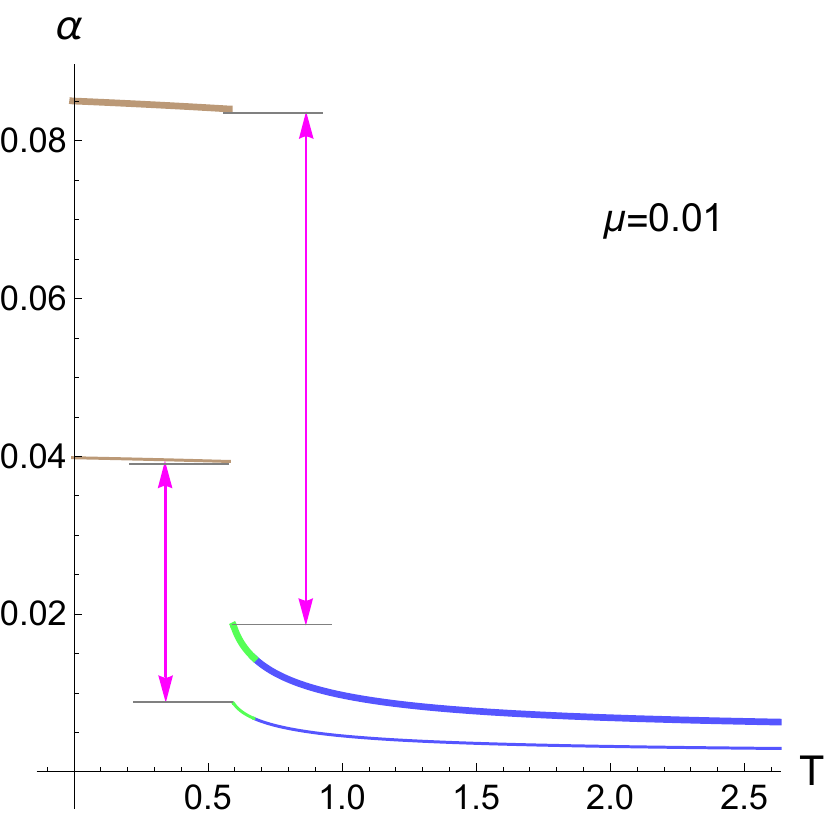} \qquad \qquad \qquad
\includegraphics[scale=0.3]{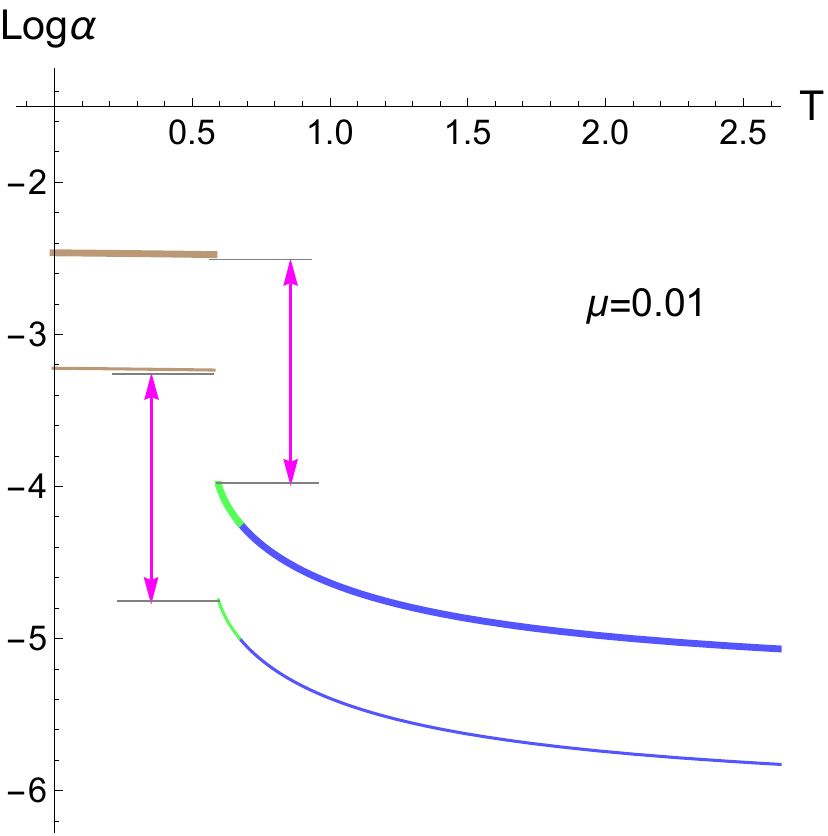}\\
 A\hspace{200pt}B \\
\includegraphics[scale=0.3]{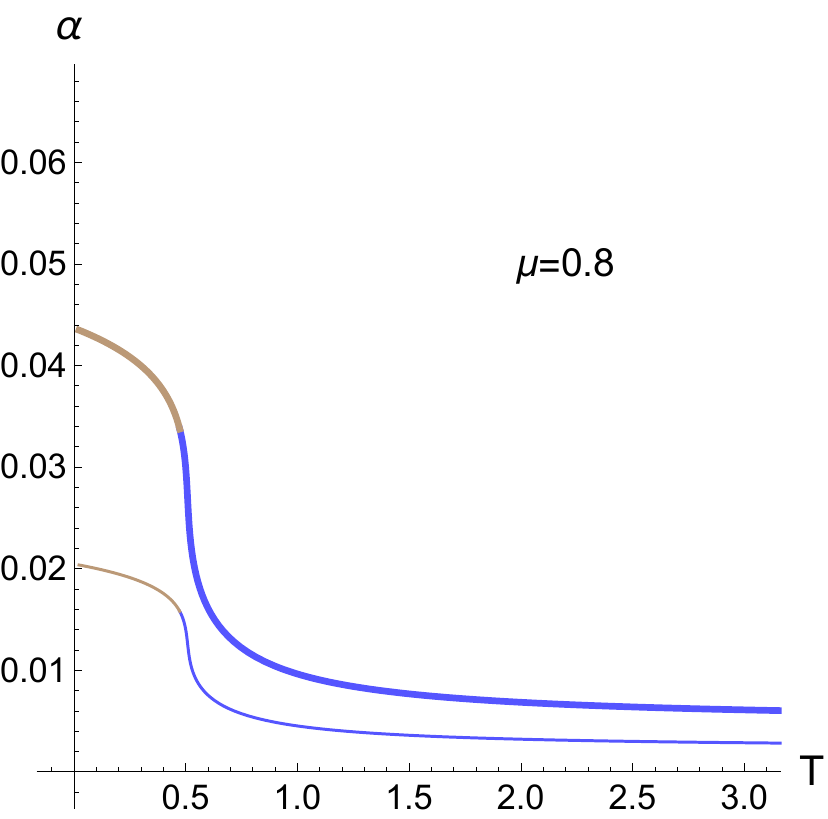} \qquad \qquad \qquad
\includegraphics[scale=0.3]{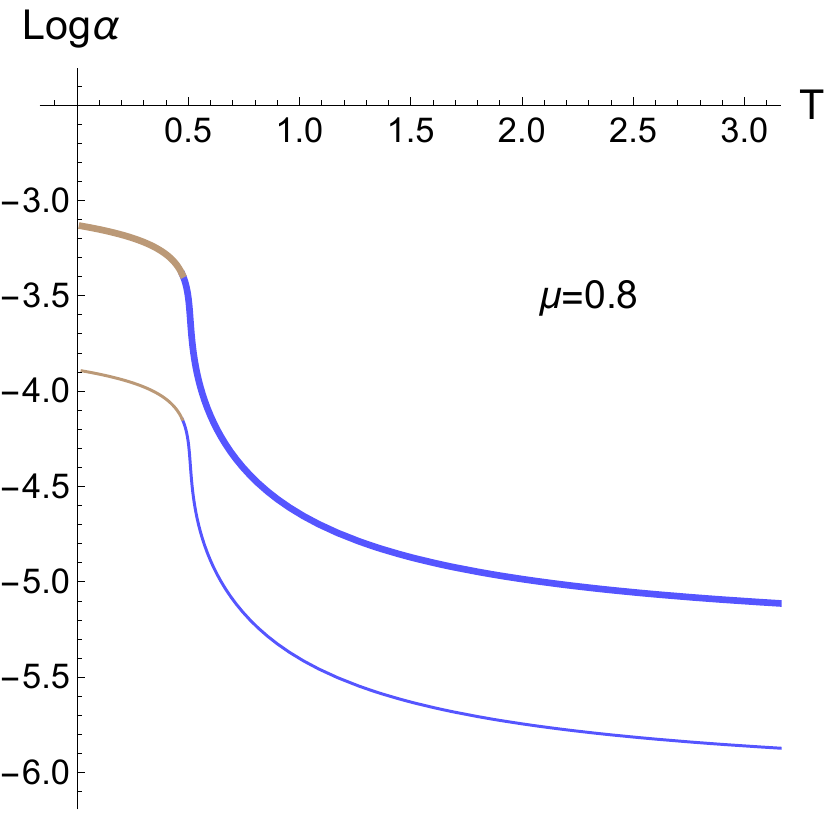}\\
 C\hspace{200pt}D 
\caption{Coupling constant $\alpha=\alpha_{\fz_{_{HQ}}}(z;\mu,T)$  and  $ \log \, \alpha_{\fz_{_{HQ}}}(z;\mu,T) $ for heavy quarks using the boundary condition \eqref{bceHQ} at fixed $\mu$ at different energy scales  $z=0$ (thin lines) and $z=0.2$ (thick lines).  Hadronic, QGP and quarkyonic phases are denoted by brown, blue and green lines, respectively. Magenta arrows in show the jumps at the 1st order phase transition; $[\mu]=[T]=[z]^{-1} =$ GeV. 
}
\label{Fig:HQ-alpha-2D21}
\end{figure}

Coupling constant $\alpha=\alpha_{\fz_{_{HQ}}}(z;\mu,T)$  and  $ \log \, \alpha_{\fz_{_{HQ}}}(z;\mu,T) $ for heavy quarks using the boundary condition \eqref{bceHQ} at fixed $\mu=0.01$ GeV (A,B), $\mu=0.8$ GeV (C,D)  at different energy scales  $z=0$ GeV${}^{-1}$ (thin lines) and $z=0.2$ GeV${}^{-1}$ (thick lines) is shown in Fig.\,\ref{Fig:HQ-alpha-2D21}. Hadronic, QGP and quarkyonic phases are denoted by brown, blue and green lines, respectively. Magenta arrows in show the jumps at the 1st order phase transition. In addition, at fixed $\mu$ the magnitude of jump is larger for larger $z$ (lower energy scale $E$). In all cases at fixed $\mu$ and energy scale $z$, running coupling decreases by increasing $T$. 

\begin{figure}[h!]
  \centering
\includegraphics[scale=0.37]{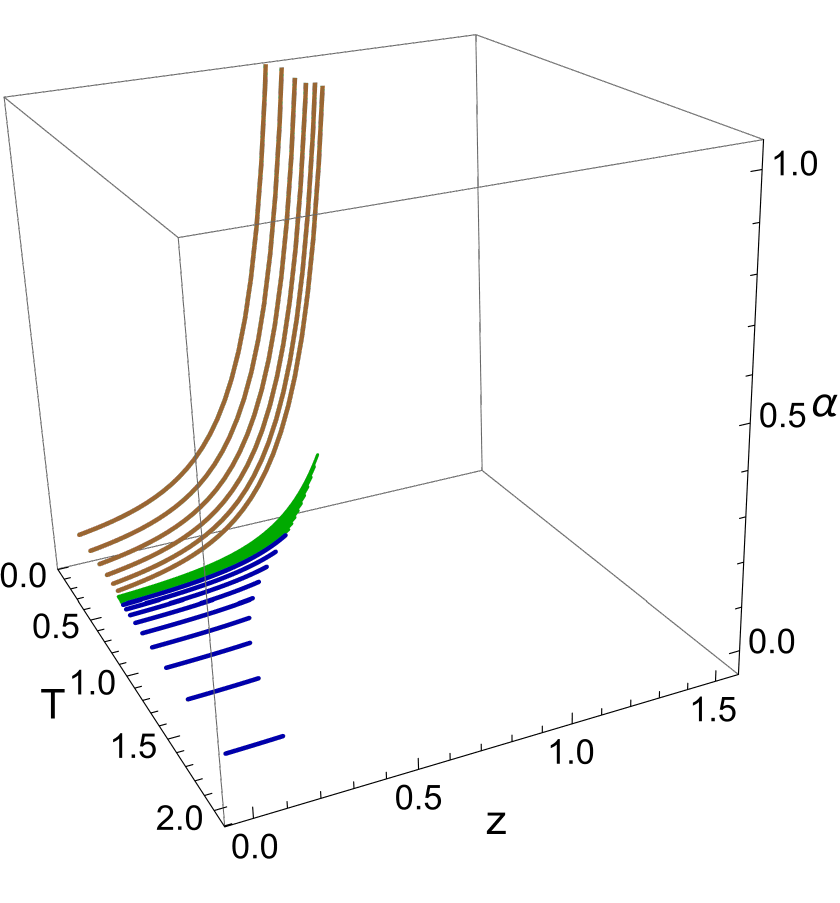} \qquad 
\includegraphics[scale=0.37]{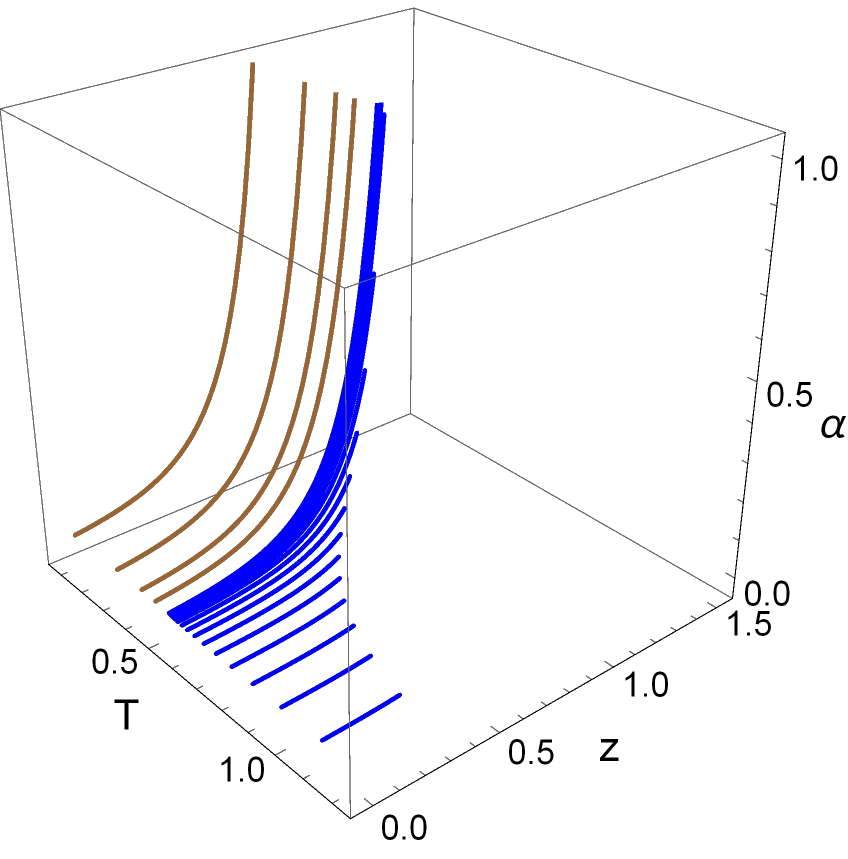} \\
A\hspace{200pt}B
 \caption{3D-plots of coupling constant $\alpha=\alpha_{\fz}(z;\mu,T)$ for heavy quarks at $\mu=0.3$ (A) and $\mu=0.8$ (B).  Hadronic, QGP and quarkyonic phases are denoted by brown, blue and green lines, respectively; $[\mu]=[T]=[z]^{-1} =$ GeV. 
 }
  \label{Fig:HQ-alpha-z-T}
\end{figure}

3D-plots of coupling constant $\alpha=\alpha_{\fz}(z;\mu,T)$ for heavy quarks model are presented at $\mu=0.3$ GeV in Fig.\,\ref{Fig:HQ-alpha-z-T}A and $\mu=0.8$ GeV in Fig.\,\ref{Fig:HQ-alpha-z-T}B. Hadronic, QGP and quarkyonic phases are denoted by brown, blue and green lines, respectively. At fixed $\mu=0.3$ GeV $\alpha$ changes continuously between quarkyonic and QGP, i.e. phase transition occurs with no jump, while there is a jump between hadronic and quarkyonic phases where the 1st order phase transition occurs for any range of the energy scale $z$. It shows that the coupling constant is sensible to the 1st order phase transition. 
For each phases in Fig.\,\ref{Fig:HQ-alpha-z-T}A at fixed $T$, decreasing the energy (increasing $z$) coupling constant increases and vice versa that is compatible with running coupling of QCD \cite{ModernTextBook}. 
At fixed $\mu=0.8$ GeV there is a continuous phase transition between  just two phases, i.e. QGP and hadronic phases.

\newpage

\subsubsection{Running coupling versus energy scale $E$ for heavy quarks model} \label{gbcEHQ}

For heavy quarks model, considering the formula \eqref{BBH}, similar to the  light quark model one can cover UV
limit of QFT for small values of $z$ and IR limit for larger values of $z$.

\begin{figure}[h!]
\centering
\includegraphics[scale=0.53]{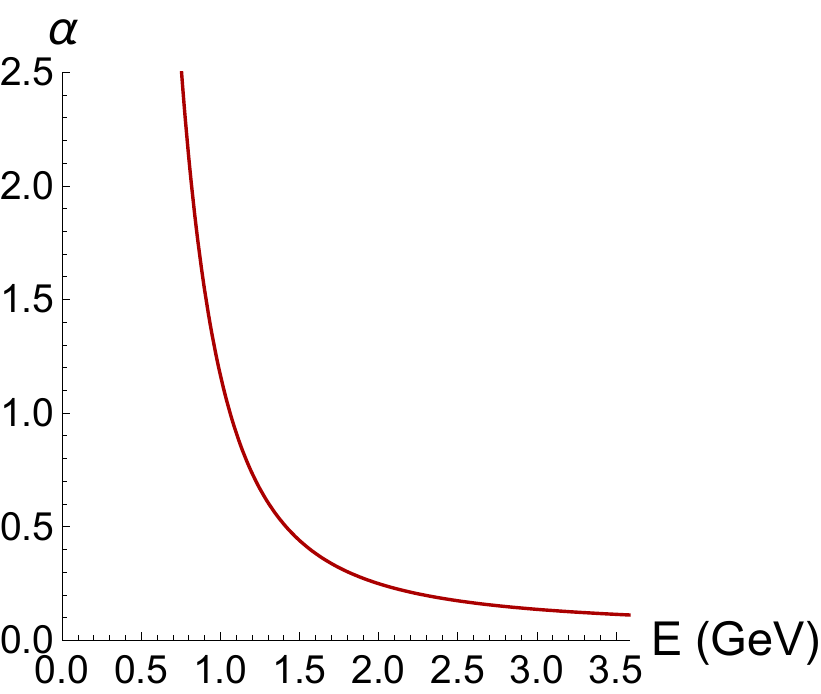} \quad
\includegraphics[scale=0.54]{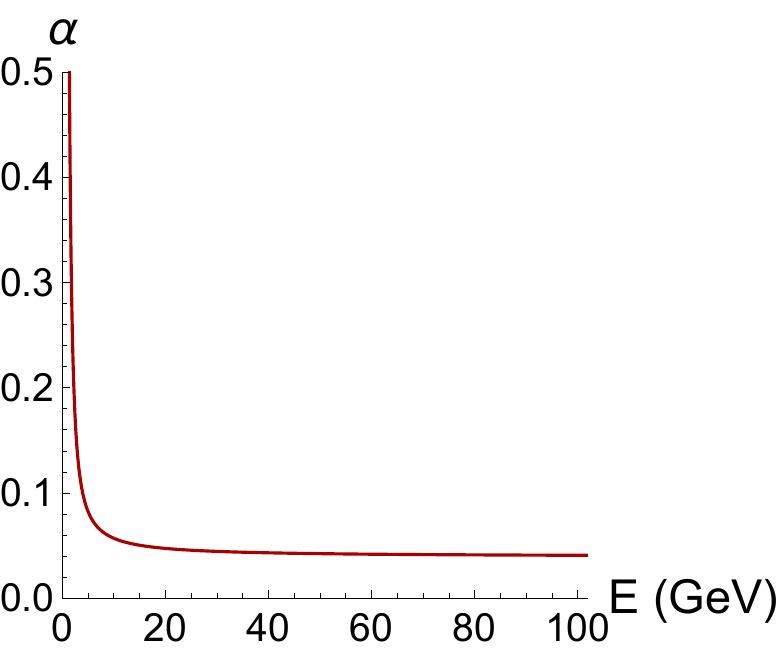}\\
 A\hspace{210pt}B
\caption{Running coupling $\alpha=\alpha_{_{HQ}}(E;\mu,T)$ for heavy quarks model as a function of the energy scale $E$ in QFT at fixed $T = 0.737$ and $\mu = 0.01$. Panel\,B presents panel\,A in different plot range; $[\mu]=[T]=$ GeV.}
    \label{alphavsEHQ}
\end{figure}

Running coupling $\alpha=\alpha_{_{HQ}}(E;\mu,T)$ as a function of the energy scale $E$ in QFT for heavy quarks at fixed $T = 0.737$ GeV and $\mu = 0.01$ GeV is depicted in Fig.\,\ref{alphavsEHQ}. The Fig.\,\ref{alphavsEHQ}B presents the Fig.\,\ref{alphavsEHQ}A in different plot range. 
The Fig.\,\ref{alphavsEHQ}A shows the monotonic behavior of running coupling $\alpha(E)$ as a decreasing function that qualitatively is compatible with the perturbation results \cite{ParticleDataGroup:2022pth,Deur:2016tte}.
Interestingly, the Fig.\,\ref{alphavsEHQ}B shows that although we can reach to the higher values of the energy scale $E$, but we cannot cover the very small values of the running coupling $\alpha$ in our holographic model for heavy quarks. In other words, our model automatically covers the strongly coupled regime of the QFT and some regions of UV and can not describe ultra-UV regime for heavy quarks model.

\newpage

\section{Conclusion and Discussion} \label{sec:concl}

In the paper we have  considered the running coupling constant $\alpha$  in  holographic models supported by Einstein-dilaton-Maxwell action for heavy and light quarks.  We obtained a significant dependence of the running coupling constant $\alpha$  on mass of quarks, chemical potential and temperature. At the 1st  order phase transitions,  $\alpha$  undergoes jumps depending on temperature and chemical potential. For heavy quarks these jumps depend more sharply on thermodynamic parameters. The magnitudes of the jumps increase with decreasing energy scales both for light and heavy quarks.\\

 It is very important to note that  different boundary conditions lead to different physical consequences. Only the second boundary condition $z_0=\fz(z_h)$, i.e. \eqref{phi-fz-LQ} for light quarks and \eqref{bceHQ} for heavy quarks, leads to physical results in accordance with lattice calculations for non-zero temperature and zero chemical potential \cite{Kaczmarek:2007pb}. For this reason, we call the second boundary condition physical. However, the phase structure of light and heavy quarks models is independent of choosing different boundary conditions for the dilaton field.
\\

Note that to obtain the physical results for light and heavy quarks models for coupling constant $\alpha_{\fz}(z;\mu,T)$ as a function of different parameters of the theory  we considered the physical domains of different phases for light and  heavy quarks presented in   Fig.\,\ref{Fig:PhL2D} and Fig.\,\ref{Fig:HQphase-colors-Е}.

To summarize our findings:

\begin{itemize}
\item The dependence of running coupling $\alpha$ for light quarks model on thermodynamic parameters $\mu$ and $T$ on the solution corresponding to physical boundary condition \eqref{phi-fz-LQ} is presented in the Sect.\,\ref{gbc10}.  We see that 
    \begin{itemize}
\item the running coupling decreases monotonically by increasing the energy scale $E$, at fixed $\mu$ and $T$;\ 
\item in the region of the hadronic phase, the running coupling decreases very fast with increasing temperature, at fixed $\mu$ and energy scale $E$; 
\item in the region of the QGP phase, the running coupling decreases significantly slower than the hadronic phase with increasing temperature, at fixed $\mu$ and energy scale $E$;
\item in the region of the hadronic and QGP, by increasing chemical potential the running coupling decreases very slowly and does not experience significant change, at fixed $T$ and energy scale $E$;
\item the slope of decreasing of the coupling constant as a function of $T$ is larger for higher energy scales $E$, at fixed $\mu$;
\item on the 1st order transition line, the running coupling has a jump. The magnitude of the jump is zero at the CEP and increases with increasing $\mu$ to a value that remains almost the same along the entire line of the 1st order phase transition;
\item the magnitude of the jump increases with decreasing the energy scale $E$.
\end{itemize}

\item The dependence of running coupling $\alpha$  for heavy quarks model on thermodynamic parameters $\mu$ and $T$ is presented in Sect.\,\ref{HQ-nbc11}.
\end{itemize}
We see that 
    \begin{itemize}
\item the running coupling decreases monotonically by increasing the energy scale $E$, at fixed $\mu$ and $T$;
\item in the region of the hadronic phase, the running coupling decreases slowly with increasing temperature, at fixed $\mu$ and energy scale $E$;
\item in the region of the QGP phase, the running coupling decreases significantly faster than the hadronic phase with increasing temperature, at fixed $\mu$ and energy scale $E$;
\item in the region of the hadronic phase, by increasing chemical potential the running coupling decreases very fast in comparison to the QGP phase that does not change significantly by changing chemical potential, at fixed $T$ and energy scale $E$;
\item the slope of decreasing in the coupling constant as a function of $T$ is larger for larger energy scales $E$, at fixed $\mu$;

\item on the 1st order transition line, the running coupling has a jump.
The magnitude of the jump is greatest at $\mu=0$ and decreases along the line of the 1st order transition to zero in the CEP;
\item the magnitude of the jump increases with decreasing the energy scale $E$.
\end{itemize}


In this research we investigated the dependence of running coupling on the holographic coordinate $z$ that is related to the energy scale $E$ in the boundary field theory, corresponding to the warp factor, i.e. equation \eqref{warp-factor}.
In particular, it would be interesting to consider the  holographic isotropic models of  light and heavy quarks to study the running coupling as a function of the energy scale $E$ and try to fit the results of holographic models with experiments such as \cite{Galow:2009kw,Pirner:2009gr}. The dilaton field plays a crucial role in the holographic approach 
to study the running coupling and it would be interesting to find out the effect of imposed proper boundary condition on the running coupling as a function of the energy scale $E$  for light and heavy quarks.
In addition, in \cite{Brodsky:2010ur} considering holographic approach and using special form of Fourier transformation from ($z$ coordinate) 5-dimensional space-time in gravity side to the 4-dimensional one, the behavior of $\alpha$ as a function of momentum transfer, $Q^2$, is studied.
We will consider these studies in a new research in future calculations.\\

We would like to emphasize that to respect the null energy condition (NEC) we should consider the blackening function and the dilaton potential for light and heavy quarks in holographic model for different values of $\mu$ and $z_h$ in such a way that $z_h$ to be less than values of second horizon $z_{h_2}$. At the second horizon $z_{h_2}$, the temperature equals to zero. Therefore, the blackening function $g(z)$ should be non-negative at the interval $0\leq z \leq z_h$.
\\

Other physical quantities also have jumps at the 1st order phase transition in isotropic and anisotropic models. Namely, entanglement entropy \cite{Dudal:2018ztm,Arefeva:2020uec}, electric conductivity and direct photons emission rate \cite{Arefeva:2022avn, Arefeva:2021jpa} and  energy loss \cite{Arefeva:2020bjk,Arefeva:2021btm} have these jumps at the 1st order phase transition.
\\

It is  known how the primary (spatial) anisotropy affects the QCD phase transition temperature
\cite{Arefeva:2018hyo,Arefeva:2018cli}.
Also that there is another type of anisotropy due
to a magnetic field and its effect on the QCD phase diagram too, see \cite{Gursoy:2017wzz,Bohra:2019ebj,Bohra:2020qom,Arefeva:2020vae, Dudal:2021jav,Jain:2022hxl,Arefeva:2022avn}. The corresponding anisotropy also changes the running coupling $\alpha$.  We study the extensions in \cite{Arefeva:2024xmg,AHNS} and $\beta$-function in separate paper \cite{Arefeva:2025xtz}.
\\

\section{Acknowledgments}
The work of IA, PS and MU, which consisted of general setting of the project and studying of thermodynamics for light quarks and running coupling for light and heavy quarks is supported by the Russian Science Foundation Grant No. 20-12–00200. The work of A.H. consisting in studying of thermodynamics for heavy quarks and comparison of running coupling for light and heavy quarks, was performed at the Steklov
International Mathematical Center  and  supported by the Ministry of
Science and Higher Education of the Russian Federation (Agreement
No. 075-15-2022-265).


\newpage 
\appendix
\appendix

\section{Potentials of light and heavy quarks models}\label{app:potentials}

The dilaton potentials $\cV(\varphi)$ and gauge kinetic functions $\ff_0(\varphi)$  of the models under consideration, (\ref{action}), where obtained by a reconstruction method from solutions (\ref{phiprime}), (\ref{Vsol}) and (\ref{wfLc}).

\begin{figure}[h!]
  \centering
  \includegraphics[scale=0.47]{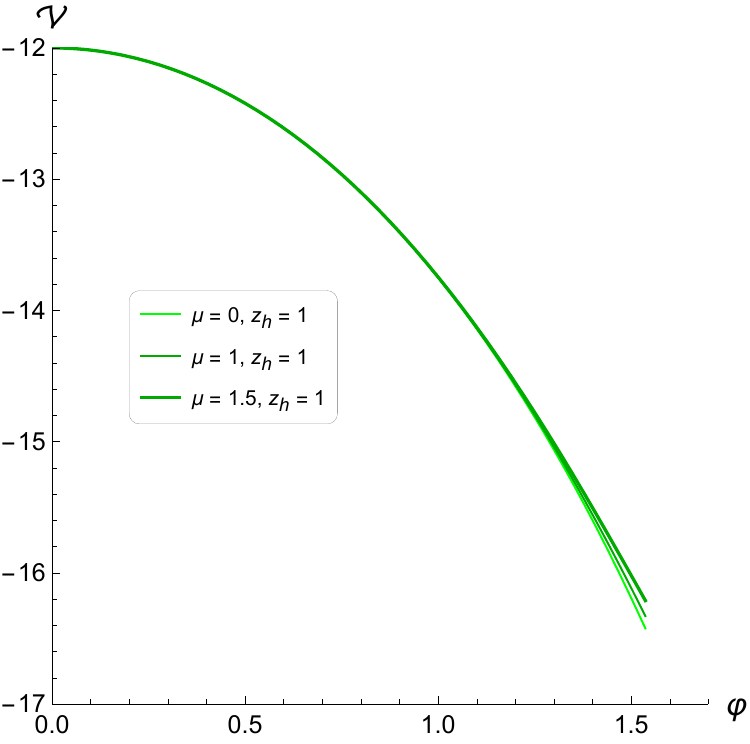} \qquad \qquad \includegraphics[scale=0.46]{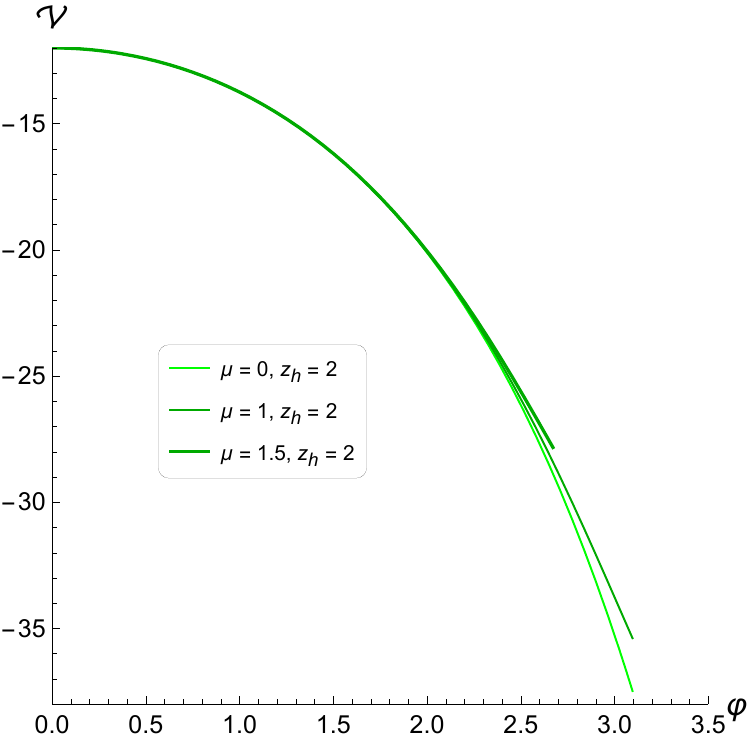}\\
A\hspace{200pt}B

\includegraphics[scale=0.47]{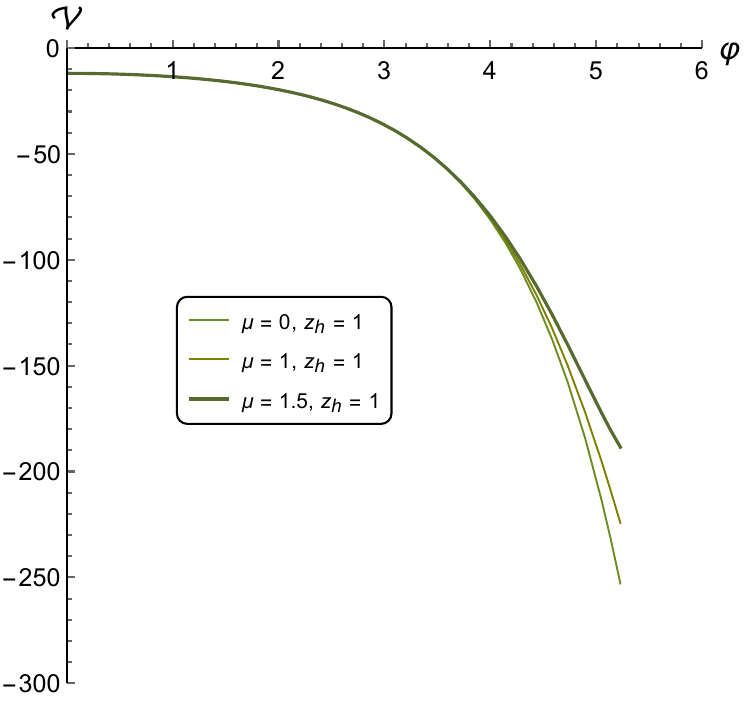} \qquad \qquad \includegraphics[scale=0.48]{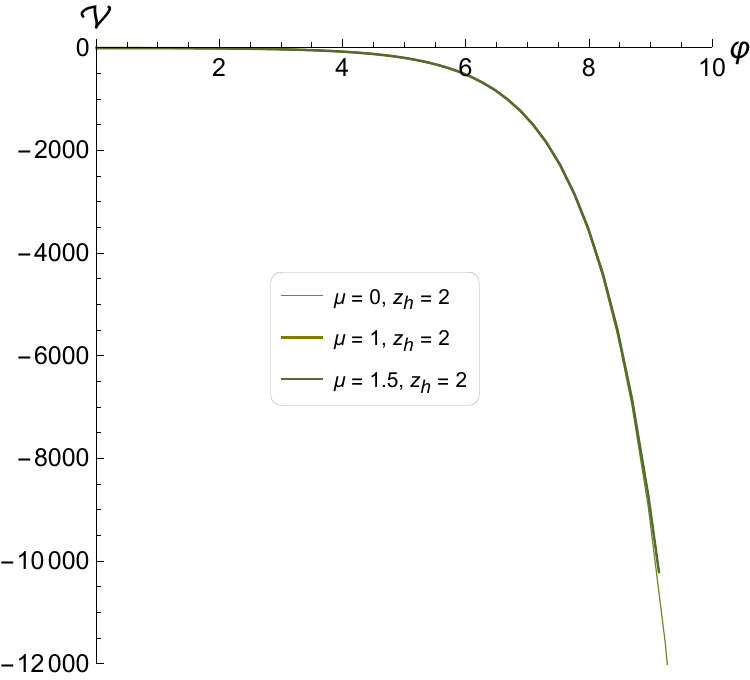}\\
C\hspace{200pt}D
 \caption{The dilaton potential for the  light quarks model (A, B) and the light quarks model (C, D) for different values of $\mu$ and $z_h$ for zero b.c. 
\label{fig:VL-H} }
\end{figure}

\begin{figure}[h!]
  \centering
  \includegraphics[scale=0.47]{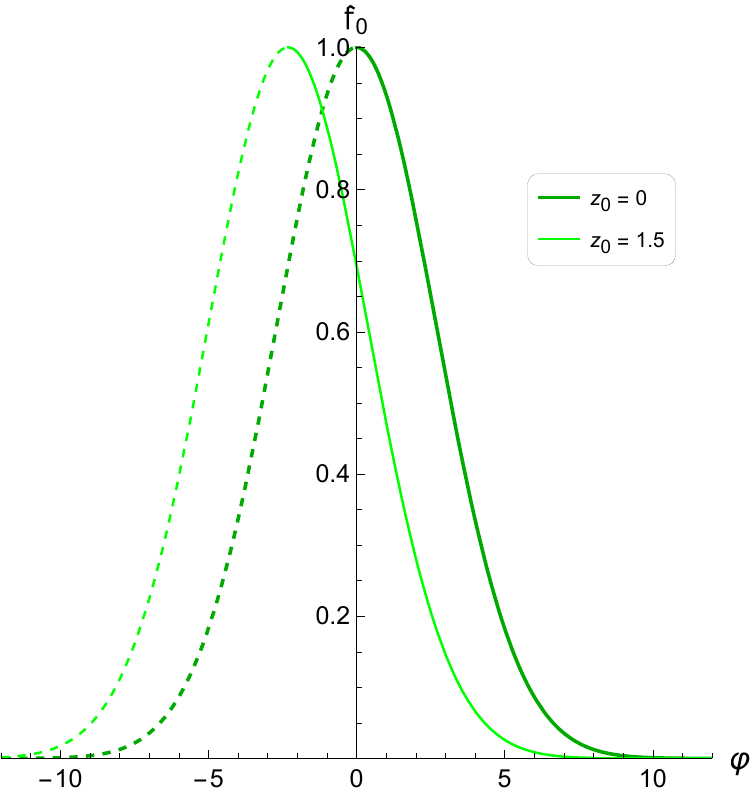}\qquad\qquad
  \includegraphics[scale=0.47]{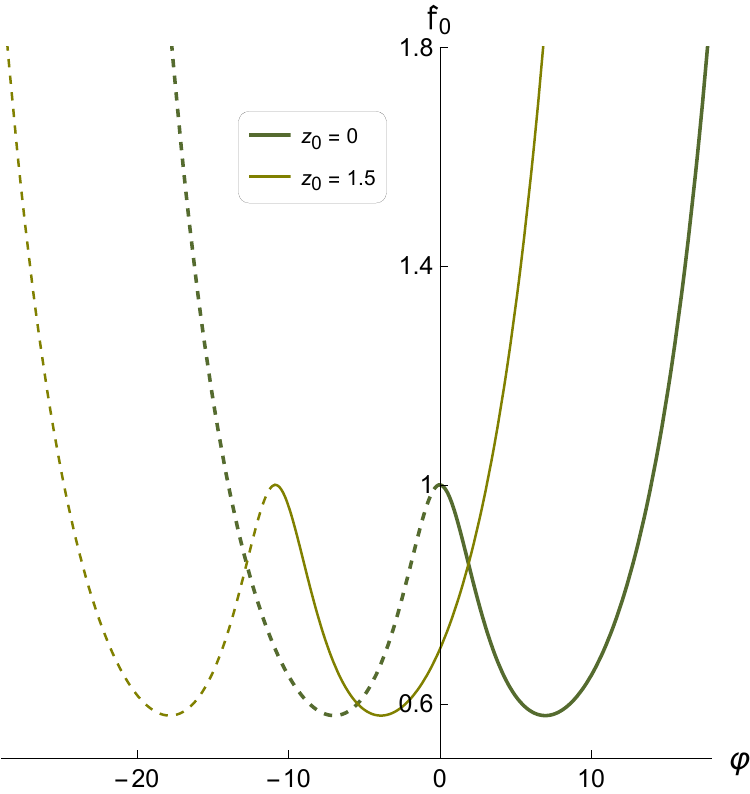}\\
  A\hspace{200pt}B\\
 \caption{The dilaton gauge kinetic function for the light quarks model (A) and  the heavy quarks model (B) for different b.c. Dashed lines define the regions where $z<0$.
\label{fig:fL-H} }
\end{figure}

For zero boundary condition,  i.e. $z_0=0$, values of the chemical potential $\mu$ and the horizon $z_h$ slightly effect both on light and heavy quarks dilaton potentials, see Fig.\,\ref{fig:VL-H}, while the gauge kinetic function  $\ff_0(\varphi)$ is independent of $\mu$ and $z_h$. In general, a change of the boundary condition $z_0$ leads only to the shift of $\cV(\varphi)$ and $\ff_0(\varphi)$  functions relative to the $\varphi$ not effecting on their qualitative behavior, e.g. see Fig.\,\ref{fig:fL-H}.

\newpage

\section{Holographic temporal Wilson loops}\label{app:WL}

To find the position of confinement/deconfinement transition line and the  Cornell potential we should calculate the expectation value of the temporal Wilson loop $W[C_\vartheta] = e^{-S_{NG}}$  in isotropic background \eqref{metric} in the string frame:

\begin{gather}
  ds^2 = G_{\mu\nu}dx^{\mu}dx^{\nu}
  = \cfrac{L^2}{z^2} \ \fb_s(z) \left[
    - \, g(z) dt^2 + d \vec{x}^2 
    + \cfrac{dz^2}{g(z)} \right], \label{eq:4.10}
\end{gather}
where
\be \label{bs}
\fb_s(z) = \exp\bigl(2 A_s(z) \bigr)=\exp\bigl(2A(z)+2\sqrt{\frac{1}{6}} \varphi(z) \bigr)
  \ee
 
Following the holographic approach we
calculate the value of the Nambu-Goto action for test string in 
background \eqref{eq:4.10}:
\begin{gather}
  S_{NG} = \frac{1}{2 \pi \alpha'} \int d\xi^{0} \, d\xi^{1} \sqrt{- \det
    h_{\alpha\beta}}, \qquad
  h_{\alpha\beta} = G_{\mu\nu} \, \partial_{\alpha}
  X^{\mu} \, \partial_{\beta} X^{\nu}. \label{eq:4.11}
\end{gather}
The world sheet is parameterized as 
\begin{gather}
  \begin{split}
     t = \xi^0, \quad
     x^1 = \xi^1, \quad
     x^2 = 0, \quad
     x^3 = 0, \quad 
     z = z(\xi^1).
  \end{split}
    \label{eq:4.12}
\end{gather}

Let us denote $\xi \equiv \xi^1$, rewrite Nambu-Goto action
(\ref{eq:4.11})
\bea
    S &=& - \, \cfrac{\tau}{2 \pi \alpha'}\,{F_{Q\bar{Q}}}, \qquad \,F_{Q\bar{Q}}= \int  _{-\ell/2}^{\ell/2}\,d\xi \
    M\left(z(\xi)\right) \ \sqrt{{\cal F}(z\left(\xi)\right) +
      \left(z'(\xi)\right)^2}, \label{eq:4.13}\\
   M\left(z(\xi)\right) &=& \cfrac{\fb_s\left(z(\xi)\right)}{z^2(\xi)}
    \, ,  \qquad \tau = \int d \xi^0, \nn \quad {\cal F}\left(z(\xi)\right) = \sqrt{g \left(z(\xi^1)\right)} \, ,
\nn
\eea
where $F_{Q\bar{Q}}$ is the energy configuration of quark and anti-quark and introduce the effective ``potential'' \cite{Andreev:2006ct,Finazzo:2013rqy,Arefeva:2016rob}
\bea
  \begin{split}
    {\cal V}\left(z(\xi)\right) 
    \equiv M\left(z(\xi)\right) \sqrt{{\cal F}\left(z(\xi)\right)} =
     \cfrac{\fb_s\left(z(\xi)\right)}{z^2(\xi)}
    \sqrt{g\left(z(\xi)\right)  }.
  \end{split}\label{eq:4.14}
\eea
The expression $F_{Q\bar{Q}}$ \eqref{eq:4.13} defines the dynamical system
with a dynamic variable $z = z(\xi)$ and time~$\xi$. This system has the first integral
\bea
\label{FI}
\frac{M(z(\xi)){\cal F}(z(\xi))}{\sqrt{{\cal
      F}(z(\xi))+(z'(\xi))^2}}={\cal I}. 
\eea
From \eqref{FI} we find the ``top'', or the turning point, ${z_{*}}$
(the closed position of the minimal surface to the horizon), where
${z'(\xi) = 0}$:
\be
M(z_{*})\sqrt{F(z_{*})}={\cal I}.
\ee
Finding  $z^{\prime}$ from \eqref{FI} one gets representations for
the string length $\ell$ and the free energy $F_{Q\bar{Q}}$ \eqref{eq:4.13}:
\bea\label{ell1}
\frac\ell2 &=&
\int_0^{z_*}\frac{1}{\sqrt{{\cal F}(z)}} \
\frac{dz}{\sqrt{\frac{{\cal V}^2(z)}{{\cal V}^2(z_*)} -1 }}, \\
\frac{F_{Q\bar{Q}}}{2} &=&
\int_\epsilon^{z_*}\frac{M(z)dz}{\sqrt{1-\frac{{\cal V}^2(z_*)}{{\cal
        V}^2(z)}}}.
\label{calS1}
\eea

Depending on the form of $M$ we have divergencies as $\epsilon \to 0$. In this case we have to perform renormalization. Performing substraction corresponding to disconnecting strings we get 
\bea
\frac{F_{Q\bar{Q}, ren}}{2} &=&
\int_\epsilon^{z_*}\Bigg(\frac{M(z)}{\sqrt{1-\frac{{\cal V}^2(z_*)}{{\cal
        V}^2(z)}}}-M(z)\Bigg)dz -\int_{z_*}^{z_h} M(z) dz.
\eea

To get $F_{Q\bar{Q}}(\ell)$ numerically we use the parametric plot.\\

 We expect Cornell potential such that
\be \label{cornel1}
F_{Q\bar{Q}}(\ell)\sim C-\frac{4}{3}\frac{\alpha_{Q\bar{Q}}}{ \ell}+\sigma_{Q\bar{Q}} \ell
\ee
The coefficients in this representation are found in the following way

\begin{itemize}
\item{\bf String tension }
\end{itemize}

To obtain the string tension we have to calculate asymptotic of $\cal
S$ for $\ell\to \infty$. If the stationary point of ${\cal V}(z)$
exists in the region $0 < z < z_h$, 
\be
  {\cal V}^\prime\Big|_{z_{DW}} = 0,\label{sp}
\ee
we call this point  a dynamical wall (DW) point and  take the top
point $z_*$ equal to the DW position, $z_*=z_{DW}$. Near this point we
get
\bea
  \ell &\underset{z\to z_{DW}}{\sim }& \frac{1}{\sqrt{F(z_{DW})}} \,
  \sqrt{\frac{{\cal V}(z_{DW})}{{\cal V}''(z_{DW})}} \, \log
  (z-z_{DW}) \underset{z\to z_{DW}}\to \infty, \\
  F_{Q\bar{Q}}&\underset{z\to z_{DW}}{\sim}& M(z_{DW}) \, \sqrt{\frac{{\cal
        V}(z_{DW})}{{\cal V}''(z_{DW})}} \log (z-z_{DW}). 
\eea
Hence  for the  string tension we get
\bea
  F_{Q\bar{Q}}&\sim&   \sigma _{DW} \cdot \ell\,,\qquad 
 \sigma _{DW} = M(z_{DW}) \,\sqrt{F(z_{DW})}. \label{ten}
\eea

Equation for DW has the form:

\bea
 &\quad &2  A'(z) + \sqrt{\cfrac23} \ \varphi'(z) +
  \cfrac{g'}{2 g} - \cfrac{2}{z} \ \Big|_{z = z_{DW}} \hspace{-15pt}
  = 0,  \label{eq:4.15x}
\eea

\begin{itemize}
    \item {\bf Coulomb part and coupling constant }
\end{itemize} 

First of all we add one more regularization in (regularization near $z=z_*$): in integrand \eqref{ell1} and \eqref{calS1} we change $z_*\to z_*-\varepsilon$

We have
\bea
\frac{\mathrm{d} F_{Q\bar{Q}}}{\mathrm{d} z_{*}}&=& \frac{2\, M(z_{*})}{\sqrt{1-\frac{{\cal V}^2(z_*)}{{\cal
        V}^2(z_*-\varepsilon)}}},
\\
\frac{\mathrm{d} \ell}{\mathrm{d} z_{*}}&=& \frac{ 2}{\sqrt{{\cal F}(z_*)}} \
\frac{1}{\sqrt{\frac{{\cal V}^2(z_*-\varepsilon)}{{\cal V}^2(z_*)} -1 }},
\eea

\be
\frac{3}{4} \ell^{2} \frac{\mathrm{d} F_{Q\bar{Q}}}{\mathrm{d} \ell}=\frac{3 \ell^{2}}{4}\frac{\mathrm{d} F_{Q\bar{Q}}}{\mathrm{d} z_{*}} \frac{\mathrm{d} z_{*}}{\mathrm{d} \ell}=\frac{3 \ell^{2}}{4}\frac{\mathrm{d} F_{Q\bar{Q}}}{\mathrm{d} z_{*}} \left(\frac{\mathrm{d} \ell}{\mathrm{d} z_{*}}\right)^{-1}=\frac{3 \ell^{2}}{4} {\cal V}(z_*(\ell)).\ee

\be \label{alpha1}
\alpha_{\mathrm{Q\bar{Q}}}\equiv
\lim _{\ell\to 0} \frac{3 \ell^{2}}{4} {\cal V}(z_*(\ell))
\ee

$$\,$$\newpage


\end{document}